\documentclass[twocolumn]{aastex631} 

\usepackage{natbib}

\usepackage{comment}
\usepackage{tabularx}
\usepackage{capt-of}
\usepackage{hhline}
\usepackage{amsmath}	
\usepackage{amssymb}	
\usepackage{textcomp}
\usepackage{upgreek}
\usepackage{graphicx}
\usepackage{txfonts}
\usepackage[]{hyperref}
\usepackage{xcolor}
\definecolor{DarkCyan}{rgb}{0,0.5,0.5}

\definecolor{teal}{HTML}{008080}

\begin{document}

   \title{Impact of initial mass function on the chemical evolution of high-redshift galaxies} 

   \author[0000-0002-4966-7450]{Boyuan Liu}
   \affiliation{Institut f\"ur Theoretische Astrophysik, Zentrum f\"ur Astronomie, Universit\"at Heidelberg, Albert Ueberle Str. 2, D-69120 Heidelberg, Germany}
   
   \author[0000-0001-8799-2548]{Michela Mapelli}
   \affiliation{Institut f\"ur Theoretische Astrophysik, Zentrum f\"ur Astronomie, Universit\"at Heidelberg, Albert Ueberle Str. 2, D-69120 Heidelberg, Germany}
   \affiliation{Interdiszipli\"ares Zentrum f\"ur Wissenschaftliches Rechnen, Universit\"at Heidelberg, D-69120 Heidelberg, Germany}
   \affiliation{INFN-Padova, Via Marzolo 8, I–35131 Padova, Italy}
   \affiliation{Dipartimento di Fisica e Astronomia Galileo Galilei, Università di Padova, Vicolo dell’Osservatorio 3, I–35122 Padova, Italy}
    \affiliation{INAF - Osservatorio Astronomico di Padova, Vicolo dell’Osservatorio 5, I-35122, Padova, Italy}

   \author[0000-0003-0212-2979]{Volker Bromm}
   \affiliation{Department of Astronomy, University of Texas, Austin, TX 78712, USA}
   \affiliation{Weinberg Institute for Theoretical Physics, University of Texas, Austin, TX 78712, USA}

    \author[0000-0002-0560-3172]{Ralf S. Klessen}
    \affiliation{Institut f\"ur Theoretische Astrophysik, Zentrum f\"ur Astronomie, Universit\"at Heidelberg, Albert Ueberle Str. 2, D-69120 Heidelberg, Germany}
    \affiliation{Interdiszipli\"ares Zentrum f\"ur Wissenschaftliches Rechnen, Universit\"at Heidelberg, D-69120 Heidelberg, Germany}
    \affiliation{Harvard-Smithsonian Center for Astrophysics, 60 Garden Street, Cambridge, MA 02138, USA}
    \affiliation{Elizabeth S. and Richard M. Cashin Fellow at the Radcliffe Institute for Advanced Studies at Harvard University, 10 Garden Street, Cambridge, MA 02138, USA}

    \author[0000-0003-3127-922X]{Lumen Boco}
    \affiliation{Institut f\"ur Theoretische Astrophysik, Zentrum f\"ur Astronomie, Universit\"at Heidelberg, Albert Ueberle Str. 2, D-69120 Heidelberg, Germany}

    \author[0000-0001-6742-8843]{Tilman Hartwig}
    \affiliation{Application Lab for AI and Big Data, German Environment Agency, Alte Messe 6, 04103 Leipzig, Germany}

    \author[0000-0001-6708-1317]{Simon C. O. Glover}
    \affiliation{Institut f\"ur Theoretische Astrophysik, Zentrum f\"ur Astronomie, Universit\"at Heidelberg, Albert Ueberle Str. 2, D-69120 Heidelberg, Germany}

    \author[0000-0002-6111-2570]{Veronika Lipatova}
    \affiliation{Institut f\"ur Theoretische Astrophysik, Zentrum f\"ur Astronomie, Universit\"at Heidelberg, Albert Ueberle Str. 2, D-69120 Heidelberg, Germany}

    \author[0000-0002-6213-6988]{Guglielmo Costa}
    \affiliation{Dipartimento di Fisica e Astronomia Galileo Galilei, Università di Padova, Vicolo dell’Osservatorio 3, I–35122 Padova, Italy}
    \affiliation{INAF - Osservatorio Astronomico di Padova, Vicolo dell’Osservatorio 5, I-35122, Padova, Italy}

    \author[0000-0003-0757-8334]{Marco Dall'Amico}
    \affiliation{Institut f\"ur Theoretische Astrophysik, Zentrum f\"ur Astronomie, Universit\"at Heidelberg, Albert Ueberle Str. 2, D-69120 Heidelberg, Germany}
    \affiliation{INFN-Padova, Via Marzolo 8, I–35131 Padova, Italy}
   \affiliation{Dipartimento di Fisica e Astronomia Galileo Galilei, Università di Padova, Vicolo dell’Osservatorio 3, I–35122 Padova, Italy}

   \author[0000-0003-0293-503X]{Giuliano Iorio}
   \affiliation{Departament de Física Quàntica i Astrofísica, Institut de Ciències del Cosmos, Universitat de Barcelona, Martí i Franquès 1, E-08028 Barcelona, Spain}

   \author[0000-0001-5231-0631]{Kendall Shepherd}
    \affiliation{INAF - Osservatorio Astronomico di Padova, Vicolo dell’Osservatorio 5, I-35122, Padova, Italy}
   \affiliation{SISSA, via Bonomea 365, I–34136 Trieste, Italy}

   \author[0000-0002-7922-8440]{Alessandro Bressan}
   \affiliation{SISSA, via Bonomea 365, I–34136 Trieste, Italy}
  
    \email{boyuan.liu.astro@gmail.com}


\begin{abstract}
   Recent observations by the James Webb Space Telescope (JWST) have found evidence for an invariant relation between stellar mass, metallicity, and star formation rate up to $z\sim 8$ and its breakdown at higher redshifts. Understanding the underlying physics driving such correlations is thus crucial. Here, we explore the impact of the initial mass function (IMF) on the chemical evolution of high-redshift galaxies. Indeed, star formation and metal enrichment in galaxies are regulated by supernova (SN) explosions and metal yields from massive stars, which are sensitive to the high-mass end of the IMF. Using the semi-analytical galaxy evolution code \textsc{a-sloth}, we follow galactic baryon cycles along merger trees built from a high-resolution cosmological simulation. Stellar feedback is modeled with up-to-date stellar evolution tracks covering the full metallicity range ($Z \sim 10^{-11} - 0.03$) and a broad stellar mass range ($m_\star\sim2 - 600\ \rm M_\odot$), including metal yields from stellar winds, core-collapse SNe, (pulsational) pair-instability SNe, and Type Ia SNe. Assuming a Kroupa-like IMF with a varying upper mass limit $m_{\max}$, we find that only models with $m_{\max} \gtrsim 200\ \rm M_\odot$ can simultaneously reproduce the observed mass-metallicity-star formation rate relation and cosmic star formation history at $z\gtrsim 4$ owing to enhanced metal yields from pair-instability SNe. Our results confirm that very massive ($\gtrsim 200\ \rm M_\odot$) stars and pair-instability SNe play an important role in the star formation and chemical enrichment histories of high-$z$ galaxies. They also have profound implications for electromagnetic transients and gravitational-wave events.
\end{abstract}


%

\section{Introduction}
\label{sec:intro}

The initial mass function (IMF) plays a pivotal role in astrophysics. It serves as a key connection between star formation and stellar evolution at small scales and galactic dynamics and baryon cycles at large scales, which makes it one of the most widely used functions in both theoretical and observational studies \citep[][]{Kroupa2021,Kroupa2024}. The high-mass end of the IMF is particularly important in galaxy evolution, since star formation and metal enrichment is strongly regulated by the feedback from massive ($\gtrsim 10\ \rm M_\odot$) stars via their radiation, winds, supernova (SN) explosions, and metal yields \citep[see reviews by, e.g.,][]{Eldridge2022,Ekstrom2025}. Besides, such massive stars, commonly formed in binary/multiple systems \citep[e.g.,][]{Sana2012,Sana2014,Moe2017}, are crucial for establishing a multi-messenger view of the Universe, since they are the progenitors of various high-energy astrophysical processes/objects including SNe, X-ray binaries, gamma-ray bursts, (binary) black holes and neutron stars and their mergers as gravitational wave sources \citep[see recent reviews by, e.g.,][]{Mapelli2021,Tauris2023,Chen2024,Marchant2024}.

Since the initial introduction of IMF by \citet{Salpeter1955} as an invariant probability density distribution function, our understanding of IMF has evolved significantly. Recent advancements in theoretical calculations \citep[e.g.,][]{Klessen2012,Schneider2012,Dopcke2013,Chon2021,Chon2024,Mathew2021,Mathew2023,Mathew2025,Guszejnov2022,Hennebelle2022,Hix2023,Liu2024,Tanvir2024} and observations \citep[e.g.,][]{Gunawardhana2011,Marks2012,Fraser2017,Ishigaki2018,Jerabkova2018,Rossi2021,Dib2023,Rusakov2023,Yan2023} provide abundant hints for the environmental dependence of IMF regarding metallicity, turbulence, cloud/cluster mass, star formation rate, gas density and temperature, radiation and magnetic fields \citep[reviewed by, e.g.,][]{Klessen2023,Hennebelle2024}. This points out the need to go beyond the assumption of a universal IMF in the mass range of $0.01-100\ \rm M_\odot$ emerging from local Galaxy-scale observations \citep[e.g.,][]{Kroupa2001,Chabrier2003}, especially for studying extragalactic stellar systems formed in diverse conditions across cosmic history. 

In fact, one explanation\footnote{There are alternative explanations with enhancements of the star formation efficiency \citep[e.g.,][]{Inayoshi2022,Dekel2023,Somerville2025,Yung2025}, variability of UV-emission/star formation \citep[e.g.,][]{Shen2023}, and stellar light-to-mass ratio \citep[by tweaking stellar evolution rather than the IMF, e.g.,][]{Donnan2025,Liu_CHE2025}, reduction of dust attenuation \citep[e.g.,][]{Ferrara2024}, and even modifications of the underlying cosmic structure formation \citep[e.g.,][]{Liu2022,Shen2024}. } to the `excess' of UV-luminous galaxies at $z\gtrsim 10$ (including those with extremely blue UV spectra) recently discovered by the James Webb Space Telescope \citep[JWST, e.g.,][]{Donnan2023full,Donnan2024,Finkelstein2023,Finkelstein2024,Harikane2023,Labbe2023,Perez-Gonzalez2023,Perez-Gonzalez2025,Adams2024,Franco2025,Naidu2025} involves enhanced abundances of massive stars through top-heavier IMFs \citep[e.g.,][]{Inayoshi2022,Cueto2024,Trinca2024,Ventura2024,Jeong2025}, which can be natural outcomes of the dense, metal-poor star-forming clouds at high $z$ \citep[e.g.,][]{Marks2012,Jerabkova2018,Chon2021}. Besides, recent JWST observations find the peculiar chemical signatures of metal enrichment from very massive ($\gtrsim 100\ \rm M_\odot$) stars such as enhanced N and C abundances in several high-$z$ galaxies \citep[e.g.,][]{Charbonnel2023,Nagele2023,Vink2023,Cameron2024,Nandal2024,Nandal2024vms,Watanabe2024,Gieles2025,Nandal2025,Schaerer2025,Naidu2025}. Low-$z$ observations and chemical evolution models also favor (top-heavy) IMFs extending to the regime of very massive stars. For instance, \citet{Goswami2021} find that the abundance ratios between $\alpha$ elements, O, and Fe in the thick disk of the MW can only be explained if the IMF extends beyond $100\ \rm M_\odot$ up to $350\ \rm M_\odot$. It is shown by \citet{Goswami2022} that a bi-modal, top-heavy IMF involving very massive stars is required to reproduce the Fe/O and N/O abundance ratios in nearby ($z\lesssim 0.03$) metal-poor, low-mass, starburst galaxies.

In light of this, we study the impact of IMF, in particular its high-mass end, on the formation and (chemical) evolution of high-$z$ galaxies using the public semi-analytical galaxy evolution code \textsc{a-sloth} \citep{Hartwig2022,Hartwig2024,Magg2022b}. {We also vary the galactic outflow parameters in \textsc{a-sloth} to shed light on their interplay with IMF. Leveraging on \textsc{a-sloth}'s computational efficiency -- a key advantage of semi-analytical models, we explore a broad parameter space for poorly understood physics of high-$z$ galaxy evolution.} 

In order to better capture bursty star formation in high-$z$ galaxies \citep[e.g.,][]{Carvajal-Bohorquez2025,Naidu2025,Rojas-Ruiz2025,Simmonds2025,Tang2025,Perry2025} and the effects of very massive stars, we update the prescriptions for star formation and stellar feedback in \textsc{a-sloth}, which are coupled to a comprehensive set of stellar evolution models (covering the metallicity range $Z \sim 10^{-11} - 0.03$ and stellar mass range $m_\star\sim2 - 600\ \rm M_\odot$) \citep[]{Costa2025,Lecroq2024}. 

As an initial step, we focus on the general trends of high-$z$ galaxy chemical evolution reflected in the scaling relation between stellar mass, (gas-phase) metallicity, and star formation rate (SFR). In particular, recent JWST observations have reached a consensus on an invariant stellar mass-metallicity-star formation rate relation \citep[MZSFR,][]{Nakajima2023,Curti2024,Sarkar2025} up to $z\sim 8$ and its breakdown at higher redshifts. Our goal is to jointly constrain the IMF and galactic outflow properties of high-$z$ galaxies using this MZSFR combined with the cosmic star formation history \citep[CSFH,][]{Bouwens2016,Donnan2023full,Donnan2023,Harikane2023} and galaxy-halo connection \citep{Tacchella2018} inferred from observations at $z\sim 4-10$. 

To do so, we run 165 models in \textsc{a-sloth} on the merger trees constructed from a high-resolution cosmological simulation \citep{Ishiyama2016} that resolves the smallest star-forming structures (with halo masses $\sim 10^6\ \rm M_\odot$) in the standard $\Lambda$CDM universe. {Therefore, while reproducing current observations biased by luminous objects at high $z$, our simulations self-consistently produce a large population of (unseen) faint, low-mass, metal-poor galaxies, whose predicted properties provide valuable implications and guidance for future multi-messenger observations of the high-$z$ Universe.}

The structure of our paper is as follows. In Section~\ref{sec:gal_evo}, we introduce our galaxy evolution model, focusing particularly on the updates of star formation (Sec.~\ref{sec:sf}) and stellar evolution (Sec.~\ref{sec:stellar}) prescriptions. In Section~\ref{sec:setup}, we explain the simulation setup and the parameter space explored. In Section~\ref{sec:obs}, we outline the observational constraints and define the corresponding likelihoods that quantify the agreement between observations and theoretical predictions. In Section~\ref{sec:res}, we show our main results. These include the likelihoods as functions of model parameters, through which we discuss the interplay between IMF and galactic outflows in shaping the high-$z$ galaxy populations (Sec.~\ref{sec:interplay}) and search for the best-fit model with the highest overall likelihood. The population of simulated galaxies in this model is then compared with the observed population in detail (Sec.~\ref{sec:best}). In Section~\ref{sec:diss}, we discuss the implications of our results on high-$z$ transient sources (Sec.~\ref{sec:imp}) and the uncertainties/caveats (Sec.~\ref{sec:caveats}). Finally, our main findings are summarized in Section~\ref{sec:summary}. 

\section{Galaxy evolution model}
\label{sec:gal_evo}
We use the public semi-analytical code \textsc{a-sloth} \citep{Hartwig2022,Hartwig2024} to model galaxy evolution through dark matter merger trees. In this Section, we highlight the updates made in this work with respect to \citet{Hartwig2022} on the prescriptions for star formation (Sec.~\ref{sec:sf}) and stellar feedback (Sec.~\ref{sec:fdbk}) with novel input stellar evolution data (Sec.~\ref{sec:stellar}). We refer the reader to the code release and calibration papers \citep{Magg2022b,Hartwig2022,Hartwig2024} and Appendix~\ref{apdx:asloth} for full details.  

\subsection{Updated star formation prescription}
\label{sec:sf}

\begin{figure}
    \centering
    \includegraphics[width=1\linewidth]{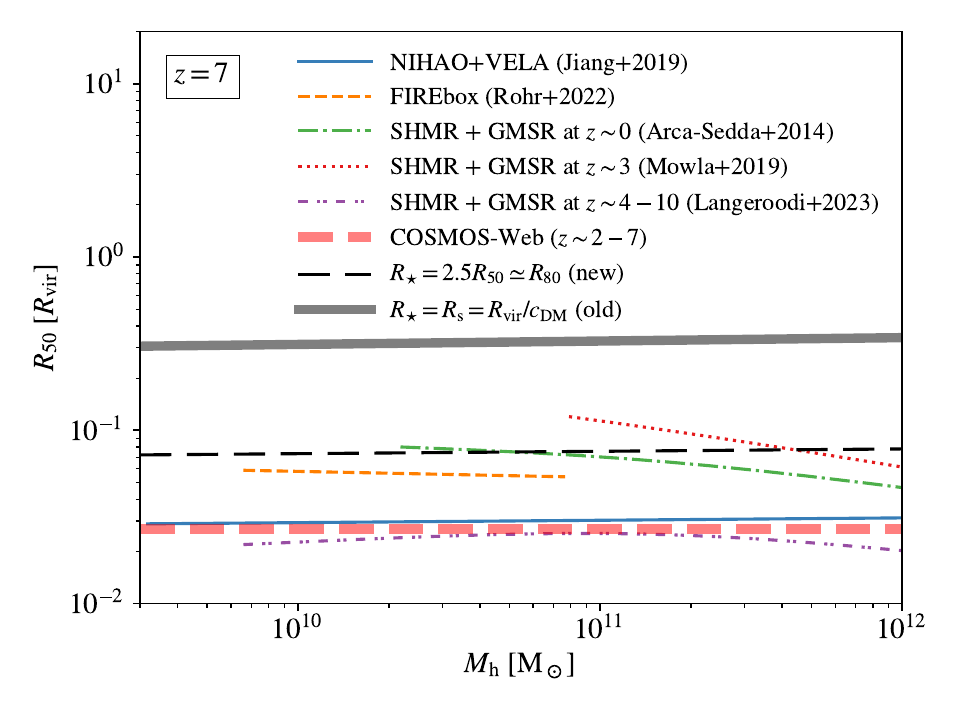}
    \caption{Relation between galaxy half-mass/light radius (in units of $R_{\rm vir}$) and halo mass at $z=7$. The solid and dashed curves show the fitting formulae of half-mass radius from the NIHAO and VELA simulations \citep{Jiang2019} and from the FIREbox simulations \citep{Rohr2022}, respectively. We also plot the empirical results for half-light radii based on the SHMR in \citet{Tacchella2018} at $z=7$ (Eq.~\ref{eq:shmr}) combined with the extrapolation of the galaxy mass-size relation from \citet[$z\sim 0$, dash-dotted]{Arca-Sedda2014}, \citet[$z\sim 3$, dotted]{Mowla2019}, and \citet[$z\sim 4-10$, dot-dash-dotted]{Langeroodi2023size}. The thick dashed line shows the median ratio of the galaxy half-light radius to halo virial radius $R_{50}/R_{\rm vir}=2.7\%$ at $z\sim 2-7$ from the COSMOS-Web survey \citep[see their fig.~10]{Yang2025}. The thick solid curve shows the old model for galaxy size $R_{\star}=R_{\rm s}$ of \citet{Hartwig2022}, while the long-dashed curve shows the new model $R_{\star}=2.5R_{50}$ adopted in this work given $R_{50}$ from \citet{Jiang2019}.}
    \label{fig:rs_mh}
\end{figure}

In \textsc{a-sloth}, star formation is modeled as an inflow cascade from the intergalactic medium (IGM) to hot gas in the halo, then to cold gas in the galaxy, and finally to stars (see Appendix~\ref{apdx:sf} for details). This process is primarily governed by a characteristic dynamical timescale $t_{\rm dyn}\sim R_\star^{3/2}/(G M_{\rm c})^{1/2}$, given the size $R_{\star}$ and mass $M_{\rm c}$ of the central region occupied by cold gas and stars. 
In the original model of \citet{Hartwig2022}, $R_{\star}$ is set to the scale radius of the halo $R_{\rm s}=R_{\rm vir}/c_{\rm DM}$, given the halo's concentration factor $c_{\rm DM}$ and virial radius $R_{\rm vir}$. However, given the same halo mass $M_{\rm h}$, $R_{\rm s}$ is much larger compared with the (stellar) half-mass/light radius $R_{50}$ of galaxies in cosmological (zoom-in) simulations and observations and (by at least a factor of 2 and up to a factor of 20 for $M_{\rm h}\sim 10^{9.5}-10^{12}\ \rm M_\odot$ at $z\sim 4-10$). 
This significant difference is illustrated in Fig.~\ref{fig:rs_mh} for $z=7$ as an example. 

To make this comparison, we consider the scaling relation $R_{50}=0.053 [R_{\rm vir}/(0.7\ \rm kpc)]^{0.894}\ kpc$ from the FIREbox simulations \citep[for $z\lesssim 5$,][]{Rohr2022} and that from the NIHAO and VELA zoom-in simulations \citep[for $z\lesssim 7$,][]{Jiang2019} $R_{50}=0.02 R_{\rm vir}(c_{\rm DM}/10)^{-0.7}(1+z)^{-0.2}$, where the factor $(1+z)^{-0.2}$ is introduced to reproduce the redshift evolution of median galaxy-halo size relation in (abundance matching) observations \citep{Somerville2018}, and $c_{\rm DM}(M_{\rm h})$ is derived using the fitting formula from \citet[see their appendix~B1]{Correa2015}. For observational constraints on the (2D) half-light radius, we adopt the stellar-halo mass relation (SHMR) from the semi-empirical model in \citet{Tacchella2018}
\begin{align}
    M_{\star,\rm SHMR}=\frac{0.05[(1+z)/5]^{-0.6}M_{\rm h}}{(M_{\rm h}/M_{\rm 0})^{-1}+(M_{\rm h}/M_{\rm 0})^{0.3}}\ , \label{eq:shmr}
\end{align}
where $M_{\rm 0}=1.6\times 10^{11}\ \rm M_\odot$. We first derive the stellar mass as $M_{\star}=M_{\star,\rm SHMR}(M_{\rm h},z)$, which is then substituted to a given galaxy mass-size relation from observations to estimate $R_{50}(M_{\star}(M_{\rm h}))$. For the latter, we consider the fitting formulae from \citet[for $z\sim 0$]{Arca-Sedda2014}, \citet[for $z\lesssim 3$]{Mowla2019}, and \citep[for $z\sim 4-10$]{Langeroodi2023size}. We also consider the recent measurements of $R_{50}/R_{\rm vir}$ from the COSMOS-Web survey \citep[for $z\sim 2-7$]{Yang2025}. Note that these observational data may not necessarily cover the high-$z$ regime ($z\sim 4-10$) that we are concerned with. In this case, we simply make extrapolations by assuming that the relation for the highest redshift bin applies to higher redshifts.

The comparison in Fig.~\ref{fig:rs_mh} indicates that $R_{\rm s}$ may not be a good choice for $R_{\star}$. Therefore, for atomic-cooling halos (with virial temperatures $T_{\rm vir}>10^{4}$~K), we instead set $R_{\star}$ to $2.5 R_{50}$ as an approximation of the radius $R_{80}$ that encloses 80\% of the total stellar mass \citep{Mowla2019}: 
\begin{align}
    R_{\star}=2.5R_{50}=0.5 R_{\rm vir}(c_{\rm DM}/10)^{-0.7}(1+z)^{-0.2}\ .\label{eq:rs}
\end{align}
Here, we adopt the fitting formula from \citet{Jiang2019} to calculate $R_{50}$ as a conservative estimate among the references considered above, which also covers the largest dynamical range of $M_{\rm h}$ and agrees well with the median ratio $R_{50}/R_{\rm vir}=0.027$ of galaxies at $z\sim 2-7$ observed in the COSMOS-Web survey \citep[see their fig.~10]{Yang2025}. 

In the new model, $R_{\star}$ is reduced by a factor of $\sim 4$ compared with the old model. The immediate consequence of this update is that the scatter around the star-formation main sequence (SFMS) is enhanced by a factor of $\sim 2$, achieving a better agreement with observations, as discussed in Appendix~\ref{apdx:sfms}.

\subsection{Galactic outflows and metal enrichment}
\label{sec:fdbk}
The feedback of individual stars with $m_{\star}>5\ \rm M_\odot$ (called `massive stars' henceforth) are tracked in \textsc{a-sloth}, including metal enrichment, photoheating from ionizing photons, and galactic outflows driven by SN explosions. Here, we briefly outline the SN feedback model and the treatment of metal enrichment that are most relevant for galaxy chemical evolution. The reader is referred to Appendix~\ref{apdx:fdbk} for details. 

Given the total energy of SN explosions $E_{\rm SNe}$ within a timestep, we remove a fraction $\sim \gamma_{\rm out}E_{\rm SNe}/E_{\rm bind}$ of gas from the galaxy as outflows, where $E_{\rm bind}$ is the binding energy of gas\footnote{In practice, we calculate the outflow fraction of cold and hot gas separately by distributing the total SN energy among these two components (see Appendix~\ref{apdx:fdbk}).}, and $\gamma_{\rm out}$ is the outflow efficiency. Following \citet{Chen2022}, we adopt a phenomenological model for $\gamma_{\rm out}$ as
\begin{align}
    \gamma_{\rm out}=(M_{\rm h,peak}/M_{\rm out0})^{\alpha_{\rm out}}\ ,\label{eq:gamma_out}
\end{align}
where $M_{\rm h,peak}$ is the peak (virial) halo mass, and the characteristic mass $M_{\rm out0}$ and power-law slope $\alpha_{\rm out}$ are free parameters to be constrained by observations together with IMF (Sec.~\ref{sec:setup}). For $\alpha_{\rm out}>0$, $M_{\rm out0}$ defines the halo mass scale above (below) which galactic outflows are suppressed (enhanced) by the deeper (shallower) potential wells of larger (smaller) halos. The strength of this halo mass dependence is governed by $\alpha_{\rm out}$. 
Since $M_{\rm out0}$ becomes meaningless in Eq.~\ref{eq:gamma_out} when $\alpha_{\rm out}=0$, we further extend this model by
\begin{align}
    \gamma_{\rm out}=10^{10}{\rm\ M_\odot}/M_{\rm out0}\quad {\rm if} \quad \alpha_{\rm out}=0\ ,\label{eq:gamma_out_0}
\end{align}
such that increasing $M_{\rm out0}$ always enhances galactic outflows for $\alpha_{\rm out}\ge 0$, and increasing $\alpha_{\rm out}$ boosts (reduces) outflows in small (large) halos with $M_{\rm h,peak}$ below (above) $M_{\rm out0}$ for $\alpha_{\rm out}>0$. 

For metal enrichment, we first add up the metal yields from dying stars and metals accreted from the IGM to obtain the total metal budget in the halo. Next, we calculate the mass of metals retained in the halo by assuming that metals are mixed uniformly into the current gas reservoir of the halo, including newly launched outflows and the remaining gas as interstellar medium (ISM) and circumgalactic medium (CGM). The retained metal mass is then divided by the total mass of remaining gas to give the average gas-phase metallicity of the galaxy $Z_{\rm ave}$. Finally, we apply a randomly generated shift factor $d\log Z$ \citep[][]{Tarumi2020} to estimate the metallicity of newly-formed stars as $Z_\star = Z_{\rm ave}\times 10^{d\log Z}$ to statistically capture the effect of inhomogeneous metal enrichment (see Appendix~\ref{apdx:fdbk} for details).

\subsection{Stellar populations}
\label{sec:stellar}
Stellar feedback is governed by the properties of massive stars, namely, the production rate of ionizing photons $\dot{q}_{\rm ion}$, lifetime $t_{\star}$, metal yield $m_Z$, and SN energy ejection $e_{\rm SN}$ (see Sec.~\ref{apdx:fdbk}). In \textsc{a-sloth}, stars are classified into two populations: extremely metal-poor/free Population~III (Pop~III) and metal-enriched Population~II (Pop~II), depending on the chemical composition of star-forming gas (see eq.~17 in \citealt{Hartwig2022}). For both populations, the stellar properties are derived from the initial stellar mass $m_\star$ using fitting formulae (for $\dot{q}_{\rm ion}$ and $t_{\star}$) and nearest-neighbor interpolation over pre-computed tables (for $m_Z$ and $e_{\rm SN}$) from single star evolution models. 

During star formation events, $m_\star$ is sampled from the IMF, modeled separately for Pop~III and Pop~II stars. 
Pop~III stars typically form in the extremely metal-poor/free regime with $Z\lesssim 10^{-5}\ \rm Z_\odot$, where $\rm Z_{\odot}=0.0142$ is the bulk solar metallicity \citep{AllendePrieto2001,Asplund2004,Asplund2009}. Their IMF is modeled as a power-law ($dN/dm_\star\propto m_\star^{-\alpha_{\rm III}}$), whose slope ($\alpha_{\rm III}$) and mass bounds ($m_{\min,\rm III}$, $m_{\max,\rm III}$) are in principle free parameters considering the diverse predictions from simulations \citep[e.g., fig.~6 in][]{Klessen2023}. In this work, we fix these parameters to the best-fit values from \citet{Hartwig2024} for simplicity: $m_{\min,\rm III}=13.6$, $m_{\max,\rm III}=197$, and $\alpha_{\rm III}=1.77$. Our focus is metal-enriched ($Z\gtrsim 10^{-5}\ \rm Z_\odot$) Pop~II stars that are inferred to be the dominant population in galaxies observed by JWST at $z\gtrsim 5$ \citep{Riaz2022,Finkelstein2023}. Unlike \citet{Hartwig2022}, where the Pop~II stars follow the \citet{Kroupa2001} IMF in a fixed mass range $m_\star\in [0.01,100]\ \rm M_{\odot}$, we allow the upper mass limit $m_{\max}$ to change between $100$ and $600\ \rm M_{\odot}$ while keeping the high-mass end slope $\alpha_{\rm II}=2.3$ and the low-mass end ($m_\star<0.5\ \rm M_\odot$) shape unchanged. For simplicity, we assume that $m_{\max}$ is invariant within each simulation, and therefore, meant to characterize the galaxy-population-averaged IMF. We defer the investigation of more complex IMF models with environmental dependence \citep[e.g.,][]{Gunawardhana2011,Marks2012,Jerabkova2018,Dib2023,Rusakov2023} to future work. We note that a shift of the IMF towards a more top-heavy nature at redshifts $\gtrsim 5$ is expected on general physical grounds, such as the elevated CMB temperature floor \citep[e.g.,][]{Larson_IMF1998}. 

In the original approach of \citet[see their sec.~2.1.3]{Hartwig2022}, one single model is used for each population of stars without any metallicity dependence within the population. In other words, there are effectively only two metallicity bins: $Z\lesssim 10^{-5}\ \rm Z_\odot$ (Pop~III) and $Z\gtrsim 10^{-5}\ \rm Z_\odot$ (Pop~II). The fitting formulae for $\dot{q}_{\rm ion}$ and $t_{\star}$ are taken from \citet{Schaerer2002} and \citet{stahler2008formation}. A universal function of $e_{\rm SN}(m_\star)$ is adopted for both populations of stars, where core-collapse SNe explode with $e_{\rm SN}=10^{51}\ \rm erg\ s^{-1}$ from stars with $m_\star\in [10,40]\ \rm M_\odot$, while stars in the mass range $m_\star\in [140,260]\ \rm M_\odot$ produce pair-instability SNe with $e_{\rm SN}=3.3\times 10^{52}\ \rm erg\ s^{-1}$. The relevant SN metal yield tables are provided by \citet{Nomoto2013} and \citet{Kobayashi2006} for Pop~III and Pop~II stars, respectively. 

In this work, we instead adopt a denser grid of stellar evolution models\footnote{\url{https://stev.oapd.inaf.it/PARSEC/}} computed with the \textsc{parsec} code v2.0 \citep{Bressan2012,Costa2019,Nguyen2022}, which covers the full metallicity range $Z \sim 10^{-11} - 0.03$ with 13 bins and a broad mass range $m_\star\sim 2 - 600\ \rm M_\odot$ \citep[see their table~1]{Costa2025}\footnote{\citet{Costa2025} also provide tracks of more massive stars with $m_\star\sim 600-2000\ \rm M_\odot$ for the lowest three metallicity bins $Z=10^{-11}$, $10^{-6}$, and $10^{-4}$. These models are not considered, as we focus on the regime with $m_{\max}\le 600\ \rm M_\odot$. Very massive ($\gtrsim 600\ \rm M_\odot$) stars beyond this regime can also be important for the (chemical) evolution of high-$z$ galaxies \citep[e.g.,][]{Nagele2023,Vink2023,Nandal2024,Nandal2024vms,Nandal2025,Schaerer2025}, which is a promising topic for follow-up studies.}. {The reader is referred to Sec.~2 of \citet{Costa2025} for a detailed description of the input physics underlying these \textsc{parsec} models.} The $Z=10^{-11}$ model is applied to metal-free ($Z=0$) Pop~III stars, while the rest describe extremely metal-poor Pop~III stars\footnote{Pop~III stars are assigned to the lowest two metallicity bins $Z=10^{-11}$ and $10^{-6}$ in \citet{Costa2025} by nearest neighbor interpolation (in the $\log Z$ space) with a middle point $Z=10^{-8.5}$. This is motivated by the fact that stellar evolution converges to the metal-free mode for $Z\lesssim 10^{-9}$ \citep[see, e.g., fig.~1 in][]{Larkin2023}. Here, our definition of Pop~III stars is based on the physics of star formation that drive the transition of IMF from the bottom-heavy Pop~II regime to the top-heavy Pop~III regime around a critical metallicity $Z\sim 10^{-7}$ \citep{Chiaki2017,Chon2021}, which is broader than completely metal-free stellar evolution.} ($Z\sim 10^{-8.5}-10^{-7}$) and Pop~II stars\footnote{Stars with $Z\gtrsim 0.006$ are usually called Population~I \citep{Costa2025}. Here, we still classified them as Pop~II for simplicity.} ($Z\gtrsim 10^{-7}$). 

The fates of these stars as well as the corresponding metal yields are derived using the methodology in \citet{Goswami2021} and \citet{Goswami2022}. Specifically, 
stars with zero-age main sequence 
mass $m_{\star}>5\ \rm M_\odot$ end their life as white dwarfs, electron-capture SNe, core-collapse SNe, failed SNe, pulsational pair-instability SNe (PPISNe), pair-instability SNe (PISNe), or direct-collapse black holes. Here, white dwarfs, failed SNe and direct-collapse black holes have negligible energy ejection ($e_{\rm SN}$) at birth. We adopt the typical values of $e_{\rm SN}$ from SN explosion simulations for the other fates: $e_{\rm SN}=10^{51}\ \rm erg\ s^{-1}$ for electron-capture SNe and core-collapse SNe, $e_{\rm SN}=3.3\times 10^{51}\ \rm erg\ s^{-1}$ for PPISNe, and $e_{\rm SN}=3.3\times 10^{52}\ \rm erg\ s^{-1}$ for PISNe. Unlike the old models of \citet{Hartwig2022} that only consider metal yields from SNe, our new models also include the contributions of stellar winds, which can also affect the chemical abundances significantly \citep[e.g.,][]{Cescutti2010,Farmer2021,Liu2021,Jeena2023,Vink2023,Higgins2023,Higgins2025,Ma2025}. 
{The metal yields are computed by adding up the wind ejecta (integrated from the zero-age main sequence to the final configuration) and the newly synthesized elements during SN events. For the latter, we use the yield tables from \citet[]{Wanajo2009} for electron-capture SNe, \citet{Limongi2003} and \citet{Chieffi2004} for core-collapse SNe, \citet[]{Woosley2017} for PPISNe, and \citet[]{Heger2002} for PISNe through interpolation in helium or carbon-oxygen core mass \citep[][see their sec.~4 for details]{Goswami2021}. } 
Our yields are mildly different from those already published by \citet[][see, e.g., their Fig. 7]{Costa2025} even if they are derived from the same \textsc{parsec} evolutionary tracks. The reason is that we calculate the compact-remnant mass during core-collapse SNe with the delayed formalism by \cite{Fryer2012}, whereas \cite{Costa2025} adopt the explodability models from \cite{Limongi2003} and \cite{Chieffi2004}. The main difference is that at intermediate and high metallicity ($Z\gtrsim{}0.005$) our models predict a larger zero-age main-sequence mass range for direct collapse black holes with respect to successful core-collapse SNe \citep{Spera2019}. This difference has negligible impact on the results presented here.

\begin{figure}
    \centering
    \includegraphics[width=1.0\linewidth]{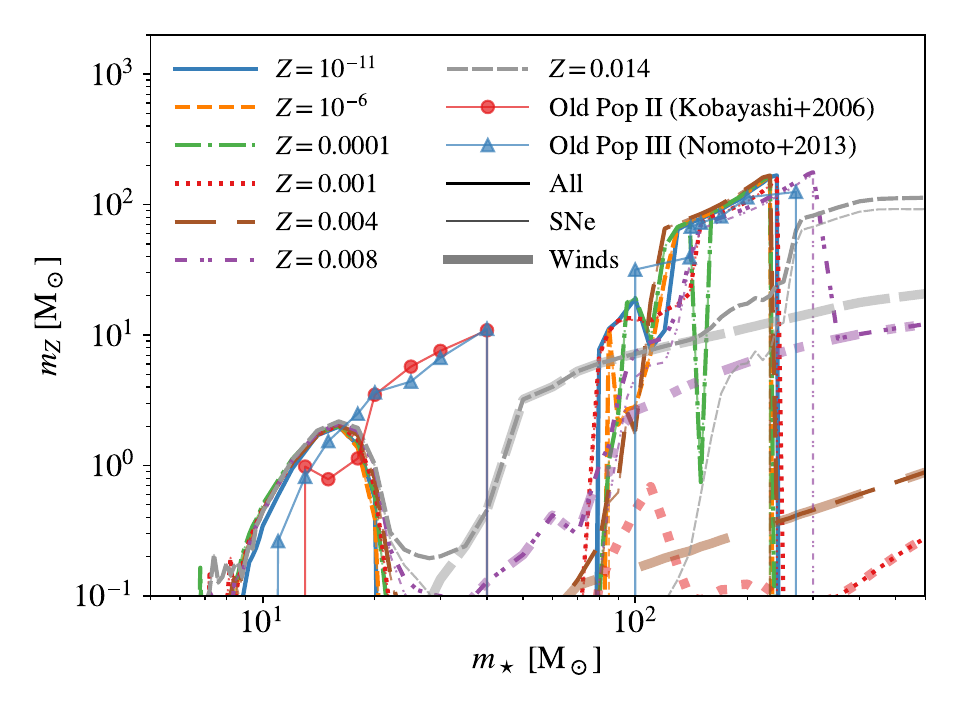}\\
    \includegraphics[width=1.0\linewidth]{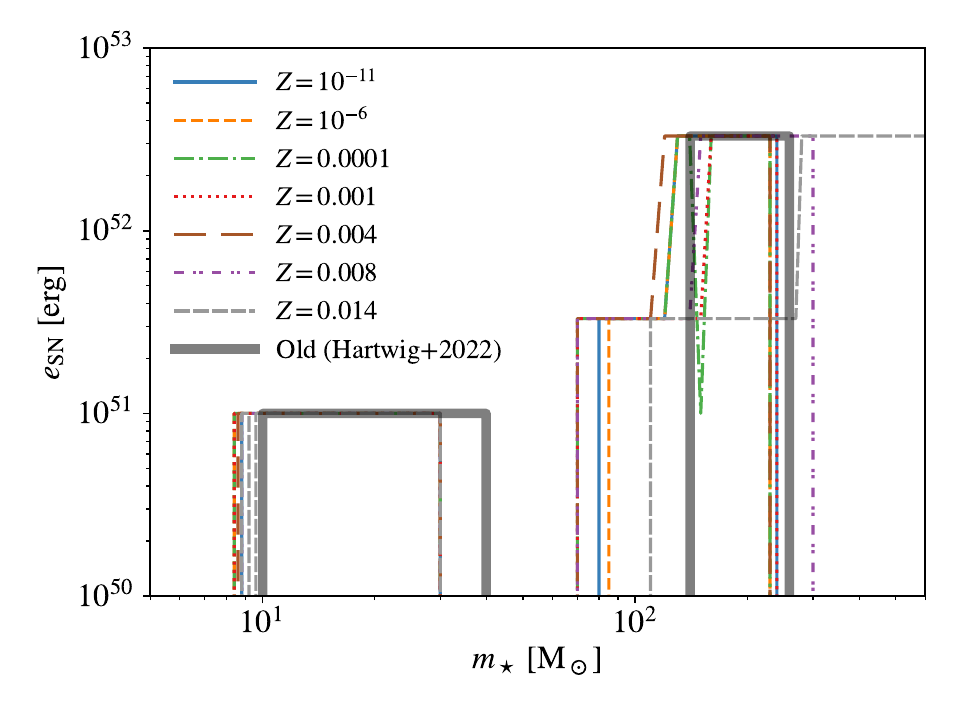}
    \caption{Metal yields (top) and SN energy (bottom) as functions of initial stellar mass for $Z=10^{-11}$ (solid), $10^{-6}$ (dashed), 0.0001 (dash-dotted), 0.001 (dotted), 0.004 (long-dashed), 0.008 (dot-dash-dotted), and 0.014 (densely-dashed). In the top panel, the yields from SNe, winds, and all combined are shown with the thin, thick, and intermediate curves. The thin curves marked by circles and triangles show the old SN yields adopted by \citet{Hartwig2022} for Pop~II and Pop~III stars from \citet[for $Z=0.001$]{Kobayashi2006} and \citet[for $Z=0$]{Nomoto2013}, respectively. Here, \citet{Kobayashi2006} only consider stars in the mass range $m_\star\sim 10-40\ \rm M_\odot$, and zero yields are assumed for stars outside this range by \citet{Hartwig2022}. 
    In the bottom panel, the old SN energy model of \citet{Hartwig2022} is shown with the thick solid curve.}
    \label{fig:mzesn_m}
\end{figure}

\begin{figure}
    \centering
    \includegraphics[width=1.0\linewidth]{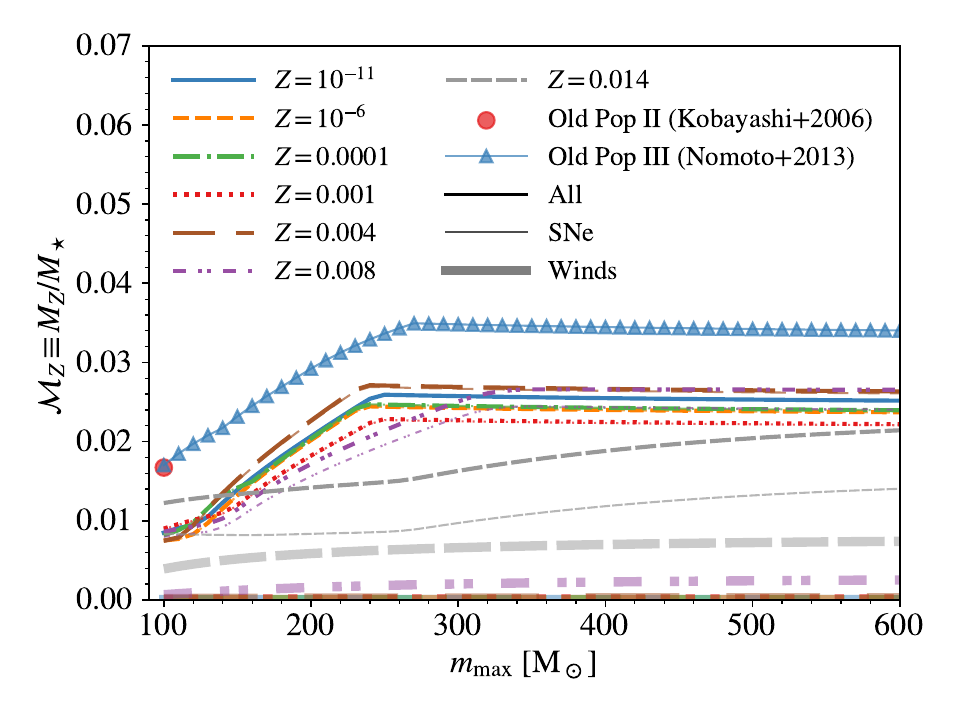}\\
    \includegraphics[width=1.0\linewidth]{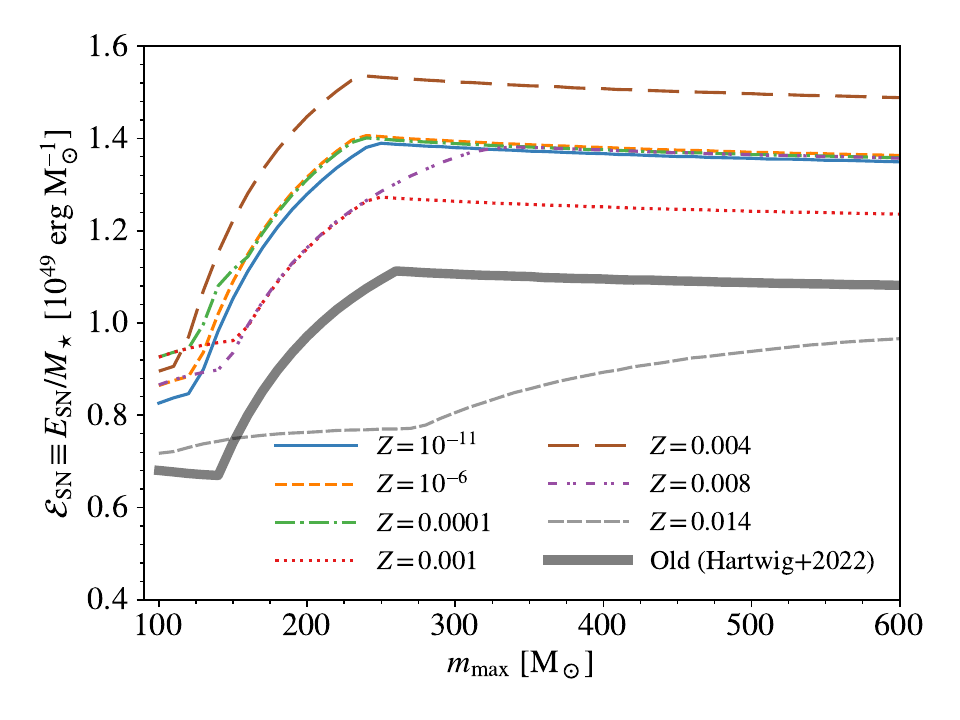}
    \caption{Total metal yield {from stars with $m_\star>5\ \rm M_\odot$} (top) and SN energy output (bottom) per unit stellar mass as functions of IMF upper mass limit for $Z=10^{-11}$ (solid), $10^{-6}$ (dashed), 0.0001 (dash-dotted), 0.001 (dotted), 0.004 (long-dashed), 0.008 (dot-dash-dotted), and 0.014 (densely-dashed). The IMF shape is assumed to follow \citet{Kroupa2001} with the high-mass end described by a power-law $dN/dm_\star\propto m_\star^{-2.3}$. In the top panel, the yields from SNe, winds, and all combined are shown with the thin, thick, and intermediate curves. The big circle shows the result for the old Pop~II model of \citet{Hartwig2022} at $m_{\max}=100\ \rm M_\odot$ based on the SN yields from \citet[for $Z=0.001$]{Kobayashi2006}. The thin curve marked by triangles shows the results for the old Pop~III model of \citet{Hartwig2022} based on the SN yields from \citet[for $Z=0$]{Nomoto2013}. In the bottom panel, the results for the old SN energy model of \citet{Hartwig2022} are shown with the thick solid line.}
    \label{fig:mzesn_mmax}
\end{figure} 

In Fig.~\ref{fig:mzesn_m}, we compare our new models for $m_Z(m_\star)$ (top panel) and $e_{\rm SN}(m_{\star})$ (bottom panel) at seven select metallicities with the old models of \citet{Hartwig2022}. Clearly, both $m_Z$ and $e_{\rm SN}$ show non-negligible metallicity dependence in our new models. Besides, the variations with $Z$ are complex and non-monotonic in the metal-poor regime relevant for Pop~III and Pop~II stars with $Z\lesssim 0.004$. For more metal-rich stars, the effects of stellar winds become important, shifting the curves to the right (larger $m_\star)$ with reduced pre-SN masses. 
The main difference is that in the new models, the metal yields from core-collapse SNe are significantly reduced by fall-back at $m_{\star}\sim 20-70\ \rm M_\odot$, while this fall-back dominated regime occurs at $m_{\star}\sim 40-100\ \rm M_\odot$ in \citet{Hartwig2022}. Besides, our models consider electron-capture SNe from $m_\star\sim 8-10\ \rm M_\odot$ and PPISNe from $m_{\star}\sim 70-120\ \rm M_\odot$ that are absent in the old models. 

To better demonstrate the difference, in Fig.~\ref{fig:mzesn_mmax}, we show the corresponding IMF-integrated total metal yield (top panel) and SN energy output (bottom panel) per unit stellar mass as functions of $m_{\max}$. By allowing $m_{\max}$ to increase beyond $100~\rm M_\odot$ and taking into account stellar winds, {the IMF-averaged metal yield $\mathcal{M}_{Z}$ can be enhanced by up to a factor of $\sim 3.4$ ($1.6$) compared with the new (old) model for $m_{\max}=100\ \rm M_\odot$. In our new models, winds contribute $\sim 1-40\%$ of the total metal yield for $Z\sim 0.004-0.014$. Note that we only consider stars with $m_\star>5\ \rm M_\odot$ throughout this work. Including smaller stars will increase the overall yield moderately by $\Delta\mathcal{M}_{Z}\lesssim 0.006$, which is much weaker than the effect of extending the IMF ($\Delta\mathcal{M}_{Z}\sim 0.02$), and therefore, not expected to change our conclusions.} The total energy output from SNe is generally higher in our models by $\sim 20-80\%$ due to the contributions of electron-capture SNe and PPISNe.  

Beyond the aforementioned improvements of the metallicity and mass grids, SN types, and wind metal yields from single star evolution, we further implement a phenomenological model for Type Ia SNe following \citet{Deng2024}, as detailed in Appendix~\ref{apdx:snia}. We adopt the updated fitting formulae for $\dot{q}_{\rm ion}$ and $t_{\star}$ from \citet{Klessen2023} following \citet{Hartwig2024} to consider more detailed metallicity dependence. The new models are similar to the old ones adopted in \citet[see their sec.~2.1.3]{Hartwig2022} in most cases. We have verified by numerical experiments that the updates in $\dot{q}_{\rm ion}$ and $t_{\star}$ have minor effects on our results, but we plan to improve 
the treatments of these parameters in future work, based on the results of \citet{Lecroq2024}. 

\section{Simulation setup}
\label{sec:setup}

We apply \textsc{a-sloth} to the merger trees constructed from the cosmological simulation by \citet{Ishiyama2016} in a co-moving volume of $V_{\rm com}\simeq (8\ h^{-1}\rm cMpc)^{3}\simeq 1750\ cMpc^{3}$ with a dark matter mass resolution of $5000\ h^{-1}\rm M_{\odot}$. The merger trees span the redshift range $z\sim 4.5-30$ in which the box is marginally large enough to be cosmologically representative. We are particularly interested in the regime $z\sim 4-10$ probed by recent JWST observations. In our simulations, this redshift range is covered by 30 snapshots with global timesteps $\Delta t_{j}\sim 17-46~\ \rm Myr$. 
Our simulation box is rather small as a tradeoff for the high resolution required to capture the first star-forming halos ($M_{\rm h}\sim 10^{6}\ \rm M_\odot$), such that the statistics of the most massive/luminous objects that favor over-dense regions is poor. Therefore, we focus on the scaling relations between galaxy properties and cosmic integrated quantities (e.g., star formation rate density) rather than the distributions of specific galaxy properties (e.g., UV luminosity function and stellar mass function), as discussed in Sec.~\ref{sec:obs}. 

\begin{table}[htbp]
    \centering
    \caption{Summary of key free parameters in \textsc{a-sloth}. The first section shows the parameters fixed throughout this work based on the best-fit values from \citet[see their table~1]{Hartwig2024}. The second shows the parameters explored in plausible ranges to reproduce observations of high-$z$ galaxies. }
    \begin{tabular}{ccc}
    \hline
        Parameter & Description & Value(s)/range \\
    \hline
        $m_{\max,\rm III}\ [\rm M_\odot]$ & Max. mass of Pop III stars & 197 \\
        $m_{\min,\rm III}\ [\rm M_\odot]$ & Min. mass of Pop III stars & 13.6 \\
        $\alpha_{\rm III}$ & Pop~III IMF slope & 1.77 \\
        $\eta_{\rm III}$ & Pop~III SFE & 8.15 \\
        $\eta_{\rm II}$ & Pop~II SFE & 0.237 \\
        $v_{\rm sv}/\sigma_{\rm sv}$ & Baryon streaming velocity & 0.8 \\
        $c_{\rm ZIGM}$ & IGM $Z$ clumping factor & 3.32 \\
    \hline 
        $\alpha_{\rm out}$ & Slope of outflow efficiency & 0, 0.5, 1 \\
        $M_{\rm out0}\ [\rm M_\odot]$ & Norm. of outflow efficiency &  $[1,5]\times 10^9$ \\
        $m_{\max}\ [\rm M_\odot]$ & Max. mass of Pop II stars & $[100, 600]$ \\
    \hline
    \end{tabular}
    \label{tab:param}
\end{table}

Throughout this work, we adopt the cosmological parameters from \citet{planck}{: $\Omega_{m}=0.3089$, $\Omega_{b}=0.0486$, 
and $H_{0}=100h\ \rm km\ s^{-1}\ Mpc^{-1}$ with $h=0.6774$}. The streaming velocity between baryons and dark matter $v_{\rm sv}$ is set to the most probable value $0.8\sigma_{\rm sv}$, where $\sigma_{\rm sv}$ is the root-mean-square velocity \citep{Schauer2019}. The key free parameters for star formation and feedback are summarized in Table~\ref{tab:param}. {Here, we focus on three parameters that directly regulate star formation and chemical evolution of high-$z$ galaxies: $\alpha_{\rm out}$, $M_{\rm out0}$, and $m_{\max}$, while keeping the other parameters fixed to the best-fit values from \citet{Hartwig2024}.}

We run $3\times5\times11=165$ simulations in total on a grid defined by $\alpha_{\rm out}=0$, 0.5, and 1, $M_{\rm out0}\in [1,5]\times10^{9}\ \rm M_\odot$ with linear spacing $\Delta M_{\rm out0} = 10^{9}\ \rm M_\odot$, and $m_{\max}\in [100,600]\ \rm M_\odot$ with $\Delta m_{\rm max}=50\ \rm M_\odot$. The values of $\alpha_{\rm out}$ and $M_{\rm out0}$ considered here differ significantly from those ($\alpha_{\rm out}\sim 1.8-4$, $M_{\rm out0}\sim 6-11\times 10^{9}\ \rm M_\odot$) preferred by observational constraints found by \citet{Hartwig2024}. This is partially due to the updates of star formation and stellar feedback schemes in our work (see Sec.~\ref{sec:sf} and \ref{sec:fdbk}). With these updates, in principle, a new systematic calibration of all parameters is required, which is beyond the scope of this paper. 

The main reason for the different choices of $\alpha_{\rm out}$ and $M_{\rm out0}$ is that our work focuses on observables of galaxies/halos at $z\gtrsim 4$ (see below), while most (6 out of 9) observables considered by \citet{Hartwig2024} come from the Milky Way at $z=0$. We have found by numerical experiments that the favored values of $\alpha_{\rm out}$ and $M_{\rm out0}$ in \citet{Hartwig2024} cannot reproduce the high-$z$ observations considered here, even for the old star formation model given maximum metal yields (with $m_{\max}\sim 250\ \rm M_\odot$). They generally fail to form enough stars and produce enough metals. This implies that galactic outflow parameters can evolve significantly from $z\gtrsim 4$ to $z\sim 0$, as high-$z$ dwarf galaxies (at least those detected by JWST) are less vulnerable to stellar feedback than their local counterparts due to their compact nature. In fact, it is implied by our simulation results (see Sec.~\ref{sec:best} and Appendix~\ref{apdx:csfh_alt}) that redshift evolution of outflow parameters at $z\lesssim 6$ is required to fully reproduce the cosmic star formation history inferred from observations, which favor increasing (decreasing) outflow efficiency at lower redshifts in relatively small (large) halos with $M_{\rm h}\lesssim(\gtrsim)\ M_{\rm out0}$.

\section{Observational constraints}
\label{sec:obs}
We consider observational constraints from three aspects: cosmic star formation history (Sec.~\ref{sec:sfh}), galaxy chemical evolution (Sec.\ref{sec:mzsfr}), and galaxy-halo connection (Sec.\ref{sec:shmr}). Each aspect includes one or two observables, characterized by their likelihoods $\mathcal{L}_j$. We define the geometric mean of the likelihoods of individual observables as the likelihood of the aspect. Each observable $j$ includes several observations (for individual galaxies or redshift bins). Their likelihoods $\mathcal{L}_i$ are combined with weights $w_i$ (to be discussed in detail below) to estimate $\mathcal{L}_{j}$ as:
\begin{align}
    \log\mathcal{L}_{j}=\frac{\sum_i w_i\log\mathcal{L}_{i}}{\sum_{i}w_i}\ .\label{eq:lh_obs}
\end{align}
Following \citet{Hartwig2024}, the likelihood of an observation $i$ is given by
\begin{align}
    \mathcal{L}_i(x_{{\rm obs},i},\sigma_i,x_{{\rm sim},i})=\frac{1}{\sqrt{2\pi\sigma_i^2}}\exp\left[-\frac{(x_{{\rm obs},i}-x_{{\rm sim},i})^2}{2\sigma_i^2}\right]\ ,\label{eq:lh}
\end{align}
where $x_{{\rm obs},i}$ is the observed value, $x_{{\rm sim},i}$ is the simulated value, and $\sigma_{i}$ is the uncertainty, assuming that the error $\Delta x_i\equiv x_{{\rm obs},i}-x_{{\rm sim},i}$ follows a normal distribution. 

The overall likelihood $\mathcal{L}_{\rm all}$ is defined as the geometric mean of the likelihoods of individual aspects:
\begin{align}
    \mathcal{L}_{\rm all}=(\mathcal{L}_{\rm CSFH}
    \mathcal{L}_{\rm MZSFR}
    \mathcal{L}_{\rm SHMR})^{1/3}\ .\label{eq:lh_all}
\end{align}
In the following subsections, we will explain the observables underlying the three terms in this formula and how we derive $x_{{\rm sim},i}$ from simulations and obtain $x_{{\rm obs},i}$ and $\sigma_i$ from the observational literature. Once $\mathcal{L}_{\rm all}$ is known, we search for the `best-fit' model(s) on the grid of $\alpha_{\rm out}$--$M_{\rm out0}$--$m_{\max}$ that achieve the highest value(s) of $\mathcal{L}_{\rm all}$, as discussed in Sec.~\ref{sec:res}. This process is analogous to the least-squares method of fitting, as $-\log \mathcal{L}_{\rm all}$ is proportional to a weighted sum of the squares of residuals $(x_{{\rm obs},i}-x_{{\rm sim},i})^{2}$ of individual observations. 


\subsection{Cosmic star formation history}
\label{sec:sfh}

The first aspect, cosmic star formation history (CSFH), is described by two observables: star formation rate density (SFRD) and cosmic stellar mass density (CSMD) $\rho_\star$, such that $\mathcal{L}_{\rm CSFH}=\sqrt{\mathcal{L}_{\rm SFRD}\mathcal{L}_{\rm CSMD}}$. 

For SFRD, we collect the measurements at 12 redshifts in the range $z\sim 5-16$ based on integration of UV luminosity functions down to magnitude $M_{\rm UV}=-17$ \citep{Bouwens2016,Donnan2023full,Donnan2023,Harikane2023}, as listed in Table~\ref{tab:sfrd}. This is of course not a complete collection of existing SFRD measurements \citep[see also, e.g.,][]{Madau2014,Finkelstein2015,McLeod2016,Oesch2018,Bhatawdekar2019,Bouwens2021,Bouwens2023,D'Silva2025,Chemerynska2025}. The adopted ones are selected such that they have consistent definitions and moderate uncertainties, and delineate a smooth increase of SFRD with decreasing redshift at $z\gtrsim 5$. As shown by \citet{Donnan2023full}, the evolution at $z\sim 7.5-15$ is well captured by a linear relation between $\log\rm SFRD$ and $z$:
\begin{align}
    \log&({\rm SFRD\ [M_\odot\ yr^{-1}\ cMpc^{-3}]})\notag\\
    &=(-0.231\pm 0.037)z-0.439\pm 0.3\ ,\label{eq:sfrd}
\end{align}
given the canonical conversion factor between SFR and UV luminosity $L_{\rm UV}$ from \citet{Madau2014}: $\kappa_{\rm UV}\equiv {\rm SFR/(M_\odot\ yr^{-1})}/[L_{\rm UV}/(\rm erg\ s^{-1}\ Hz^{-1})]=1.15\times~10^{-28}$. To be consistent with the definition of SFRD in observations, we only consider galaxies with $\rm SFR>10^{-0.5}\sim 0.3\ \rm M_\odot\ yr^{-1}$ corresponding to $M_{\rm UV}\lesssim-17$ according to $\kappa_{\rm UV}$, where the SFR is measured on a timescale of $t_{\rm SF}=10$~Myr. The SFRs of these galaxies are summed up and divided by $V_{\rm com}$ to give the `instantaneous' SFRD. Since the UV luminosity of a galaxy typically probes the SFR in a longer timescale $t_{\rm SF}\sim 100$~Myr, we further convolve this `instantaneous' SFRD with a time window of 100~Myr to obtain the final simulated SFRD to be compared with observations\footnote{These considerations highlight the importance of proper comparison between simulations and observations. The SFRD in simulations is usually measured in simulation timesteps and includes the contributions of all galaxies, like in \citet{Hartwig2024}. However, this is often not directly comparable to the SFRD inferred from observations with SFR tracers at different timescales and incomplete samples of faint objects. }. 

We evaluate the likelihood (Eq.~\ref{eq:lh}) in terms of $x_i\equiv\log(\rm SFRD\ [M_\odot\ yr^{-1}])$ and interpolate the simulation results linearly to the redshifts of observations. The uncertainty $\sigma_i$ is set to the observational uncertainty (see Table~\ref{tab:sfrd}). To reduce the effects of small-sample statistics, we only consider the redshift bins in which there are at least 3 galaxies with $\rm SFR>10^{-0.5}\ \rm M_\odot\ yr^{-1}$ in the simulation box, and the corresponding likelihoods $\mathcal{L}_{i}$ are weighted evenly ($w_i=1$) to calculate $\mathcal{L}_{\rm SFRD}$ by Eq.~\ref{eq:lh_obs}. {We further consider the SFRD measurements at $z\sim 7-15$ from the GLIMPSE survey \citep[][]{Chemerynska2025,Korber2025} that includes fainter galaxies down to $M_{\rm UV}\sim-13$, i.e., $\rm SFR\sim0.01\ \rm M_\odot\ yr^{-1}$, for cross-validation.}

\begin{table}[htbp]
    \centering
    \caption{Compilation of SFRD measurements at $z\sim 5-16$. The first section shows the pre-launch results from HST and ALMA. The second section shows the results from JWST.} 
    \begin{tabular}{ccc}
    \hline
        $z$ & $\log(\rm SFRD\ [M_\odot\ yr^{-1}])$ & Reference \\
    \hline
        4.9 & $-1.387\pm 0.124$ & \citet{Bouwens2016}\\
        5.9 & $-1.640\pm 0.130$ & \citet{Bouwens2016}\\
        6.8 & $-1.883\pm 0.076$ & \citet{Bouwens2016}\\
        7.9 & $-2.213\pm 0.065$ & \citet{Bouwens2016}\\
    \hline
        8 & $-2.31\pm 0.06$ & \citet{Donnan2023full,Donnan2023} \\
        9 & $-2.52_{-0.10}^{+0.08}$ & \citet{Donnan2023full,Donnan2023} \\
        10.5 & $-2.80_{-0.20}^{+0.13}$ & \citet{Donnan2023full,Donnan2023} \\
        11.2 & $-2.95_{-0.36}^{+0.20}$ & \citet{Donnan2023} \\
        13.25 & $-3.62_{-0.25}^{+0.16}$ & \citet{Donnan2023full,Donnan2023} \\
        9 & $-2.61_{-0.16}^{+0.18}$ & \citet{Harikane2023} \\
        12 & $-3.23_{-0.27}^{+0.29}$ & \citet{Harikane2023} \\
        16 & $-3.59_{-2.83}^{+0.33}$ & \citet{Harikane2023} \\
    \hline
    \end{tabular}
    \label{tab:sfrd}
\end{table}

For CSMD, we adopt the results at $z\sim 6-12$ from the empirical model in \citet[see their fig.~7 and the references therein]{Donnan2025}, derived by integrating the galaxy stellar mass function over the mass range $10^{8}-10^{13}\ \rm M_\odot$. This model captures the median value among various literature results \citep[see also fig.~4 in][]{Bosi2025} and can be described by the fit:
\begin{align}
    \log(\rho_{\star}\ [{\rm M_\odot\ cMpc^{-3}}])=-0.195z-0.022z^{2}+8.728\ .\label{eq:csmd}
\end{align}
The discrepancy between different CSMD measurements and their largest uncertainty are around 0.5 dex. Therefore, we define the CSMD likelihood with $x_i\equiv\log(\rho_{\star}\ [{\rm M_\odot\ cMpc^{-3}}])$ and adopt a constant uncertainty $\sigma_i=0.5$~dex. To be consistent with the observed galaxy sample, we only consider galaxies with $M_{\star}>10^{8}\ \rm M_\odot$ to derive the simulated CSMD for comparison. Similar to the case of SFRD, we combine the likelihoods of simulation redshift bins within $z\in [6,12]$ that have at least 3 galaxies with $M_{\star}>10^{8}\ \rm M_\odot$ using equal weights in Eq.~\ref{eq:lh_obs} for $\mathcal{L}_{\rm CSMD}$.

\subsection{Galaxy chemical evolution}
\label{sec:mzsfr}
High-$z$ galaxies are unresolved point sources in our approach, their overall chemical evolution is captured by the relation between stellar mass $M_\star$, metallicity $Z$, and SFR. Recent observations by JWST have measured these properties of a few hundred galaxies at $z\sim 4-10$ \citep{Langeroodi2023,Heintz2023,Nakajima2023,Chakraborty2024,Curti2024,Li2025,Li2025apjs,Nishigaki2025mzsfr,Rowland2025,Sarkar2025}. In particular, the results from spectroscopy programs of ERO, GLASS, CEERS, JADES, and the JWST-PRIMAL Legacy Survey \citep{Nakajima2023,Curti2024,Sarkar2025} have reached a consensus that there is a universal $M_{\star}$--$Z$-SFR relation (MZSFR) at $z\sim 4- 8$ consistent with that observed at $z\lesssim 3$ \citep{Andrews2013}. This relation is also found to hold at $z\sim 6-8$ by the REBELS ALMA large program \citep{Rowland2025}. On the other hand, the relation tends to break down at higher redshifts where galaxies are systematically more metal-poor \citep[by $\sim 0.27$~dex,][]{Nishigaki2025mzsfr,Sarkar2025}{, likely caused by strong pristine gas inflows diluting the metal abundances during early galaxy assembly \citep[e.g.,][]{Pollock2025}}. Considering that the samples of both observed and simulated galaxies are small at $z\gtrsim 8$, we focus on the JWST-concordance (invariant) relation at $z\sim 4-8$ to search for the best-fit model. The systematic decrease of metallicity at $z\sim 8-10$ is used for cross-validation. The underlying galaxy sample of \citet{Nakajima2023}, \citet{Curti2024}, and \citet{Sarkar2025} is referred to as `the JWST galaxies' henceforth.

In these observations, the SFR is typically derived from the H$\beta$ line luminosity (except for JADES, which uses spectral fitting) as a good indicator for ongoing ($\sim 10\ \rm Myr$) star formation. Note that the SFR of simulated galaxies is calculated on a short timescale of $t_{\rm SF}=10\ \rm Myr$ consistent with that of the SFR indicator in observations, which ensures fair comparison. Besides, our analysis only includes simulated galaxies with $\rm SFR\gtrsim 10^{-0.5}\ \rm M_\odot\ yr^{-1}$ to be consistent with the observed sample. Note that high-$z$ observations tend to be biased by luminous objects, especially when compared with simulations in a much smaller volume. One could apply a more complicated treatment for incompleteness and cosmic variance. As an exploratory approach, we instead apply a simple SFR cut and make corrections for observational biases when necessary by comparing the median values of SFR from the simulated and observed galaxies (see below). 

The observed (gas-phase) metallicity of a galaxy is characterized by the absolute oxygen abundance, $\rm \log(O/H)+12$, measured via nebular emission lines from star-forming regions. Considering that these lines are mostly powered by young massive stars, we define the metallicity of a simulated galaxy as the mass-weighted average oxygen abundance of massive stars ($m_{\star}>5\ \rm M_\odot$) with ages less than $10\ \rm Myr$. We further normalize the metallicity by the solar value and thus consider $\rm [O/H]\equiv\log(O/H)+12-8.69$, which turns out to be a good tracer of (the logarithm of) bulk metallicity $\log(Z/\rm Z_\odot)$: {Throughout our simulations, the difference between $\log(Z/\rm Z_\odot)$ and $\rm [O/H]$ is typically $\sim 0.03-0.07$~dex and remains below 0.15~dex for typical galaxies covered by current JWST spectroscopy surveys with $M_\star\gtrsim 10^{7}\ \rm M_\odot$ and $\rm SFR\gtrsim 0.3\ \rm M_\odot\ yr^{-1}$ at $z\gtrsim 4$.} 

To fully extract information from observations, we look into the observed invariant MZSFR in two ways: the fundamental metallicity relation (FMR) and stellar mass-metallicity relation (MZR), such that $\mathcal{L}_{\rm MZSFR}=\sqrt{\mathcal{L}_{\rm FMR}\mathcal{L}_{\rm MZR}}$. The former, originally based on $z\lesssim 3$ observations \citep{Andrews2013}, can be written as:
\begin{align}
    {\rm [O/H]_{\rm FMR}}(\mu_{\alpha})=0.43\mu_{\alpha} -4.11\ ,\quad \alpha=0.66\ ,\label{eq:fmr}
\end{align}
where $\mu_{\alpha}\equiv \log(M_\star\ [\rm M_\odot])-\alpha\log(SFR\ [M_\odot\ yr^{-1}])$, and $\alpha=0.66$ is chosen to minimize the scatter in [O/H] at a given $\mu_{\alpha}$ down to $\sigma_{\rm FMR}=0.013$~dex. To derive the likelihood $\mathcal{L}_{\rm FMR}$, we apply Eq.~\ref{eq:lh_obs} to all galaxies with $\rm SFR>10^{-0.5}\ \rm M_\odot\ yr^{-1}$ from the redshift bins within $z\sim 4-8$. All galaxies from a redshift bin $j$ are assigned with the same weight as the co-moving volume of the corresponding shell in the lightcone for an observer at $z=0$, i.e., $w_i=4\pi\int_{z_{j}}^{z_{j+1}}D_{\rm C}^{2}(z)|dD_{\rm C}/dz|dz$, where $D_{\rm C}(z)$ is the co-moving distance. For each simulated galaxy, the likelihood (Eq.~\ref{eq:lh}) is calculated from the predicted metallicity $x_{{\rm sim},i}=\rm [O/H]_{\rm sim}$ and that expected from the FMR $x_{{\rm obs},i}=\rm [O/H]_{\rm FMR}(\mu_{\rm 0.66,sim})$ given $\sigma_i=\sigma_{\rm FMR}=0.013$~dex, where $\mu_{\rm 0.66,sim}$ is derived from the simulated SFR and $M_{\star}$. Here, we adopt $\sigma_i=\sigma_{\rm FMR}=0.013$~dex as an optimistic assumption for the tightness of FMR. In fact, we estimate the scatter of JWST galaxies at $z\sim 4-10$ \citep[see their fig.~6]{Sarkar2025} around the invirant FMR (Eq.~\ref{eq:fmr}) as $\sigma_{\rm FMR, JWST}\simeq 0.27$~dex, much larger than the value at $z\lesssim 3$. Considering potentially large uncertainties in the measurements of SFR, $M_\star$, and [O/H] in high-$z$ observations (see Sec.~\ref{sec:caveats}), the intrinsic scatter will be smaller and closer to the low-$z$ value. 

For MZR, we consider the results from \citet[see their table~1]{Sarkar2025} for two observational redshift bins:
\begin{align}
    {\rm [O/H]}_{\rm MZR}=\begin{cases}
    0.28\log(M_\star\ [\rm M_\odot]) - 3.12\ ,& z\sim 4-6\ ,\\
    0.23\log(M_\star\ [\rm M_\odot]) - 2.70\ ,& z\sim 6-8\ ,
    \end{cases}\label{eq:mzr}
\end{align}
which are consistent with the earlier results of \citet{Nakajima2023} and \citet{Curti2024}. We take the uncertainty/scatter in $\rm [O/H]$ from \citet{Nakajima2023} as $\sigma_{\rm MZR}=0.28$ and 0.23~dex for $z\sim 4-6$ and $z\sim 6-8$, respectively. Individual simulated galaxies (with $\rm SFR>10^{-0.5}\ \rm M_\odot\ yr^{-1}$) are weighted in the same manner as for FMR to first calculate the likelihood $\mathcal{L}_{{\rm MZR},[z_{\rm low},z_{\rm up}]}$ of each observational redshift bin $[z_{\rm low},z_{\rm up}]$ by Eq.~\ref{eq:lh_obs}. The final likelihood is given by $\log\mathcal{L}_{\rm MZR}=(W_{[4,6]}\log\mathcal{L}_{\rm MZR,[4,6]}+W_{[6,8]}\log\mathcal{L}_{\rm MZR,[6,8]})/(w_{[4,6]}+w_{[6,8]})$, where $W_{[z_{\rm low},z_{\rm low}]}=\sum_{i}w_{i}$ is the sum of weights for galaxies within $z_i\in [z_{\rm low}, z_{\rm up}]$. 

We further apply a correction to the original observed MZR in Eq.~\ref{eq:mzr}, considering the fact that observations (of metal lines) are biased by luminous galaxies with systematically higher SFR (and therefore more metal-poor) than the simulated galaxies in our small simulation volume. The median $\rm \log SFR$ of observed galaxies at a given stellar mass is typically larger than that of our simulated galaxies by $\Delta\log \rm SFR\sim 0.6$~dex, as shown in Appendix~\ref{apdx:sfms}. Assuming that the FMR (Eq.~\ref{eq:fmr}) holds, reducing the median SFR by $\Delta\log\rm SFR$ will cause an increase of metallicity in MZR by ${\rm \Delta[O/H]}_{\rm MZR}=0.43\times0.66\times \Delta\log\rm SFR\sim 0.16$~dex. In light of this, we define $x_{{\rm obs},i}={\rm [O/H]_{\rm MZR}}(M_{\star,\rm sim})+\Delta{\rm [O/H]}_{\rm MZR}$ to calculate the likelihood (Eq.~\ref{eq:lh}) of a simulated galaxy using $\sigma_i=\sigma_{\rm MZR}$, given the simulated stellar mass $M_{\star,\rm sim}$ and metallicity $x_{{\rm sim},i}=\rm [O/H]_{\rm sim}$. In this way, we find that the constraints from FMR and MZR are generally consistent with each other.

Whether the invariant FMR (Eq.~\ref{eq:fmr}) better describes the galaxy chemical evolution than MZR {beyond or even at the local Universe ($z\sim 0$), especially for low-mass ($M_\star\lesssim 10^9\ \rm M_\odot$) galaxies (that are most relevant here), is still in debate \citep{KorhonenCuestas2025,Kotiwale2025,Laseter2025,Stanton2025}}. For instance, it is shown by \citet{KorhonenCuestas2025} that for galaxies from the Keck Baryonic Structure Survey at $z\sim 2.3$, introducing SFR to the metal scaling relation via $\mu_{\alpha}$ hardly reduces the scatter in [O/H], and the anti-correlation between SFR and the residual of [O/H] from MZR is {very weak -- weaker than that implied by the invariant FMR \citep[see also][]{Kotiwale2025}}. This is consistent with the large scatter of JWST galaxies at $z\sim 4-10$ around the invairant FMR $\sigma_{\rm FMR, JWST}\simeq 0.27$~dex, which is comparable to the scatter in MZR $\sigma_{\rm MZR}\sim 0.23-0.28$~dex. We also notice that there is a small systematic offset $\Delta\rm [O/H]_{\rm FMR}\sim -0.07$~dex between the JWST galaxies of \citet{Sarkar2025} and the invariant FMR \citep[see also fig.~11 in][]{KorhonenCuestas2025}. We can approximately incorporate these effects into our analysis {by applying a correction of $\Delta\rm [O/H]_{\rm FMR}\sim -0.07$~dex to the FMR at $z\sim 4-8$ with a larger FMR scatter of $\sigma_{\rm FMR, JWST}\sim 0.27$~dex and $\rm \Delta[O/H]_{\rm MZR}=0$. Doing so does not change our main conclusions according to numerical experiments.} 

\subsection{Galaxy-halo connection}
\label{sec:shmr}

The last aspect, i.e., the galaxy-halo connection, is relatively more difficult to probe in observations and, therefore, is expected to have weaker constraining power. 
Since it is directly related to the underlying halo merger trees -- backbones of our simulations, we still take it into account for completeness, focusing on the stellar-halo mass relation (SMHR). This relation is normally derived from semi-empirical models based on abundance matching \citep[e.g.,][]{Tacchella2018,Behroozi2019,Stefanon2021,Zaritsky2023} whose results still show large discrepancies up to $\sim 1$~dex at $z\sim 4-10$ (see Sec.~\ref{sec:best}). Here, we choose the results of \citet{Tacchella2018} as our target since they appear to be the `median' case, which can be described by the fit formula for $M_{\star,\rm SHMR}(M_{\rm h},z)$ in Eq.~\ref{eq:shmr}. The scatter around this relation is found to be around $0.14-0.18$~dex by \citet{Tacchella2018}. Considering the large discrepancies among different models, we instead use a larger uncertainty in $\log M_{\star}$ as $\sigma_{\rm SHMR}=0.31$ based on observations of local dwarf galaxies \citep{Zaritsky2023}. 

Similar to the case of MZR, we focus on galaxies with $\rm SFR>10^{-0.5}\ \rm M_\odot\ yr^{-1}$ which are divided into two redshift bins for $z\sim 4-6$ and $z\sim 6-8$ where the likelihoods are calculated separately and then combined to give $\log\mathcal{L}_{\rm SHMR}=(W_{[4,6]}\log\mathcal{L}_{\rm SHMR,[4,6]}+W_{[6,8]}\log\mathcal{L}_{\rm SHMR,[6,8]})/(W_{[4,6]}+W_{[6,8]})$. The likelihood of a simulated galaxy $i$ at the simulation redshift bin $z_j$ is derived using Eq.~\ref{eq:lh} with $x_{{\rm obs},i}=\log(M_{\star,\rm SHMR}(M_{\rm h,sim},z_j)\ [\rm M_\odot])$, $x_{{\rm sim},i}=\log(M_{\star,\rm sim}\ [\rm M_\odot])$, and $\sigma_i = \sigma_{\rm SHMR}=0.31$~dex given the stellar mass $M_{\star,\rm sim}$ and halo mass $M_{\rm h,sim}$. 

\section{Results}
\label{sec:res}

\begin{figure}
    \centering
    \includegraphics[width=1\linewidth]{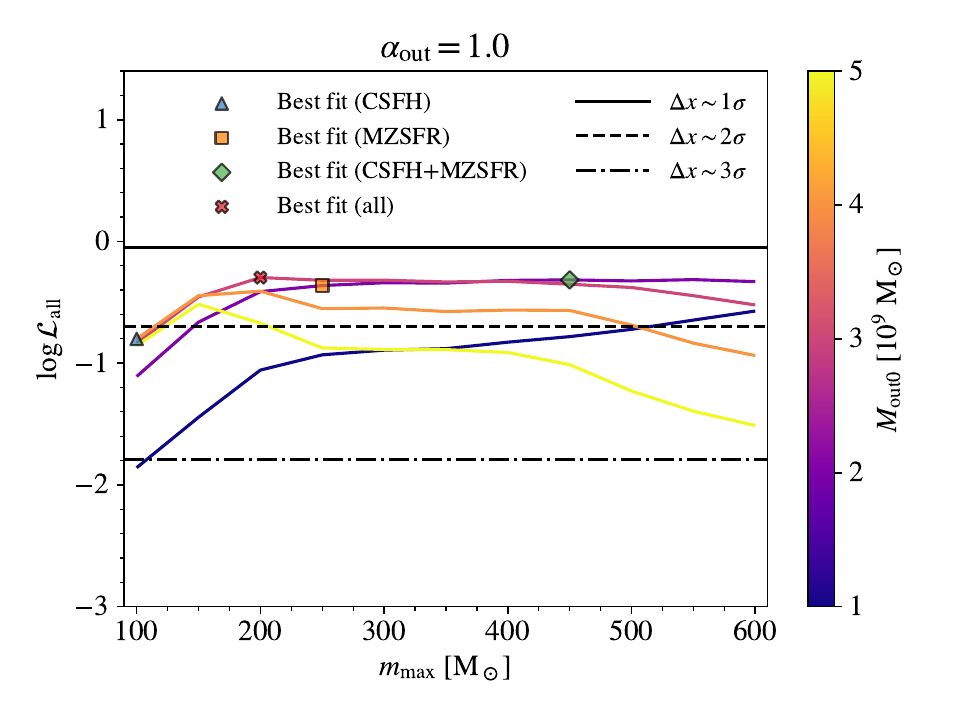}\\
    \includegraphics[width=1\linewidth]{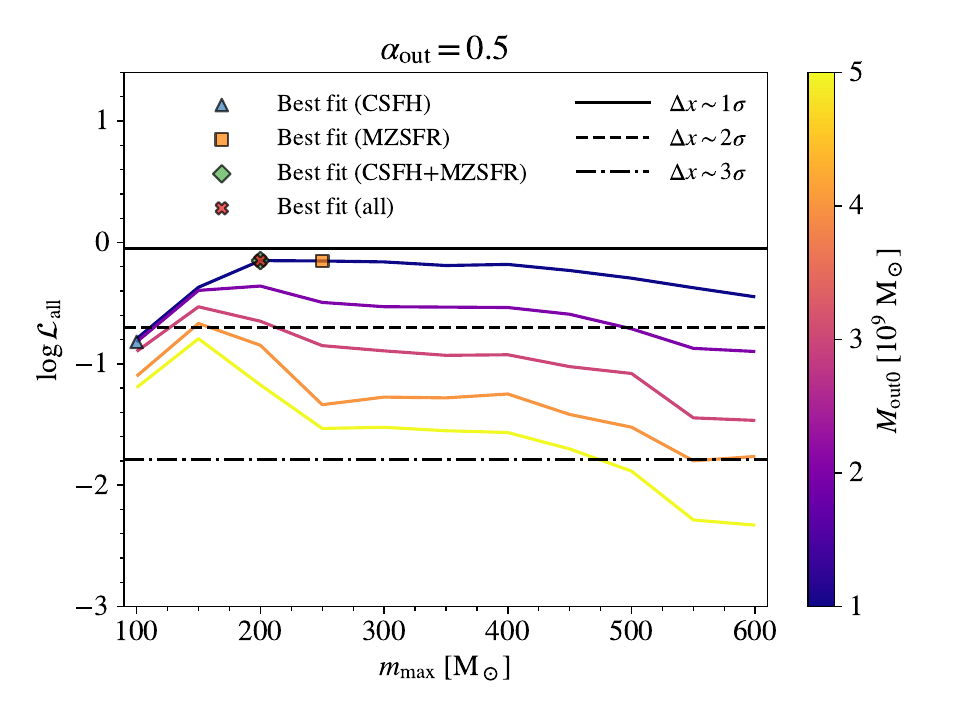}\\
    \includegraphics[width=1\linewidth]{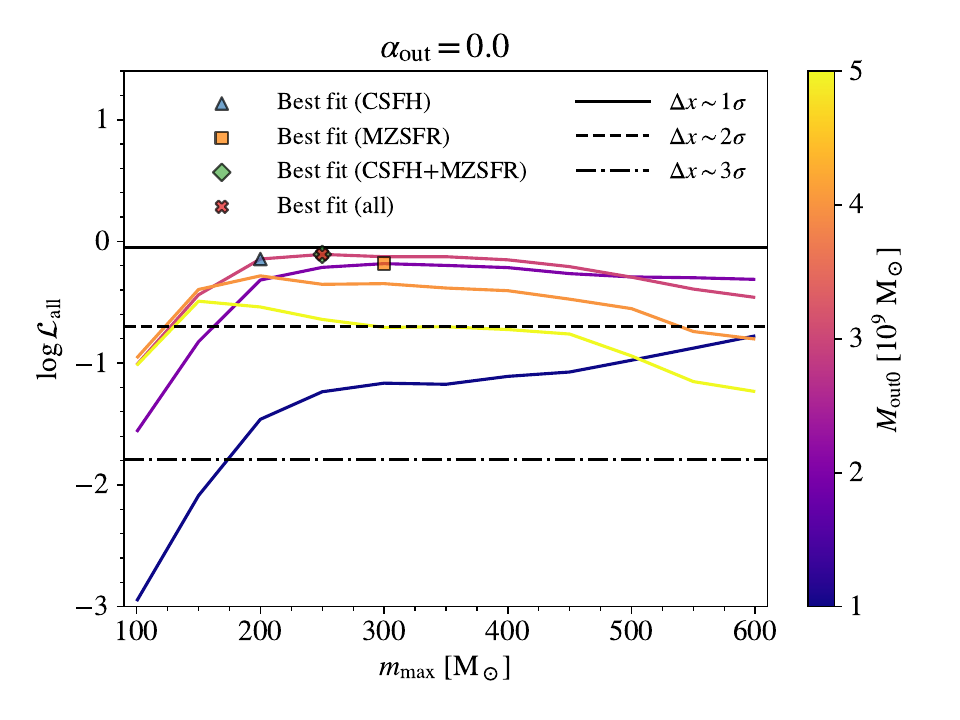}
    \caption{The overall likelihood $\mathcal{L}_{\rm all}$ (Eq.~\ref{eq:lh_all}) combining all observational constraints (CSFH+MZSFR+SHMR) as a function of $m_{\max}$ ($x$-axis) and $M_{\rm out0}$ (colorbar). The variation of $\mathcal{L}_{\rm all}$ with $m_{\max}$ is shown by lines color-coded by $M_{\rm out0}$ where darker colors correspond to smaller $M_{\rm out0}$. The top, middle, and bottom panels show the results for $\alpha_{\rm out}=1.0$, 0.5, and 0, respectively. The data points mark the best-match models considering 4 combinations of constraints: CSFH alone (triangle), MZSFR alone (square), CSFH+MZSFR (diamond), and all (cross). The solid, dashed, and dash-dotted horizontal lines show the estimated likelihood values for 1, 2, and $3\sigma$ deviations, respectively. } 
    \label{fig:lh_all}
\end{figure}

\begin{figure}
    \centering
    \includegraphics[width=1\linewidth]{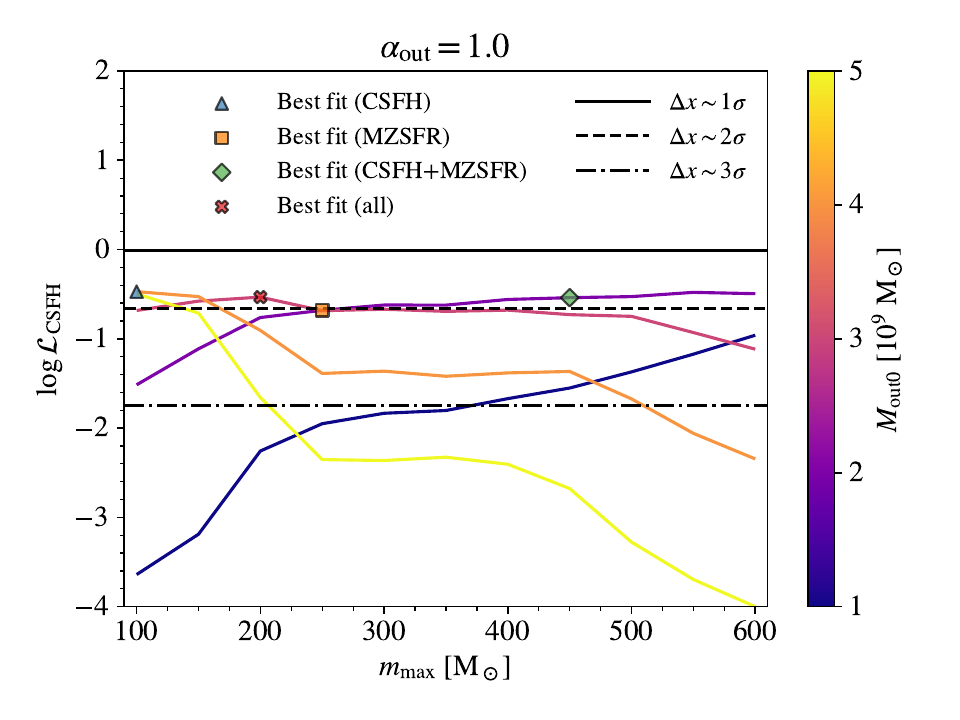}\\
    \includegraphics[width=1\linewidth]{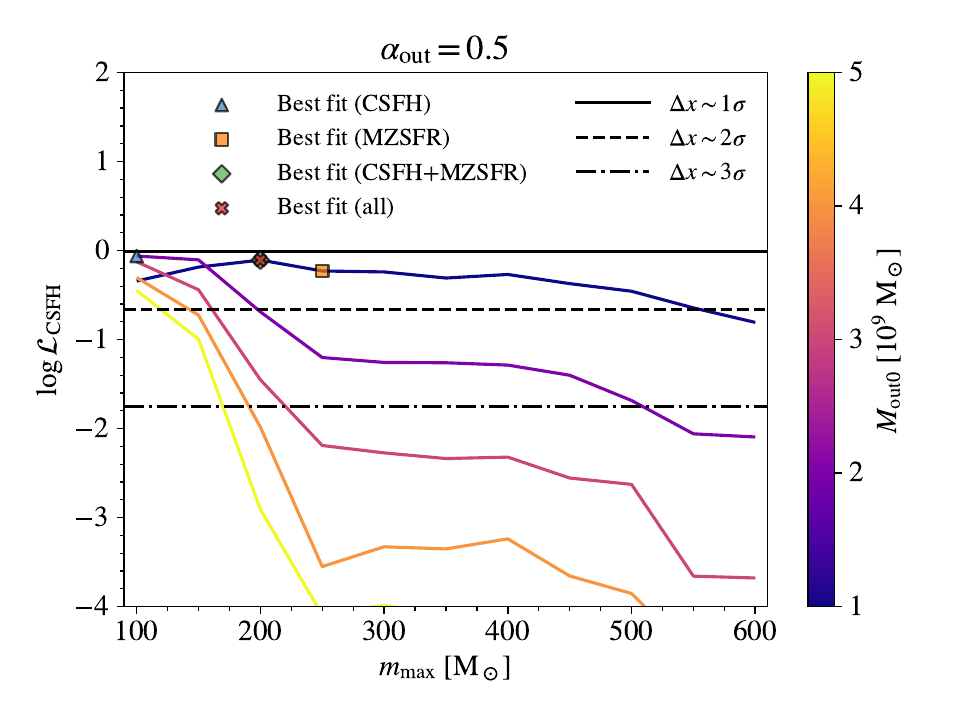}\\
    \includegraphics[width=1\linewidth]{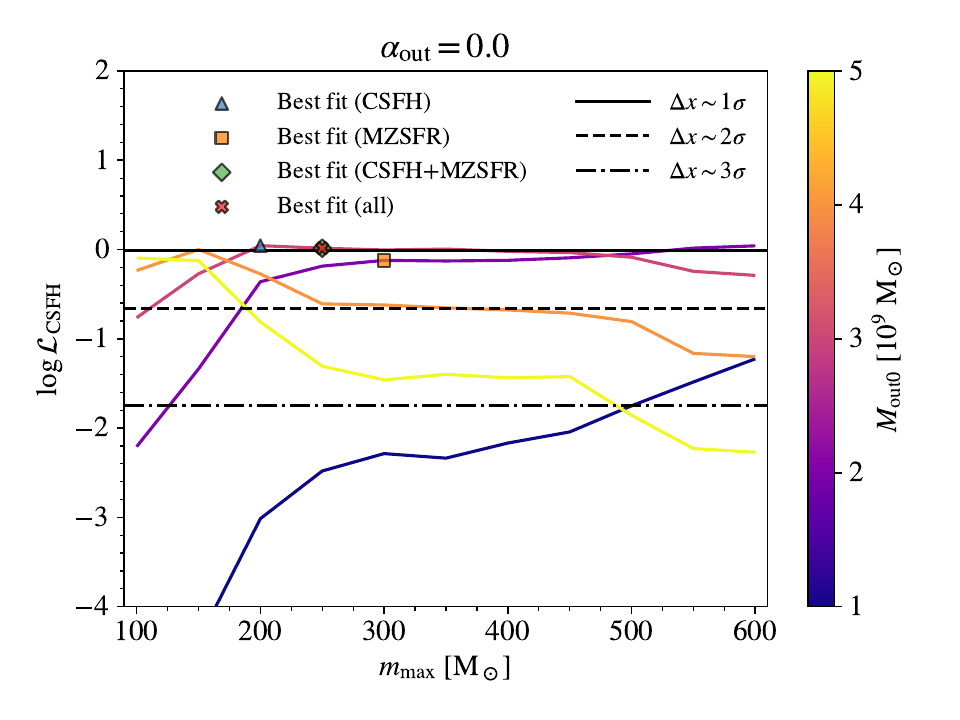}
    \caption{Same as Fig.~\ref{fig:lh_all} but for the likelihood of CSFH alone.}
    \label{fig:lh_csfh}
\end{figure}

\begin{figure}
    \centering
    \includegraphics[width=1\linewidth]{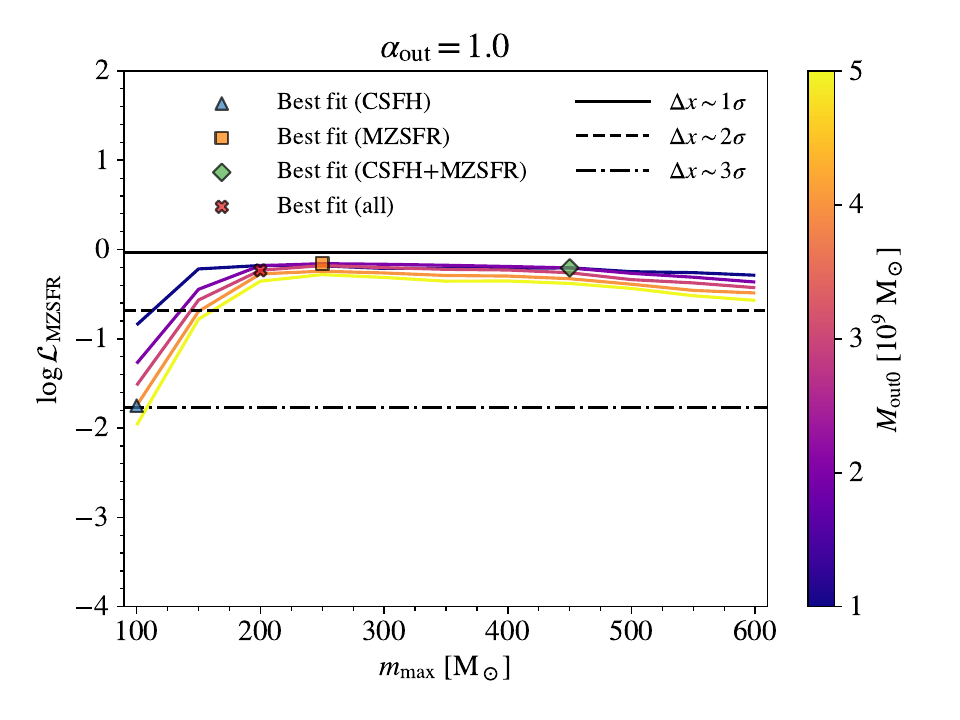}\\
    \includegraphics[width=1\linewidth]{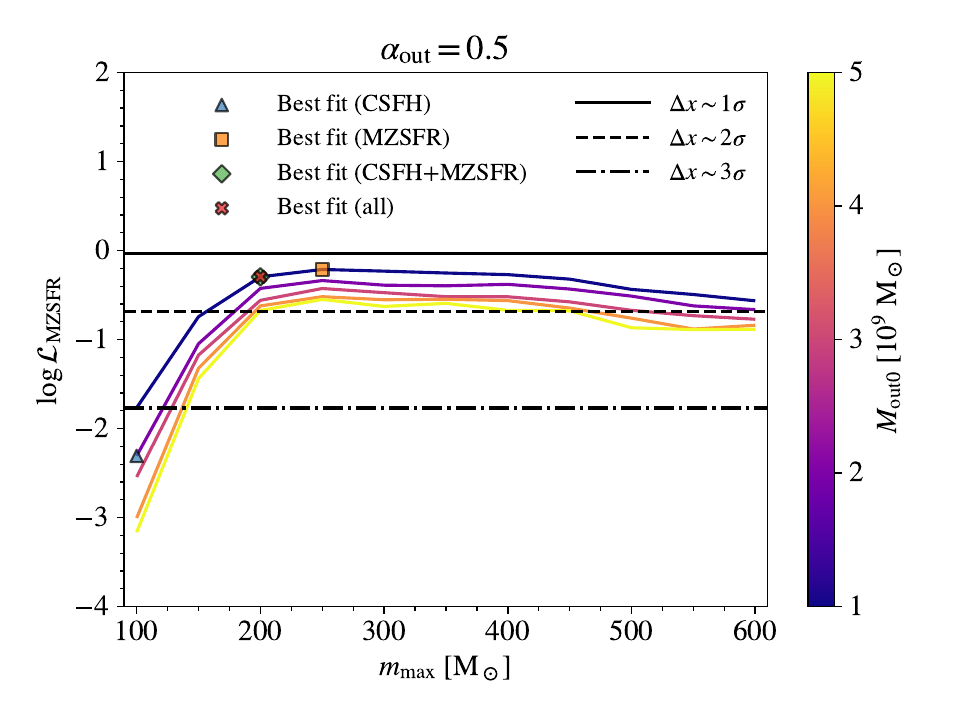}\\
    \includegraphics[width=1\linewidth]{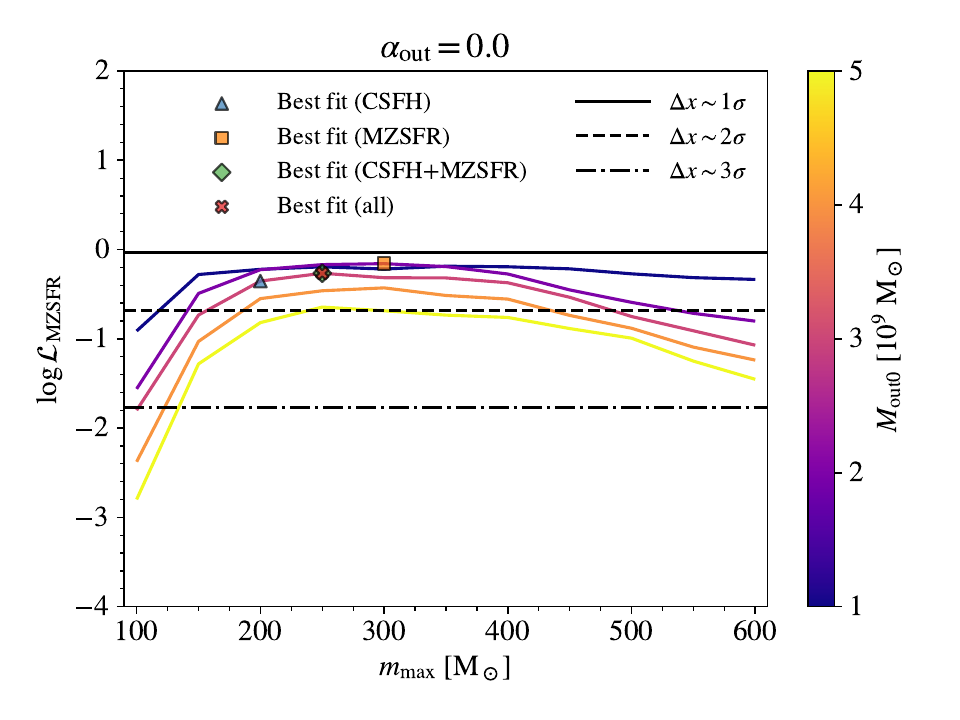}
    \caption{Same as Fig.~\ref{fig:lh_all} but for the likelihood of MZSFR alone.}
    \label{fig:lh_mzsfr}
\end{figure}

\begin{figure}
    \centering
    \includegraphics[width=1\linewidth]{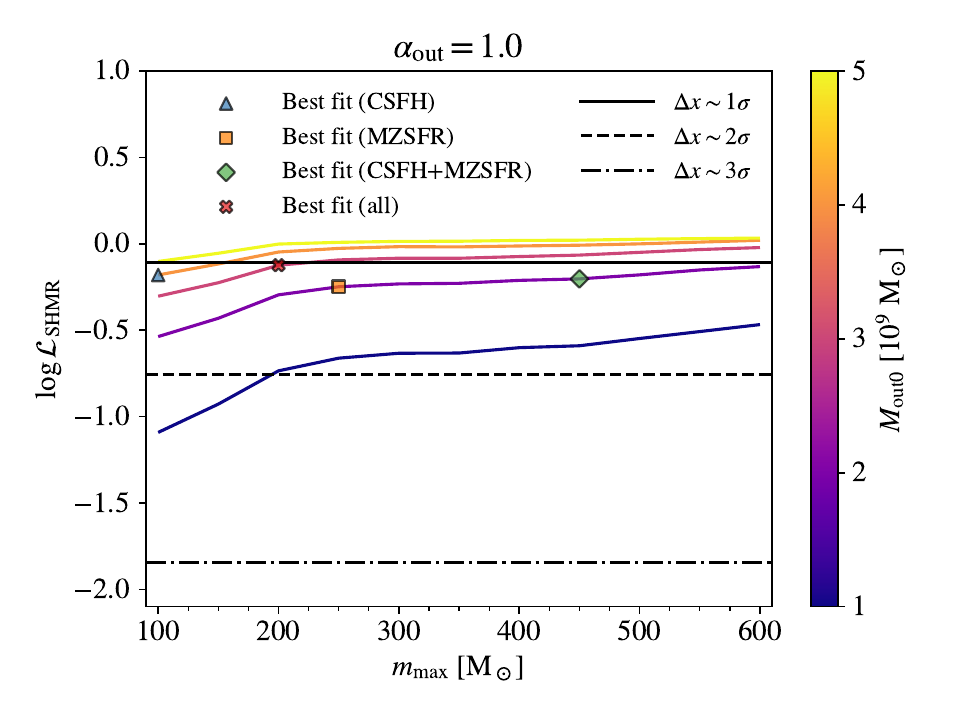}\\
    \includegraphics[width=1\linewidth]{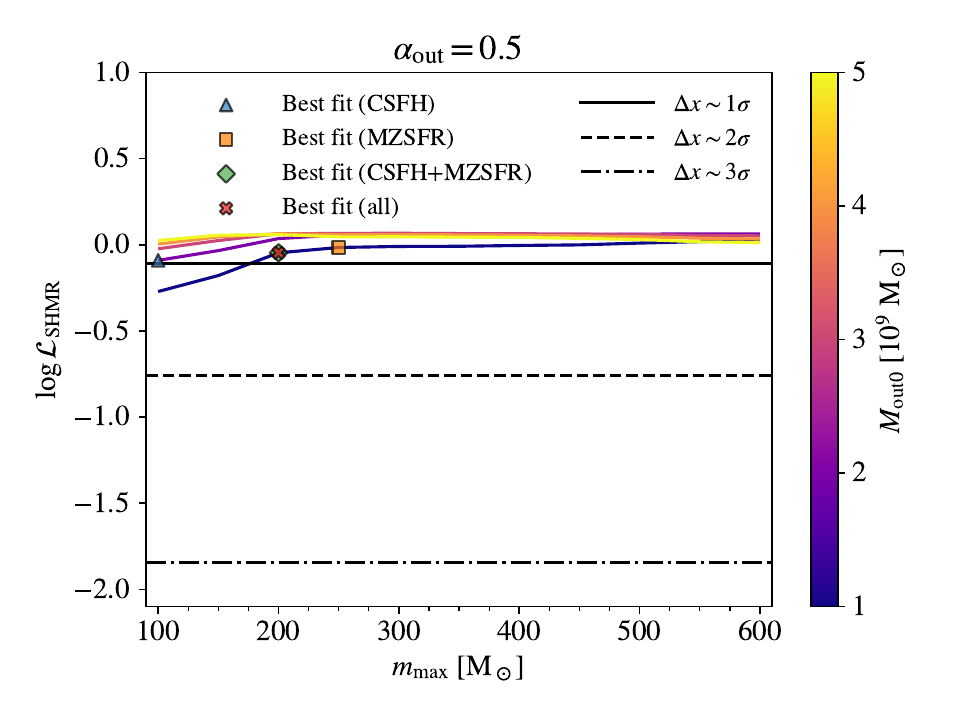}\\
    \includegraphics[width=1\linewidth]{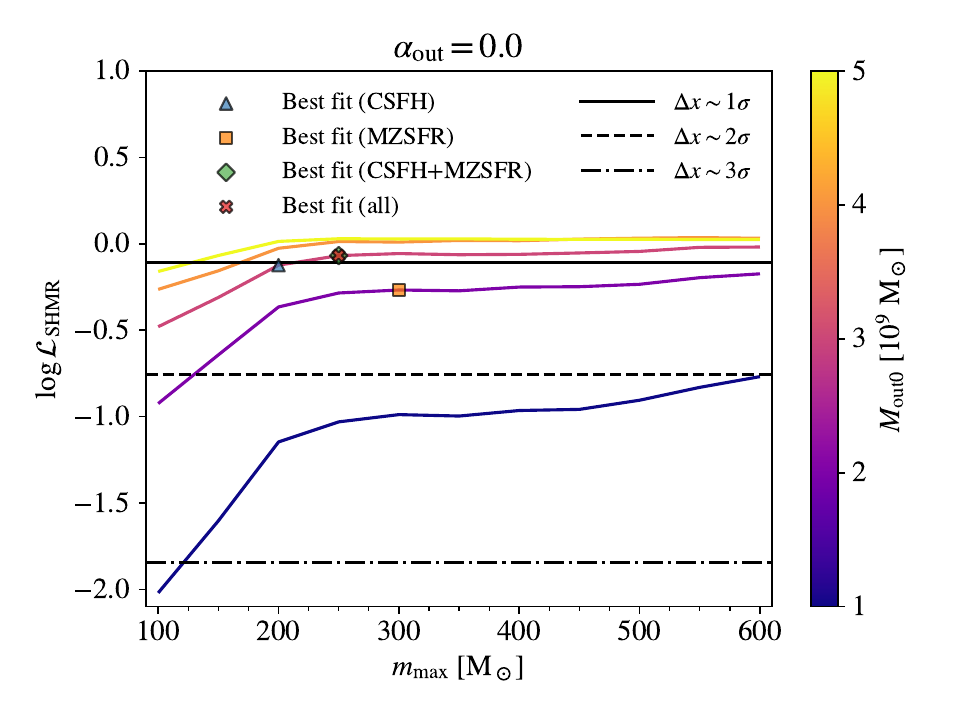}
    \caption{Same as Fig.~\ref{fig:lh_all} but for the likelihood of SHMR alone.}
    \label{fig:lh_shmr}
\end{figure}

In this section, we present our simulation results in the context of the observational constraints described in the previous section. First, we evaluate the overall likelihood $\mathcal{L}_{\rm all}$ and the likelihoods for individual observables as functions of model parameters ($\alpha_{\rm out}$, $M_{\rm out0}$, and $m_{\max}$) to elucidate the interplay between IMF and galactic outflow parameters in star formation and chemical evolution of high-$z$ galaxies. Next, we focus on the best-fit model with the highest $\mathcal{L}_{\rm all}$ for which we compare theoretical predictions with observed quantities/scaling relations in detail to show how well observations can be reproduced by our model and what can be learned about the underlying physics (Sec.~\ref{sec:best}). 

\subsection{Interplay between IMF and galactic outflow parameters}
\label{sec:interplay}

In Fig.~\ref{fig:lh_all}, we plot the overall likelihood $\log\mathcal{L}_{\rm all}$ as a function of the maximum mass of a Pop~II star $m_{\max}$ under different values of the galactic outflow parameters (see Eqs.~\ref{eq:gamma_out}--\ref{eq:gamma_out_0}): the normalization $M_{\rm out0}$ (with different line colors) and the power-law slope $\alpha_{\rm out}$ in the outflow efficiency--halo mass relation (in separate panels). Similar plots for the likelihoods of the cosmic star formation history (CSFH, Sec.~\ref{sec:sfh}), the mass-metallicity-star formation rate relation (MZSFR, Sec.~\ref{sec:mzsfr}), and the stellar-halo mass relation (SHMR, see Eq.~\ref{eq:shmr} and Sec.~\ref{sec:shmr}) are shown in Figs.~\ref{fig:lh_csfh}, \ref{fig:lh_mzsfr}, and \ref{fig:lh_shmr}, respectively. In these plots, we also show the likelihood values corresponding to $k\sigma$ ($k=1$, 2, and 3) deviation between simulation results and observations, estimated by substituting $\Delta x=x_{{\rm sim,}i}-x_{{\rm obs},i}=k\sigma_i$ in Eq.~\ref{eq:lh}.

To understand the roles played by individual observables in constraining our model parameters, for each of the three $\alpha_{\rm out}$ values considered, we identify the best-fit models for individual (combinations of) observational constraints considering four cases: (1) CSFH alone, where $\mathcal{L}_{\rm CSFH}$ is maximized, (2) MZSFR alone, where $\mathcal{L}_{\rm MZSFR}$ is maximized, (3) CSFH+MZSFR, where $\sqrt{\mathcal{L}_{\rm CSFH}\mathcal{L}_{\rm MZSFR}}$ is maximized, and finally (4) all combined (CSFH+MZSFR+SHMR), where $\mathcal{L}_{\rm all}$ is maximized. These models are labeled by (1) triangles, (2) squares, (3) diamonds, and (4) crosses in Figs.~\ref{fig:lh_all}--\ref{fig:lh_shmr}. 

As a starting point, we check the consistency between the best-fit models in the aforementioned four cases. Given an outflow efficiency slope $\alpha_{\rm out}=0$, the four best-fit models are generally consistent with each other with $M_{\rm out0}\sim 2-3\times 10^{9}\ \rm M_\odot$, and $m_{\max}\sim 200-300\ \rm M_\odot$. However, this is not the case for $\alpha_{\rm out0}=0.5$ and 1, which shows a (mild) tension between the requirements for reproducing the observed CSFH and MZSFR. To be specific, when MZSFR is taken into account in cases (2), (3), (4), $m_{\max}\sim 200-450\ \rm M_\odot$ is favored, while for case (1), CSFH alone, the best-fit model has $m_{\max}=100\ \rm M_\odot$, and the corresponding $M_{\rm out0}$ is also higher. 

To better understand these outcomes, we look into the likelihoods for individual observables:  
\begin{itemize}
    \item \textbf{CSFH}: As shown in Fig.~\ref{fig:lh_csfh}, the dependence of $\mathcal{L}_{\rm CSFH}$ on model parameters is complex: For a given $\alpha_{\rm out}$, multiple combinations of $m_{\max}$ and $M_{\rm out0}$ can achieve the highest values of $\mathcal{L}_{\rm CSFH}$. The reason is that the observed CSFH prefers some intermediate outflow efficiency achievable in different ways. Indeed, the strength of galactic outflows is boosted by increasing either $m_{\max}$ (for $m_{\max}\sim 100-250\ \rm M_\odot$) or $M_{\rm out0}$ (see Fig.~\ref{fig:mzesn_mmax} and Eq.~\ref{eq:gamma_out}). Meanwhile, it is reduced by decreasing $\alpha$ for relatively massive halos with $M_{\rm h}\gtrsim M_{\rm out0}$, which dominate cosmic star formation at $z\lesssim 8$ in our simulations. The maximum of $\mathcal{L}_{\rm CSFH}$ is slightly (significantly) higher for $\alpha_{\rm out}=0$ than for $\alpha_{\rm out}=0.5$ (1). In the latter case, the minimum discrepancy between simulations and observations can be up to $2\sigma$. This results from the fact that the outflow efficiency is suppressed at $M_{\rm h}\gtrsim M_{\rm out0}$ by Eq.~\ref{eq:gamma_out} given $\alpha_{\rm out}>0$, such that the rise of SFRD with decreasing redshift is too rapid compared with observations at $z\sim 6-13$ (see Appendix~\ref{apdx:csfh_alt}). Therefore, $\alpha_{\rm out}=0$ is favored by the observed CSFH (and all constraints combined as well).
    \item \textbf{MZSFR:} Fig.~\ref{fig:lh_mzsfr} shows that the observed MZSFR is always best reproduced by $m_{\max}\sim 200-300\ \rm M_\odot$ regardless of the choices of $M_{\rm out0}$ and $\alpha_{\rm out}$, because such IMFs provide maximum metal yields (see Fig.~\ref{fig:mzesn_mmax}). Besides, all models with $m_{\max}=100\ \rm M_\odot$ show $>2\sigma$ tensions with the observed MZSFR, and the same is true for all observables combined (Fig.~\ref{fig:lh_all}). For a given $m_{\max}$, smaller $M_{\rm out0}$ is generally preferred although the trend is weak/unclear at $M_{\rm out0}\lesssim3\times 10^{9}\ \rm M_\odot$ for $\alpha_{\rm out}=1$ and 0. The maximum of $\mathcal{L}_{\rm MZSFR}$ is similar for different values of $\alpha_{\rm out}$. In general, reproducing the observed MZSFR requires the highest metal yields and the weakest outflows covered by our simulations, as stronger outflows not only suppress star formation (and metal production) but also reduce the fraction of metals retained in galaxies. 
    \item \textbf{SHMR:} The constraints from SHMR have minor impact on $\mathcal{L}_{\rm all}$, as the best-fit models for CSFH+MZSFR and all combined (CSFH+MZSFR+SHMR) are identical for $\alpha_{\rm out}=0$ and 0.5. For $\alpha_{\rm out}=1$, the best-fit models are different but the corresponding values of $\mathcal{L}_{\rm all}$ are very similar. It is shown in Fig.~\ref{fig:lh_shmr} that $\mathcal{L}_{\rm SHMR}$ favors $m_{\max}\gtrsim 200\ \rm M_\odot$ like $\mathcal{L}_{\rm MZSFR}$. However, $\mathcal{L}_{\rm SHMR}$ is generally larger for increasing $M_{\rm out0}$, in contrast to the case of $\mathcal{L}_{\rm MZSFR}$. Moreover, the models preferred by the target SHMR (Eq.~\ref{eq:shmr}) from \citet{Tacchella2018} with $M_{\rm out 0}\sim 4-5\times 10^{9}\ \rm M_\odot$ and $m_{\max}\gtrsim 200\ \rm M_\odot$ have already been ruled out by the observed CSFH by $\gtrsim3\sigma$. As a result, our best-fit models do not achieve the maximum possible value of $\mathcal{L}_{\rm SHMR}$. As mentioned in Sec.~\ref{sec:shmr}, the SHMR inferred from observations is highly uncertain and model-dependent. So, the seeming tension between the constraints from SHMR and CSFH(+MZSFR) does not necessarily mean a failure of our simulations. 
\end{itemize}

\begin{figure*}
    \centering
    \includegraphics[width=0.495\linewidth]{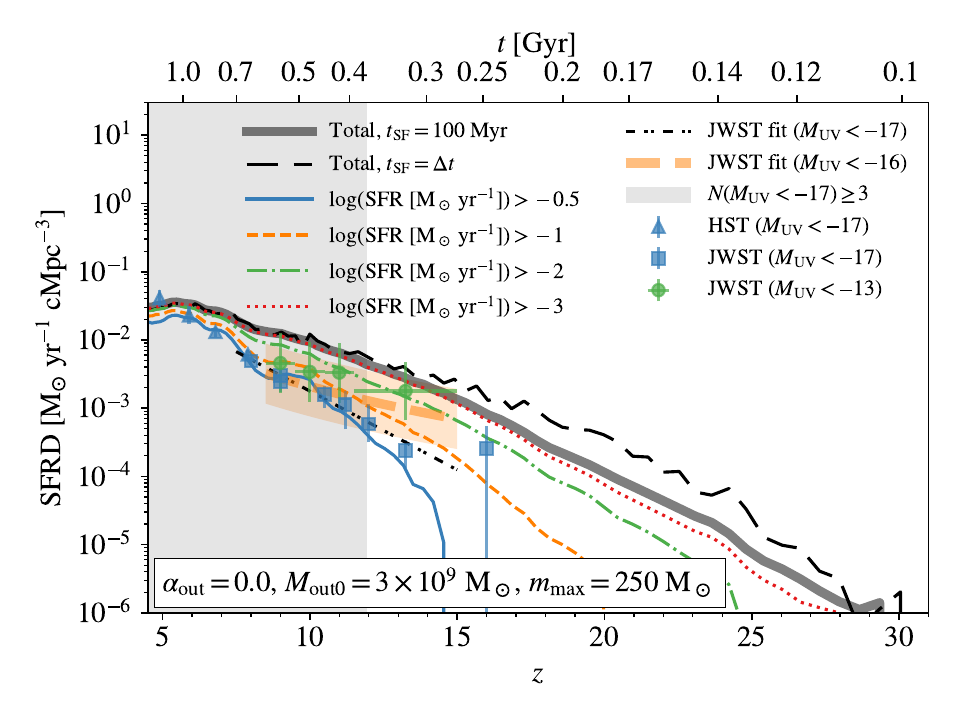}
    \includegraphics[width=0.495\linewidth]{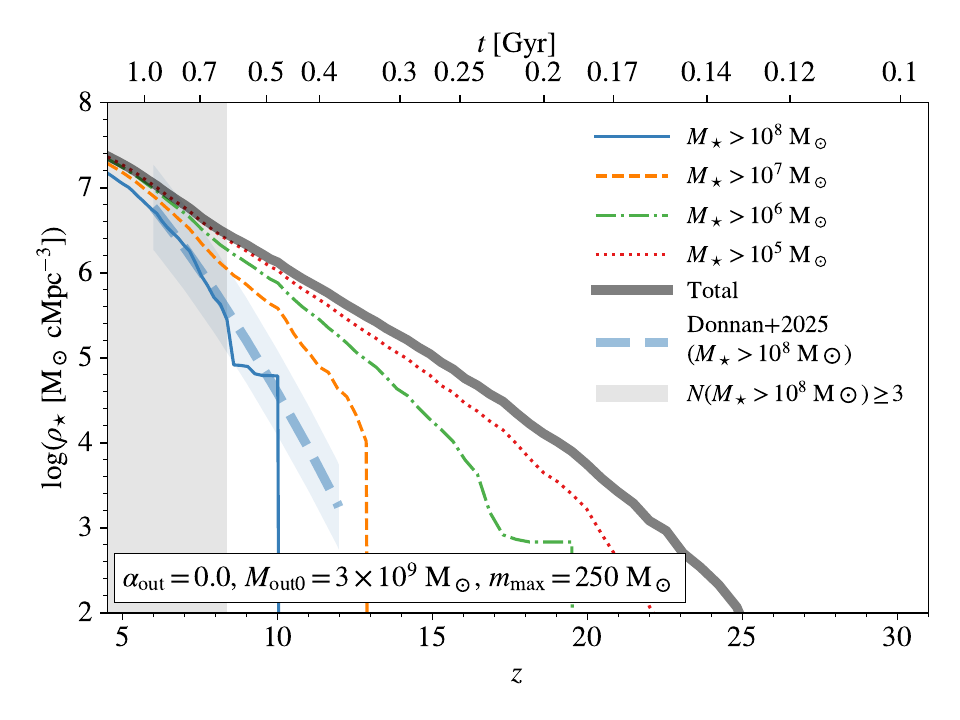}
    \caption{SFRD (left) and CSMD (right) in our simulation for the best-fit model with $\alpha_{\rm out}=0$, $M_{\rm out0}=3\times 10^{9}\ \rm M_\odot$, and $m_{\max}=250\ \rm M_\odot$. In the left panel, the thick solid curve shows the total SFRD measured on a timescale of $t_{\rm SF}=100$~Myr, while the solid, dashed, dash-dotted, and dotted curves show the contributions from galaxies with $\rm \log( SFR\ [M_\odot\ yr^{-1}])>-0.5,\ -1,\ -2$, and $-3$, respectively. The long dashed curve shows the total SFRD measured in simulation timesteps. {The curve for $\log\rm (SFR\ [M_\odot\ yr^{-1}])>-0.5$ should be compared with the observational results from HST and JWST for luminous galaxies with $M_{\rm UV}<-17$ embodied by the triangles and squares (see Table~\ref{tab:sfrd}). Here, the JWST results satisfy a linear relation between $\log\rm SFR$ and $z$ (dot-dash-dotted, Eq.~\ref{eq:sfrd}) at $z\sim 7.5-15$ \citep{Donnan2023full}. We also plot the cross-validation SFRD measurements at $z\sim 8.5-15$ from the GLIMPSE survey \citep{Chemerynska2025} including fainter galaxies with $M_{\rm UV}<-16$ and $M_{\rm UV}<-13$ as the thick dashed curve surrounded by a shaded region for $1\sigma$ uncertainty and the squares, which should be compared with the simulation results for $\log\rm (SFR\ [M_\odot\ yr^{-1}])>-1$ and $\log\rm (SFR\ [M_\odot\ yr^{-1}])>-2$, respectively. }
    In the right panel, the thick solid curve shows the total CSMD, while the solid, dashed, dash-dotted, and dotted curves show the contributions from galaxies with $M_\star>10^{8}$, $10^{7}$, $10^{6}$, and $10^{5}\ \rm M_\odot$, respectively. The curve for $M_\star>10^{8}\ \rm M_\odot$ should be compared with the observational results from \citet{Donnan2025} shown as the thick dashed curve (see Eq.~\ref{eq:csmd}) surrounded by a shaded region for $1\sigma$ scatter. The left shaded region shows the redshift range in which {the likelihood is calculated}. This means that there are at least 3 galaxies in the simulation box with $\log\rm (SFR\ [M_\odot\ yr^{-1}])>-0.5$ for SFRD (left), and with $M_\star>10^{8}\ \rm M_\odot$ for CSMD (right).}
    \label{fig:csfh}
\end{figure*}

In conclusion, the highest overall likelihood $\log\mathcal{L}_{\rm all}\simeq-0.11$ is achieved by the best-fit model for $\alpha_{\rm out}=0$ with $M_{\rm out0}=3\times 10^{9}\ \rm M_\odot$ and $m_{\max}=250\ \rm M_\odot$. For $\alpha_{\rm out}=0.5$ (1), the best-fit model has $\log\mathcal{L}_{\rm all}\simeq-0.15\ (-0.30)$, $M_{\rm out0}=1\ (3)\ \times10^{9}\ \rm M_\odot$, and $m_{\max}=200\ \rm M_\odot$. All these models agree with observations within $2\sigma$ ($\log\mathcal{L}_{\rm all}\sim -0.7$), and the best-fit model for $\alpha_{\rm out}=0$ almost achieves $1\sigma$ agreement ($\log\mathcal{L}_{\rm all}\simeq -0.05$).

\subsection{Best-fit model}
\label{sec:best}

\begin{figure*}
    \centering
    \includegraphics[width=0.495\linewidth]{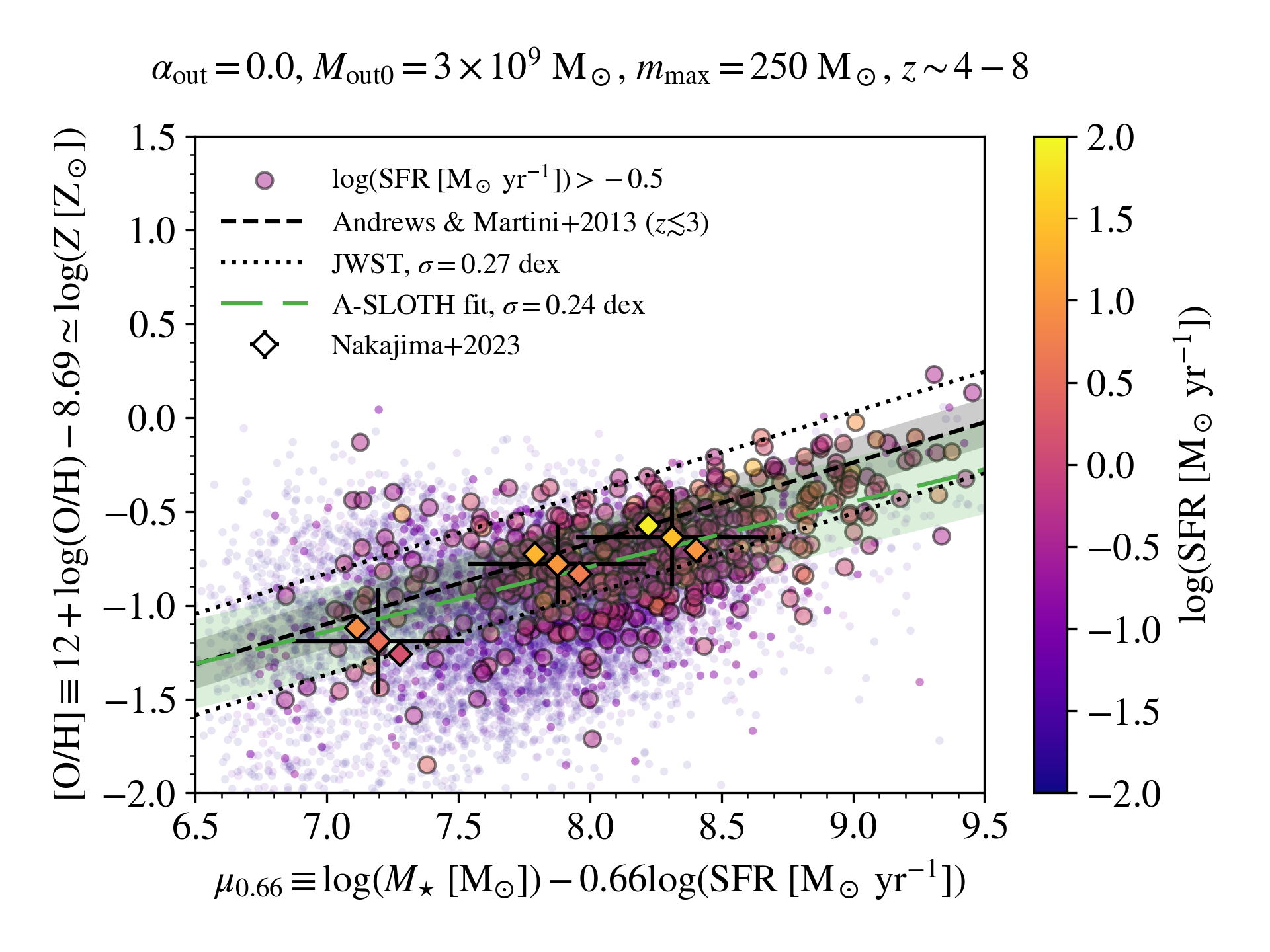}
    \includegraphics[width=0.495\linewidth]{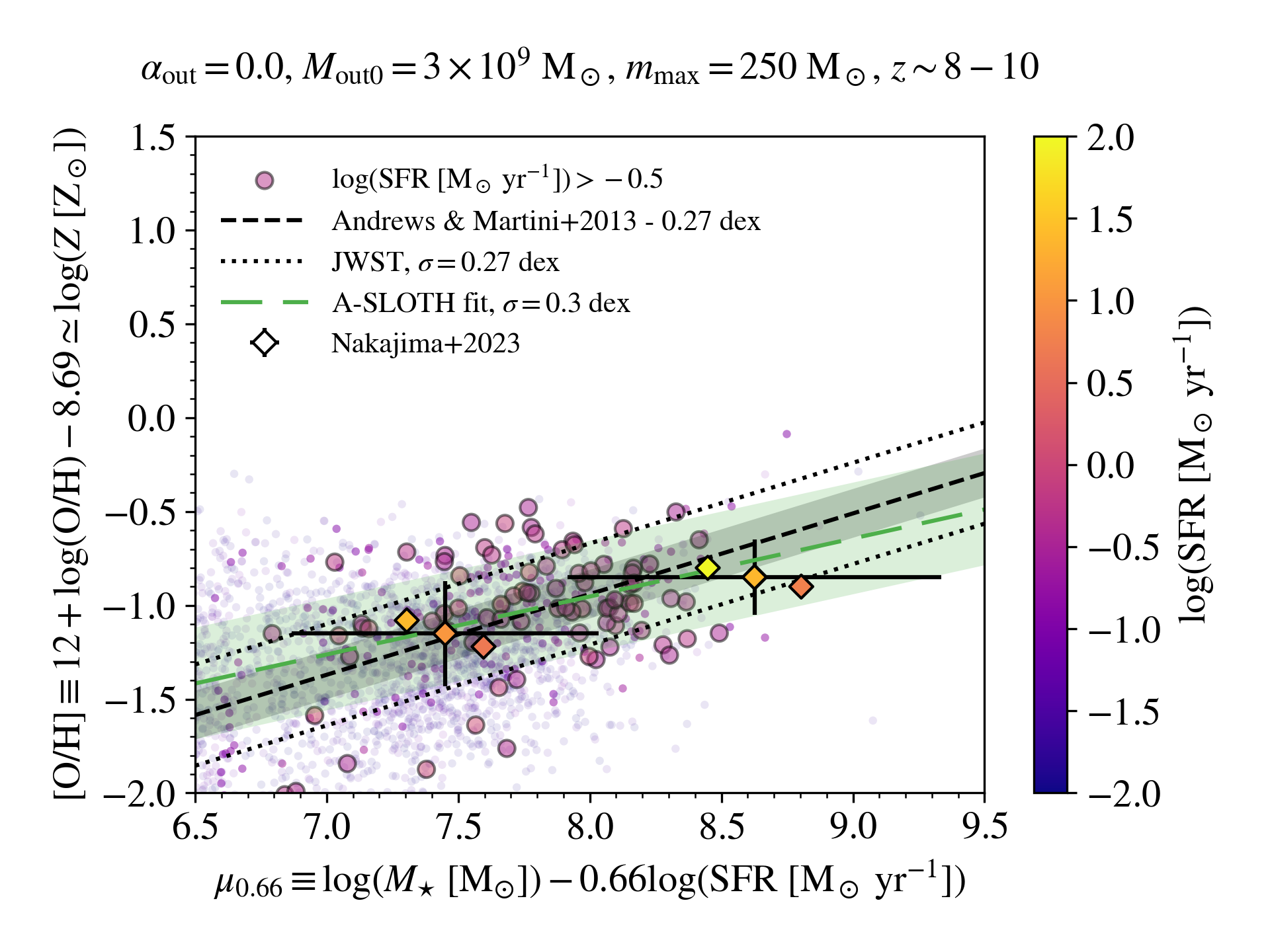}
    \caption{FMR for the best-fit model with $\alpha_{\rm out}=0$, $M_{\rm out0}=3\times 10^{9}\ \rm M_\odot$, and $m_{\max}=250\ \rm M_\odot$. Simulated galaxies with $\log\rm (SFR\ [M_\odot\ yr^{-1}])>-2$ are shown as dots, color-coded by SFR. The detectable ones with $\log\rm (SFR\ [M_\odot\ yr^{-1}])>-0.5$ are highlighted by large circles, based on which we derive a linear fit (in log-log space) as the long dashed line and green shaded region (for $1\sigma$ scatter). The left panel shows the results for $z\sim 4-8$, where observed galaxies follow the same FMR (Eq.~\ref{eq:fmr}) found at $z\lesssim 3$ by \citet[]{Andrews2013}, as shown by the dashed line. The gray shaded region around it shows the ($1\sigma$) scatter (0.13~dex) reported by \citet[]{Andrews2013}, while the dotted lines show the scatter (0.27~dex) derived from the JWST galaxies at $z\sim 4-10$ compiled by \citet[see their fig.~6]{Sarkar2025}. The right panel shows the results for $z\sim 8-10$. Here, the normalization of FMR is shifted down by 0.27~dex compared with that at $z\lesssim 8$ according to JWST observations \citep{Sarkar2025}. The shifted observational FMR is again shown by the dashed line. The average values of $\mu_{0.66}$, [O/H], and SFR of the JWST galaxies of \citet[see their table~2]{Nakajima2023} are shown with the diamonds. Each data point involves three diamonds color-coded by the mean and $1\sigma$ upper and lower limits of SFR. The relevant uncertainties in $\mu_{0.66}$ and [O/H] are shown by the errorbars.}
    \label{fig:fmr}
\end{figure*}

\begin{figure}
    \centering
    \includegraphics[width=1\linewidth]{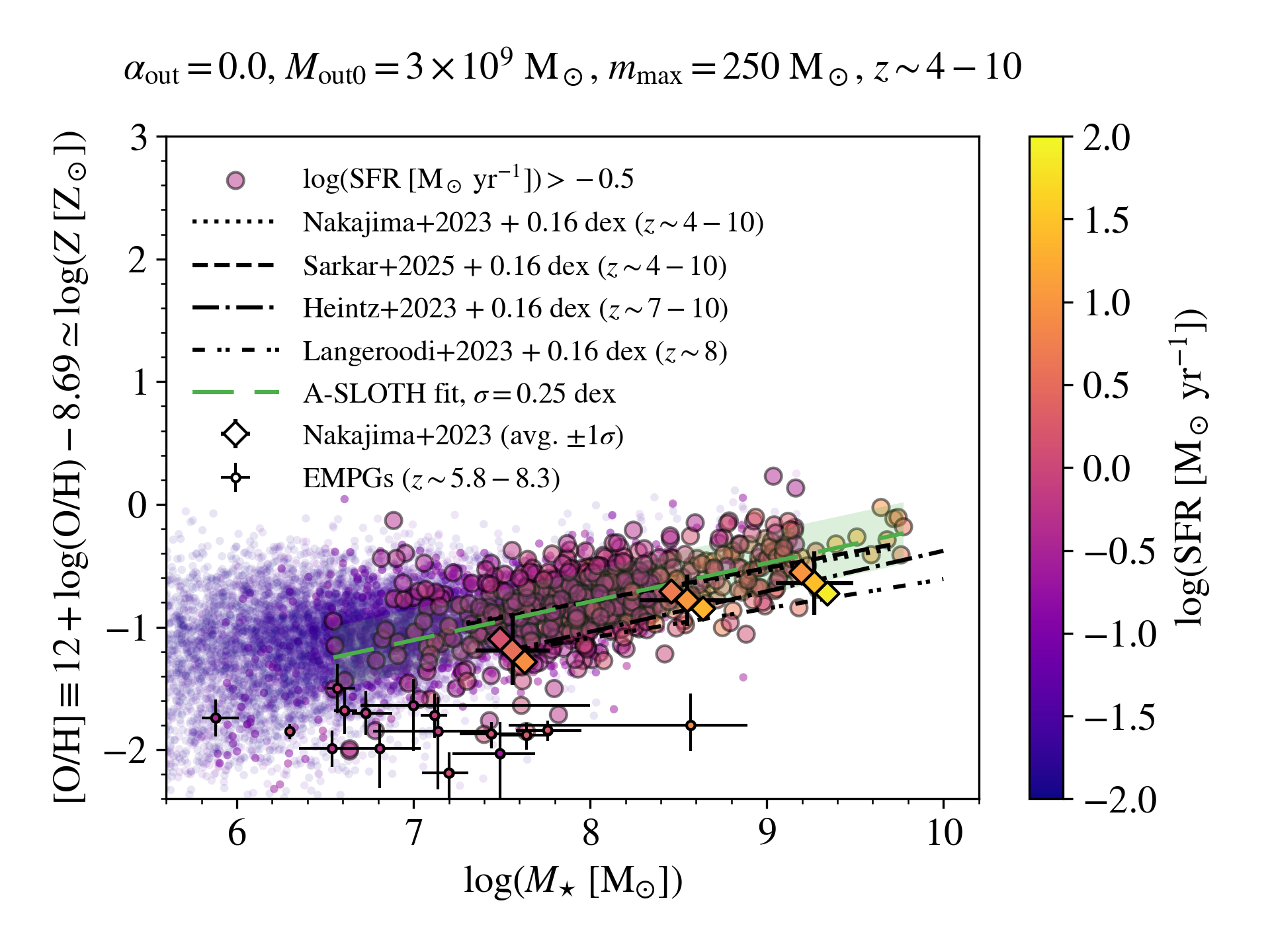}
    \caption{MZR at $z\sim 4-10$ for the best-fit model with $\alpha_{\rm out}=0$, $M_{\rm out0}=3\times 10^{9}\ \rm M_\odot$, and $m_{\max}=250\ \rm M_\odot$. Simulated galaxies with $\log\rm (SFR\ [M_\odot\ yr^{-1}])>-2$ are shown as dots color-coded by SFR. The detectable ones with $\log\rm (SFR\ [M_\odot\ yr^{-1}])>-0.5$ are highlighted by large circles, from which we derive the MZR as the long dashed line. The dotted and dashed lines show the MZR of JWST galaxies at $z\sim 4-10$ from \citet{Nakajima2023} and \citet{Sarkar2025} (which almost coincide with each other and with the simulated MZR), while the dash-dotted and dot-dash-dotted lines show the fits from \citet[for $z\sim 8$]{Heintz2023} and \citet[for $z\sim 7-10$]{Langeroodi2023}. Here, every observed MZR has been corrected for the SFR bias (see Sec.~\ref{sec:mzsfr} and Appendix~\ref{apdx:sfms}). The original average values (without correction) of $\log M_\star$, [O/H], and SFR of the JWST galaxies from \citet[see their table.~2]{Nakajima2023} are shown with the diamonds. Each data point involves three diamonds color-coded by the mean and $1\sigma$ upper and lower limits of SFR. The relevant scatter in $\log M_\star$ and [O/H] are shown by errorbars. The small dots with errorbars denote the extremely metal-poor galaxy candidates ($\rm [O/H]\lesssim-1.7$) observed by JWST at $z\sim 5.8-8.3$ \citep{Chemerynska2024,Mowla2024,Cullen2025,Hsiao2025} color-coded by the best-fit SFR value measured from observations.}
    \label{fig:mzr}
\end{figure}

The likelihood analysis in the previous subsection identifies the best-fit model \textit{among all simulations} as $\alpha_{\rm out}=0$, $M_{\rm out0}=3\times 10^{9}\ \rm M_\odot$, and $m_{\max}=250\ \rm M_\odot$. Here, we assess the consistency of the predictions of this particular model (referred to as `our simulation') with observations, and highlight its distinct features compared with the other models.

In the left panel of Fig.~\ref{fig:csfh}, we plot the SFRD of simulated galaxies above different SFR thresholds on top of the observational data summarized in Table~\ref{tab:sfrd} { used in the likelihood analysis and those for cross-validation from \citet{Chemerynska2025}. In particular, the thin solid curve shows the SFRD of galaxies with $\rm SFR> 10^{-0.5}\ M_\odot\ yr^{-1}$ corresponding to the UV magnitude limit $M_{\rm UV}\lesssim -17$ that are \textit{detectable} in most existing surveys.} For these galaxies, our predictions are excellently consistent with observations within $\lesssim 1\sigma$ at $z\sim 6-13$. The SFRD measurement from \citet{Harikane2023} at $z\sim 16$, which is highly uncertain anyway, is not reproduced, likely due to our limited simulation volume. The SRFD at $z\sim 5$ is underpredicted by a factor of $\sim 2$. It is discussed in Appendix~\ref{apdx:csfh_alt} that this discrepancy can be avoided if we use $\alpha_{\rm out}\sim 0.5$ and $M_{\rm out0}\sim 3\times10^{9}\ \rm M_\odot$ at $z\lesssim 6$. 

The total SFRD (thick solid) is larger than that of typical detectable galaxies, as fainter ($M_{\rm UV}\gtrsim -17$) galaxies become more important at higher redshifts. 
{Considering such fainter galaxies, our results agree well with the SFRD measurements from the GLIMPSE survey \citep{Chemerynska2025} at $z\sim 8.5-15$ for $M_{\rm UV}<-16$ and $M_{\rm UV}<-13$, corresponding to $\rm SFR\gtrsim 0.1\ \rm M_\odot\ yr^{-1}$ and $\rm SFR\gtrsim0.01\ \rm M_\odot\ yr^{-1}$, respectively, showing that the decrease of SFRD with increasing redshift is much slower compared with the case of $M_{\rm UV}<-17$. At $z\sim 7.5$, the SFRD for $\rm SFR>0.01\ \rm M_\odot\ yr^{-1}$ (accounting for $\sim 80\%$ of the total SFRD) is higher than that for $\rm SFR\gtrsim 0.3\ \rm M_\odot\ yr^{-1}$ by $\sim0.25\ \rm dex$ in our simulation, which is again consistent with the finding from the GLIMPSE survey \citep{Korber2025}. } 
{In fact, simulated galaxies with $\rm SFR< 10^{-0.5}\ M_\odot\ yr^{-1}$ comprise $\gtrsim 90\%$ of the total SFRD at $z\gtrsim 12$, whose contribution is still significant ($\sim 30-50\%$) approaching the end of reionization ($z\sim 5-7$). This is in line with the observational evidence for an unseen population of faint galaxies around [OIII] emitters in ionized bubbles at $z\sim 6$ \citep{Kakiichi2025}. Such faint galaxies could be the main driver of reionization \citep[e.g.,][]{Atek2024} and also make significant contributions to the merger rate density of binary black holes with their metal-poor ($Z\lesssim 0.1\ \rm Z_\odot$) stellar populations (see Sec.~\ref{sec:imp}).}

In the right panel of Fig.~\ref{fig:csfh}, we show the cosmic stellar mass density (CSMD) of galaxies above different stellar mass thresholds in comparison with the result for $M_{\star}>10^{8}\ \rm M_\odot$ from the empirical model in \citet[see Eq.~\ref{eq:csmd}]{Donnan2025}. Again, the observed CSMD is well reproduced at $z\lesssim 10$, while the agreement breaks down at higher $z$ due the poor sampling of massive galaxies in our small simulation box. Low-mass ($M_{\star}\lesssim 10^{8}\ \rm M_\odot$) galaxies dominate the cumulative stellar mass budget at $z\gtrsim 10$.

Regarding chemical evolution of high-$z$ galaxies, we first plot individual simulated galaxies in the $\rm [O/H]$--$\mu_{0.66}$ space in Fig.~\ref{fig:fmr} for two redshift ranges: $z\sim 4-8$ (left) and $z\sim 8-10$ (right). The universal fundamental metallicity relation (FMR) from \citet{Andrews2013}, which appears to be invariant up to $z\sim 8$, and the observational results of JWST at $z\sim 4-10$ from \citet{Nakajima2023} are shown for comparison. We derive a linear fit from the simulation data for detectable galaxies with $\rm SFR>10^{-0.5}\ \rm M_\odot\ yr^{-1}$ and calculate the scatter around this fit. In general, we find good agreement between theoretical predictions and observations. The linear fit for simulated galaxies at $z\sim 4-8$ gives $\rm [O/H]=(0.35\pm 0.02)\mu_{0.66}-3.56\pm0.19$ with a scatter of $\sigma\sim 0.24$~dex. This is consistent with the FMR at $z\lesssim 8$ in observations (Eq.~\ref{eq:fmr}) within $1\sigma$ but has a slightly smaller slope. We find that the FMR slope increases with larger $\alpha_{\rm out}$ in our simulations (reaching $\sim 0.7$ for $\alpha_{\rm out}=1$), which reflects the enhanced (reduced) ability of metal retention under weaker (stronger) outflows in massive (low-mass) halos (see Eqs.~\ref{eq:gamma_out} and \ref{eq:dmout}). If we use $\alpha_{\rm out}\sim 0.5$ and $M_{\rm out0}\sim 3\times10^{9}\ \rm M_\odot$ at $z\lesssim 6$, as supported by the observed CSFH (see Fig.~\ref{fig:csfh} and Appendix~\ref{apdx:csfh_alt}), the FMR slope becomes larger, achieving perfect agreement with the observed FMR. 
The scatter around the FMR in observations at $z\lesssim 3$ is $\sigma=0.13$~dex \citep{Andrews2013}, smaller than that in our simulation. However, we find a similar (total) scatter $\sigma\sim 0.27$~dex for the JWST galaxies at $z\sim 4-10$ \citep[see fig.~6 of][]{Sarkar2025}, although the intrinsic scatter is expected to be smaller and closer to the low-$z$ value once observational uncertainties are removed.

The breaking of the universal FMR at $z\gtrsim 8$ is well reproduced by our simulation (even if we do not consider $z> 8$ galaxies in the likelihood of MZSFR): There is a systematic offset from the universal FMR by around $-0.27$~dex for both simulated and observed galaxies at $z\sim 8-10$. Besides, the scatter around FMR generally increases at higher $z$ in our simulation as $\sigma\sim0.22$, $0.26$, and $\sim 0.3$~dex at $z\sim 4-6$, $z\sim 6-8$, and $z\sim 8-10$, respectively. This implies that the FMR can have subtle redshift evolution even at $z\lesssim 8$, as shown in cosmological hydrodynamic simulations \citep[e.g.,][]{Garcia2024,Garcia2025}. Current observations and our simulations at $z\gtrsim 4$ are still limited by galaxy sample sizes and are therefore unable to accurately characterize such redshift evolution.

Next, we consider the stellar mass-metallicity relation (MZR) at $z\sim 4-10$ in a similar manner, as shown in Fig.~\ref{fig:mzr}. Here we plot the MZR from four observational studies: \citet{Sarkar2025}, \citet{Nakajima2023}, \citet[for $z\sim 8$]{Langeroodi2023}, and \citet[for $z\sim 7-10$]{Heintz2023}. We shift the observed MZR upwards by $\Delta\rm [O/H]\simeq 0.16$~dex in each case to correct for observational biases (detailed in Sec.~\ref{sec:mzsfr} and Appendix~\ref{apdx:sfms}). The first two studies consider galaxies exactly at $z\sim 4-10$ and produce almost identical results that are in perfect agreement with the MZR derived from the detectable galaxies with $\rm SFR>10^{-0.5}\ \rm M_\odot\ yr^{-1}$ in our simulation, namely, $\rm [O/H]=(0.31\pm 0.02)\log(M_{\star}\ [\rm M_\odot])-3.3\pm 0.14$. The latter two studies only involve galaxies at $z\sim 7-10$, so the resulting MZR is systematically lower by $\sim 0.2-0.3$~dex, consistent with the trend seen in the FMR. 
The scatter in $\rm [O/H]$ around the MZR is $\sigma\sim 0.2-0.3$~dex in observations \citep{Nakajima2023}, which is consistent with the scatter $\sigma\simeq 0.25$~dex found in the simulated galaxies. Note that the scatter reported by \citet{Nakajima2023} also includes the uncertainties in metallicity and mass measurements. Subtracting such uncertainties can result in a smaller intrinsic scatter $\sim 0.16$~dex as estimated by \citet{Sarkar2025}. 

Below $M_\star\lesssim 10^8\ \rm M_\odot$, the MZR of simulated galaxies starts to flatten towards smaller $M_\star$, and meanwhile, the scatter around MZR increases. Thanks to this boosted metallicity spread, our simulation can produce extremely metal-poor galaxies with $\rm [O/H]\lesssim -1.7$ (i.e., below $\sim 2\%$ solar metallicity), $\log(M_{\star}\ [{\rm M_\odot}])\sim 6.5-8$, and $\rm SFR\sim 0.2-2\ M_\odot\ yr^{-1}$, similar to those found in recent JWST observations at $z\sim 5.8-8.3$ \citep{Chemerynska2024,Mowla2024,Cullen2025,Hsiao2025}. Like in the case of FMR, the MZR slope increases with larger $\alpha_{\rm out}$ in our simulations, so $\alpha_{\rm out}\sim 0$ is also supported by the observed slope.

\begin{figure*}
    \centering
    \includegraphics[width=1\linewidth]{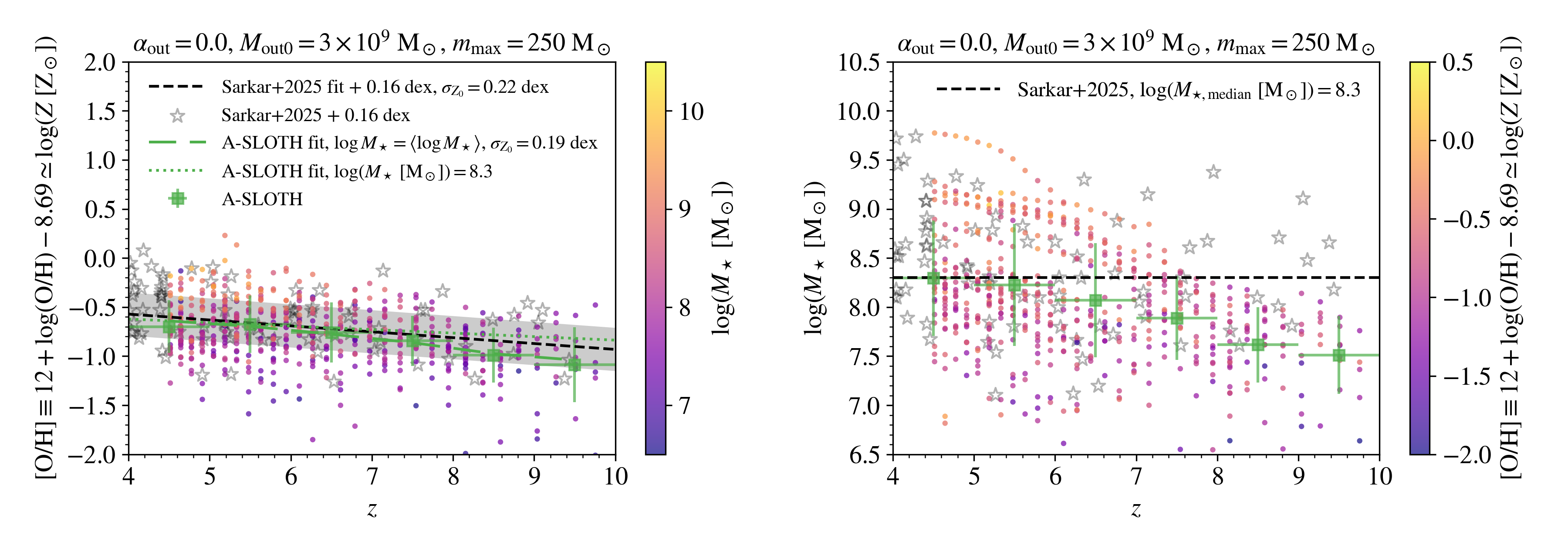}
    \caption{Redshift evolution of metallicity (left) and stellar mass (right) of detectable galaxies with $\log\rm (SFR\ [M_\odot\ yr^{-1}])>-0.5$ in our simulation for the best-fit model with $\alpha_{\rm out}=0$, $M_{\rm out0}=3\times 10^{9}\ \rm M_\odot$, and $m_{\max}=250\ \rm M_\odot$. Individual observed galaxies from \citet{Sarkar2025} are shown as empty stars. Individual simulated galaxies are shown as dots, color-coded by $\log M_{\star}$ and [O/H] on the left and right panels, respectively. The corresponding average value and $1\sigma$ scatter at 6 redshift bins (with $\Delta z=1$) are shown as the squares and errorbars. A linear fit in the form of ${\rm [O/H]}=Z_0+a\log (M_\star\ {[\rm M_\odot}])+b(1+z)$ is derived from the simulated galaxies. In the left panel, we show this fit evaluated at the average $\log M_\star$ of simulated galaxies (with weights described in Sec.~\ref{sec:mzsfr}) as the lone dashed line and at the median stellar mass of observed galaxies as the dotted line. For the former, the uncertainty in the normalization $Z_0$ is $\sigma_{Z_0}\simeq 0.19$~dex. 
    For the latter, we take the median stellar mass of galaxies observed by JWST $\log(M_{\star,\rm median}\ [\rm M_\odot])=8.3$ from \citet{Sarkar2025}, which is shown by the dashed horizontal line on the right panel. The same median stellar mass is substituted to the [O/H]-$\log M_\star$--$z$ relation derived from the JWST galaxies \citep[see their eq.~12 and fig.~8]{Sarkar2025}, as shown by the dashed line on the left panel. Here, we also shift the observational data points and relation upwards by $\Delta[\rm O/H]_{\rm MZR}\sim 0.16$~dex to correct for the bias in SFR (see Sec.~\ref{sec:mzsfr} and Appendix~\ref{apdx:sfms}). The relevant uncertainty in $Z_0$ from observations is $\sigma_{Z_0}=0.22$~dex as shown by the gray shaded region, which agrees well with that found in our simulation ($\sigma_{Z_0}\simeq 0.19$~dex). }
    \label{fig:MZz}
\end{figure*}

Figure~\ref{fig:MZz} compares the redshift evolution of metallicity (left) and stellar mass (right) of \textit{detectable} galaxies ($\rm SFR>10^{-0.5}\ M_\odot\ yr^{-1}$) in our simulation with that seen in observations \citep{Sarkar2025}. Here, the observed galaxies has a median stellar mass $M_{\star}\sim 10^{8.3}\ \rm M_\odot$ that hardly evolves with redshift. The simulated galaxies have a mass distribution around this median value similar to the observed one at $z\lesssim 7$. However, at higher $z$, the simulated galaxies are systematically less massive than the observed galaxies by $\sim 0.4-0.8$~dex. This again reflects the fact that high-$z$ observations are biased by massive, luminous objects with respect to the galaxy sample in our small simulation box. \citet{Sarkar2025} fit the observed galaxies with a plane in the [O/H]-$\log M_\star$--$z$ space as
\begin{align}
    {\rm [O/H]}=Z_0+a\log (M_\star\ {[\rm M_\odot}])+b(1+z)\ ,\label{eq:MZz}
\end{align}
with best-fit values $Z_0=-2.40\pm 0.22$, $a=0.237\pm 0.023$, and $b=-0.06\pm 0.01$. To compare this with our simulation data, we shift the normalization $Z_0$ upwards by $\rm\Delta[O/H]_{\rm MZR}=0.16$~dex to correct for the SFR bias (see Sec.~\ref{sec:mzsfr} and Appendix~\ref{apdx:sfms}), which gives $Z_0=-2.24\pm 0.22$. As shown in the left panel of Fig.~\ref{fig:MZz}, the simulated galaxies have a similar distribution in the ${\rm [O/H]}-z$ space as the observed distribution at $z\lesssim 7$, and become more metal-poor at $z\gtrsim 7$. Fitting the simulated galaxies with the same formula (Eq.~\ref{eq:MZz}) gives $Z_0=-2.77\pm 0.19$, $a=0.281\pm 0.019$, and $b=-0.036\pm 0.008$, which are generally consistent with the corrected best-fit plane in observations within $\sim 1\sigma$. The main difference is that there is stronger (weaker) dependence of metallicity on mass (redshift) in the simulated galaxies. This seems counterintuitive considering the selection bias towards more massive galaxies at higher $z$ (see the right panel of Fig.~\ref{fig:MZz}). If these galaxies are metal-richer than lower-mass galaxies, there should be an underestimation of the redshift evolution (with larger $b$) in the [O/H]-$\log M_\star$--$z$ relation from observations \citep{Sarkar2025}. However, higher-$z$ observations of metal lines are also biased towards higher-SFR galaxies to a degree larger than that of the stellar mass bias. Such galaxies can be more metal-poor despite their larger masses as indicated by the FMR (Fig.~\ref{fig:fmr}), thus reversing the trend. This highlights the importance of taking SFR into account for characterization of galaxy chemical evolution. 

\begin{figure}
    \centering
    \includegraphics[width=1\linewidth]{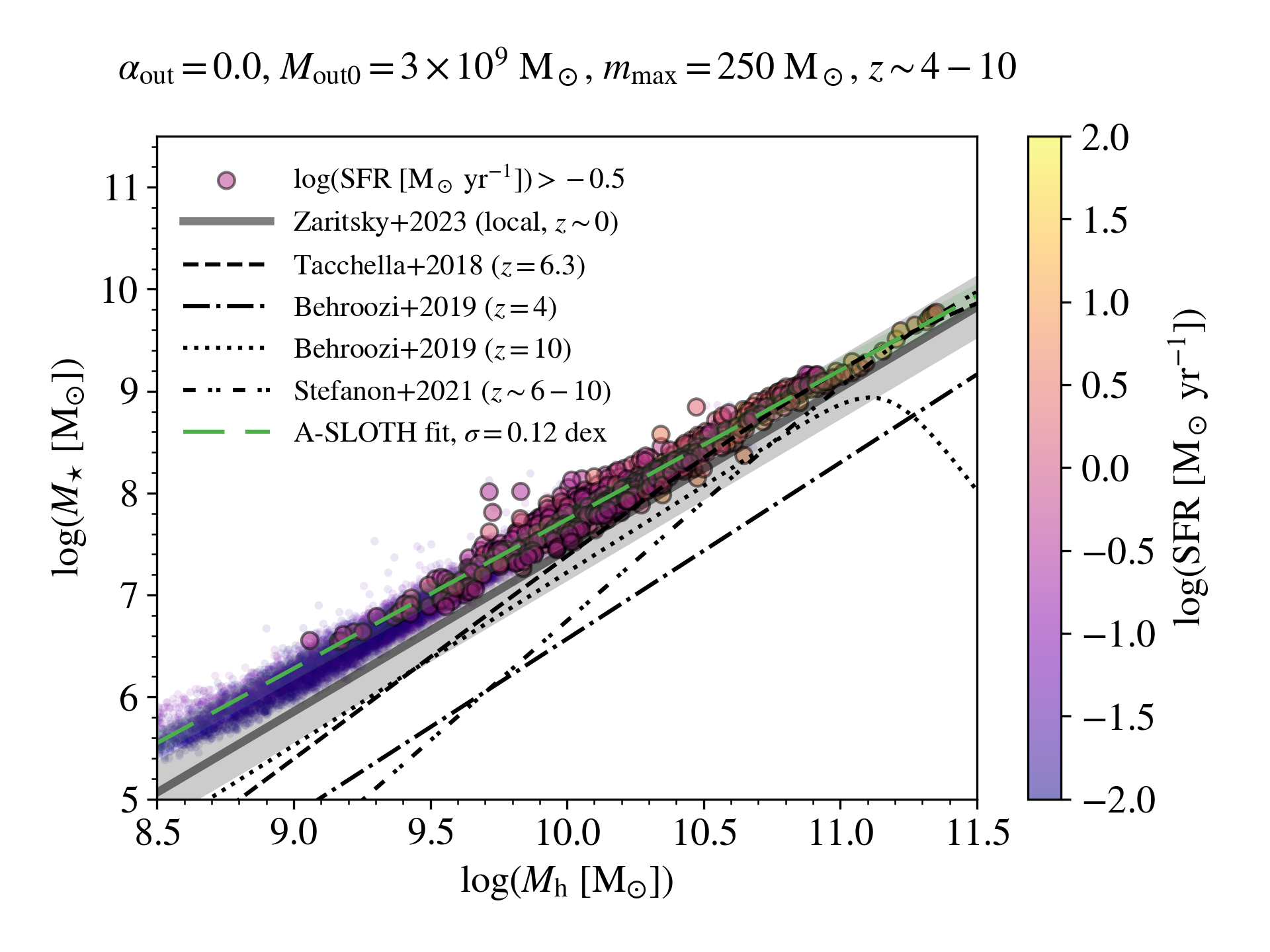}
    \caption{SHMR at $z\sim 4-10$ for the best-fit model with $\alpha_{\rm out}=0$, $M_{\rm out0}=3\times 10^{9}\ \rm M_\odot$, and $m_{\max}=250\ \rm M_\odot$. Simulated galaxies with $\log\rm (SFR\ [M_\odot\ yr^{-1}])>-2$ are shown as dots color-coded by SFR. The detectable ones with $\log\rm (SFR\ [M_\odot\ yr^{-1}])>-0.5$ are highlighted by large circles, from which we derive the SHMR as the long dashed line and green shaded region (for $1\sigma$ scatter). The results of four semi-empirical models are shown for comparison: \citet[thick solid, at $z\sim 0$]{Zaritsky2023}, \citet[dashed, at $z=6.3$ as the average redshift of detectable simulated galaxies]{Tacchella2018}, \citet[dash-dotted for $z=4$ and dotted for $z=10$]{Behroozi2019}, and \citet[dot-dash-dotted, at $z\sim 6-10$]{Stefanon2021}. The shaded region around the solid curve illustrates the scatter $\sigma_{\rm SHMR}=0.31$~dex from \citet{Zaritsky2023}.}
    \label{fig:shmr}
\end{figure}

We end this subsection with a brief discussion of the galaxy-halo connection. In Fig.~\ref{fig:shmr}, we compare the SHMR of simulated galaxies (with $\rm SFR>10^{-0.5}\ \rm M_\odot\ yr^{-1}$) at $z\sim 4-10$ with the results of four semi-empirical models: \citet[at $z=6.3$]{Tacchella2018}, \citet[at $z=4$ and 10]{Behroozi2019}, \citet[for $z\sim 6-10$]{Stefanon2021}, and \citet[in the local Universe, $z\sim 0$]{Zaritsky2023}. Although these models show large discrepancies among themselves (up to $\sim 1$~dex in stellar mass for a given halo mass), they all lie below the SHMR of our simulated galaxies. One explanation is that because massive star-forming galaxies in large-scale over-dense regions are poorly sampled in our small simulation box, the sampled halos are forced to form more stars than they would on cosmic average in order to reproduce the observed CSFH. Despite this possible bias from cosmic variance, the simulated SHMR is consistent with that inferred from local ($z\sim 0$) dwarf galaxies \citep{Zaritsky2023} within $1\sigma$ and covers a reasonable range of halo-scale cumulative SFE $\eta_{\rm halo}\equiv M_{\star}/[(\Omega_{\rm b}/\Omega_{\rm m})M_{\rm h}]\propto M_{\rm h}^{0.46}\sim 0.014-0.2$ for $M_{\rm h}\sim 10^9-10^{10.5}\ \rm M_\odot$. The scatter in $M_\star$ around the SHMR is $\sigma\sim 0.12$~dex, comparable to that ($\sigma\sim 0.14-0.18$~dex) found in \citet{Tacchella2018}.

\section{Discussion}
\label{sec:diss}

The star formation and chemical enrichment histories of high-$z$ galaxies have broad implications for the observational signatures of the early Universe beyond those considered in our likelihood analysis (Sec.~\ref{sec:obs}). However, it is challenging to make robust theoretical predictions due to uncertainties in the modeling of the rich physics involved across a large range of scales. Such uncertainties call for prudent interpretations of our findings. 
Below, we first briefly comment on the implications of our results on PISNe and binary black hole mergers as high-$z$ transient sources (Sec.~\ref{sec:imp}). Then we discuss the uncertainties/caveats in our simulations (Sec.~\ref{sec:caveats}). 

\subsection{Implications for transient rates}
\label{sec:imp}

The cosmic rate of PISNe is determined by (1) metal-dependent CSFH, (2) IMF, and (3) (binary) stellar evolution \citep[e.g.,][]{Briel2022,Tanikawa2023,Gabrielli2024,Simonato2025}. Although current observational constraints are limited to the local Universe \citep[e.g.,][]{Schulze2024}, future observations with upcoming facilities (e.g., Roman Space Telescope and Vera C. Rubin Observatory) will probe the high-$z$ regime \citep[e.g.,][]{Liu2020,Venditti2024sn}. In this work, (3) is fixed, (2) is calibrated to reproduce the observed MZSFR and overall CSFH (of bright galaxies), and (1) is predicted by the simulation. Since IMF models with $m_{\rm \max}\gtrsim 200\ \rm M_\odot$ are preferred by the observed MZSFR, we find significant contributions of Pop~II stars to PISN events at $z\lesssim 15$. The resulting PISN rate (i.e., all-sky number of events per year per redshift per unit solid angle) peaks around $z\sim 5$ as $d\mathcal{\dot{N}}_{\rm PISN}/dz\sim 10\ \rm yr^{-1}\ deg^{-2}$ in our best-fit models (Sec.~\ref{sec:interplay}) with $m_{\max}\sim 200-250\ \rm M_\odot$. On the other hand, for the models with $m_{\max}=100\ \rm M_\odot$ that marginally reproduce the observed CSFH, 
PISNe are only produced by Pop~III stars. The corresponding PISN rate reaches a much lower peak of $d\mathcal{\dot{N}}_{\rm PISN}/dz\sim 0.1\ \rm yr^{-1}\ deg^{-2}$ around $z\sim 10-15$ and decreases towards lower $z$ down to $\sim 0.01\ \rm yr^{-1}\ deg^{-2}$ at $z\sim 5$, generally consistent with the findings of previous studies considering only Pop~III PISNe \citep[e.g.,][]{Hummel2012,Magg2016,Lazar2022,Venditti2024sn,Wiggins2024}. 
Irrespective of origin (Pop~II or Pop~III), the possible nucleosynthetic PISN signature could be preserved in the high-$z$ IGM, which in turn could be probed through future absorption spectroscopy with bright gamma-ray burst afterglows as background sources \citep{Wang_PISN2012}.

Our predictions on metal-dependent CSFH also have interesting implications for gravitational wave transients such as binary black hole mergers. Binary population synthesis studies show that the formation efficiency of binary black hole mergers from isolated binary stellar evolution increases significantly (by $\sim 1-4$ orders of magnitude) when the stellar metallicity decreases from solar to below $\sim 0.1\ \rm Z_\odot$ \citep[e.g.,][]{Klencki2018,Giacobbo2018,Spera2019,Broekgaarden2022,Iorio2023,Li2023}. As a result, the cosmic merger rate density of binary black holes is highly sensitive to the SFRD of metal-poor stars \citep{Boco2019,Boco2021,Chruslinska2019,Chruslinska2021,Briel2022,Chruslinska2024,Sgalletta2024,vanson2025}. As discussed by \citet{Sgalletta2024}, most studies derive the metal-dependent SFRD from the observed total SFRD and galaxy scaling relations under various assumptions, which suffer from large uncertainties regarding the contribution of low-mass ($M_\star\lesssim 10^8\ \rm M_\odot$) galaxies that typically host metal-poor stars but are difficult to detect at high $z$. 

With the well-calibrated semi-analytical galaxy evolution model of \textsc{a-sloth}, our simulations naturally predict the metal-dependent SFRD, self-consistently capturing the contribution of the smallest galaxies in the standard $\Lambda$CDM universe (with $M_\star\sim 1000\ \rm M_\odot$ and $M_{\rm h}\sim 10^6 \rm M_\odot$) thanks to the high resolution. It is found in our best-fit model that the SFRD of metal-poor stars with $\rm [O/H]<-1$ is $\sim 0.005-0.01\ \rm M_\odot\ yr^{-1}\ cMpc^{-3}$ at $z\sim 4.5-8$, which slowly decreases towards lower $z$ and remains within the range ($0.001-0.02\ \rm M_\odot\ yr^{-1}\ cMpc^{-3}$) found in semi-empirical models \citep[e.g.,][]{Chruslinska2019,Chruslinska2021}. Moreover, the metal-poor SFRD dominates the total SFRD of simulated galaxies at $z\gtrsim 8$ that are mostly very faint with $\rm SFR\lesssim 0.1\ M_\odot\ yr^{-1}$ (see Fig.~\ref{fig:csfh}). The predicted metal-poor SFRD also exceeds the SFRD of UV-luminous ($M_{\rm UV}\lesssim-17$) galaxies inferred from observations (Table~\ref{tab:sfrd}) by up to one order of magnitude at $z\sim 8-13.5$. 

This highlights the important role of faint, low-mass, metal-poor galaxies in producing high-$z$ binary black hole mergers, which will be observed by 3rd-generation gravitational wave detectors to bring us additional constraints on early galaxy evolution \citep{Abac2025}. We plan to explore the rates and detectability of binary black hole mergers and PISNe in detail in a follow-up study.

\subsection{Caveats}
\label{sec:caveats}

Although our best-fit model reproduces observations very well (Sec.~\ref{sec:best}), the particular IMF and galactic outflow properties required in this model are sensitive to the assumptions made in the galaxy evolution prescription (Sec.~\ref{sec:gal_evo}) and the numerical setup of simulations (Sec.~\ref{sec:setup}). Below, we summarize the key underlying caveats/uncertainties in our modeling (in descending order of the scales involved) and evaluate their possible effects on our results. 

As a tradeoff for high resolution, our simulation volume is rather small ($\sim\rm 2000\ cMpc^{3}$) and cannot capture large-scale ($\gtrsim\rm 10~cMpc$) perturbations, leading to poor sampling of the relevant over-dense regions hosting massive/high-SFR galaxies. On the other hand, high-$z$ observations (of metal lines) are biased towards luminous objects. The observational data from JWST spectroscopy considered in our analysis mostly involve galaxies with $M_{\rm UV}\lesssim -17$ (corresponding to $\rm SFR\gtrsim 10^{-0.5}\ \rm M_\odot\ yr^{-1}$) at $z\sim 4-10$. As an attempt to achieve fair comparison, we only focus on such detectable galaxies and consider their cumulative (rather than statistical) properties and scaling relations. However, the actual number of galaxies with $\rm SFR> 10^{-0.5}\ \rm M_\odot\ yr^{-1}$ in our simulation volume is only around $\sim 3-40$ at $z\sim 4-10$, which is lower than the number of observed galaxies by at least a factor of a few. Therefore, our results may suffer from uncertainties of small-sample statistics\footnote{In practice, the evolution tracks of simulated galaxies are `observed' at many snapshots to produce a large galaxy sample that is used to compare with observations. This approach can capture the internal stochastic processes in galaxy evolution. A large number of evolution tracks is still needed to fully capture the diversity of halo assembly histories and environmental factors (e.g., external metal enrichment and ionization feedback).}. Beyond statistical uncertainties, there are also systematic differences between the simulated and observed galaxies under the same detection limit/SFR threshold. The latter generally have a higher SFR (at a given stellar mass) by $\sim 0.6$~dex (Appendix~\ref{apdx:sfms}), and are also more massive at $z\gtrsim 7$ (Fig.~\ref{fig:MZz}). 
Although we manage to make a correction to MZR through FMR {(assuming that an invariant, tight FMR as that observed by \citealt{Andrews2013} at $z\lesssim 3$ holds up to $z\sim 8$ with clear anti-correlation between SFR and metallicity)} to ensure their consistency (Sec.~\ref{sec:mzsfr}), it is non-trivial to evaluate the effects of selection biases on galaxy scaling relations{ \citep{Lewis2025}. Without this correction, the best-fit models still have $m_{\max}\ge 200\ \rm M_\odot$. However, the constraining power of MZSFR becomes weaker, such that the models with $m_{\max}<200\ \rm M_\odot$ are also consistent with the observed MZSFR within $2\sigma$.}

Accurate measurements of SFR, stellar mass, and metallicity are challenging in observations{, especially for high-$z$ galaxies \citep[e.g.,][]{Lonoce2025,Stanway2025}}. The observational results considered here are mostly based on indicators/calibrations established in the low-$z$ Universe, which may cause large uncertainties and even systematic biases when applied to high-$z$ galaxies{, considering their compact, clumpy, metal-poor, and volatile nature \citep[][]{Choustikov2025,Kramarenko2025,Vijayan2025}}. For instance, it is shown in \citet{Wang2025sfhobs} that state-of-the-art inference methods fail to capture SFR fluctuations on tens of Myr timescales, thus the stellar mass is typically underestimated by $\sim 0.15$~dex in bursty systems, while the opposite, i.e., overestimation of stellar mass (by $\sim 60\%$ up to a factor of 5), is found by \citet{Choe2025}. Moreover, large discrepancies (up to $\sim 0.4$~dex) exist in the metallicity measurements from different calibrations \citep[e.g.,][see their fig.~5]{Chemerynska2024}, which, mixed up with selection biases, can shift the metallicity scaling relations around \citep[e.g.,][]{KorhonenCuestas2025}. In fact, the MZR at $z\sim 3-10$ obtained by \citet{Chakraborty2024} using direct-$T_{\rm e}$-based metallicities is $\sim 0.2$~dex below the MZR adopted in our analysis from earlier studies using line-ratio metallicity indicators \citep{Nakajima2023,Curti2024,Sarkar2025}. If we attribute this outcome to a systematic overestimation of metallicities by the canonical line-ratio method, less metal yields will be required to reproduce the observed MZSFR, corresponding to $m_{\max}\sim 100-200\ \rm M_\odot$ according to our stellar population model (Sec.~\ref{sec:stellar}). On the other hand, the direct-$T_{\rm e}$ method also suffers from uncertainties in ISM photo-ionization models of the interstellar medium \citep[e.g.,][]{Cameron2023,Hayes2025} and can underestimate the metallicity {(by up to $\sim 0.7$~dex) if a low electron density ($n_{\rm e}\sim 100\ \rm cm^{-3}$) suitable for local galaxies is assumed while higher electron densities ($n_{\rm e}=10^3-10^6\ \rm cm^{-3}$) are typically found in high-$z$ galaxies \citep[e.g.,][]{Mingozzi2022,Martinez2025}}. 
    
We adopt a simple model for metal enrichment in which metals are uniformly mixed into the gas reservoir of the halo 
(Sec.~\ref{sec:fdbk}). 
In reality, the distribution of metals in different components of gas (e.g., cold gas, hot gas, accreted gas, and outflows) can be more complex than this `well-mixed' picture. For instance, it is found by \citet[][]{Nishigaki2025} that a multi-phase metallicity model that treats metallicities in $\rm H_2$ and HI regions separately is needed to spontaneously reproduce metal scaling relations in star-forming regions, HI regions, and the CGM at $z\lesssim 5$. In particular, they find that the fraction of metals mixed into the star-forming regions decreases towards higher $z$ and smaller halo mass. In our work, we focus on a higher-$z$ regime, and only consider the metallicity of star-forming (HII) regions that is traced by nebular lines in observations and young ($\lesssim 10$~Myr) massive ($\gtrsim 5\ \rm M_\odot$) stars in our simulations. We use a stochastic model based on cosmological simulations \citep{Tarumi2020} to capture the difference between this metallicity and the `well-mixed' average gas-phase metallicity. Therefore, any uncertainties in the simulation results underlying this model will propagate into our results. Besides, our approach may fail to fully capture the redshift and halo mass dependence of the metal distribution fractions in star-forming and HI regions if its effects are not degenerate to those of galactic outflows explored in our simulations (which in principle only control the distribution of metals between the IGM and ISM/CGM). 
    
Our prescription for star formation and stellar feedback may be oversimplified \citep{Boardman2025}. For instance, the SFE is fixed throughout our simulations (Appendix~\ref{apdx:sf}), while analytical models and cloud-scale simulations of star formation typically find higher SFE for a higher cloud mass and/or gas surface density \citep[e.g.,][]{Federrath2012,Kim2018,He2019,Lancaster2021,Menon2024,Polak2024}, which is also confirmed in observations \citep{Rawat2025}. Taking this into account leads to higher SFE in higher-$z$ galaxies, which helps explain the `excess' of UV-luminous galaxies observed by JWST \citep[e.g.,][]{Inayoshi2022,Dekel2023,Dhandha2025,Kar2025,Somerville2025,Yung2025}. The model for SN-driven galactic outflows can also be validated/refined based on the outflow properties inferred from observations \citep[e.g.,][]{Xu2022,Xu2025,Xu2025out,Lyu2025,Birkin2025}, 
which we defer to future work. 
In general, our results are model-dependent and should be interpreted with caution \citep[for alternative semi-analytical galaxy evolution models, see sec.~1.2 of][]{Hartwig2022}. 
It has also been found in cosmological hydrodynamic simulations that the detailed features of MZSFR, i.e., the normalizations, slopes, and scatters of MZR and FMR as well as their redshift evolution, are highly sensitive to the sub-grid models for star formation and stellar feedback \citep[e.g.,][]{Garcia2024,Garcia2025}. Current samples of observed galaxies at $z\gtrsim 4$ are still too small to characterize such detailed features accurately, and the same is true for the galaxy populations in our work with a limited simulation volume. 
    
We assume that the IMF is an invariant function at $z > 4.5$, thus focusing on the galaxy-population-averaged IMF. In reality, this IMF can still vary with redshift due to the variation of star formation condition and the environmental dependence of IMF that has been commonly seen in simulations \citep[e.g.,][]{Chon2021,Mathew2021,Mathew2023,Mathew2025,Guszejnov2022,Hennebelle2022,Hix2023,Chon2024,Liu2024,Tanvir2024} and observations \citep[e.g.,][]{Gunawardhana2011,Marks2012,Jerabkova2018,Dib2023,Rusakov2023}. In fact, we find tentative evidence for evolution of galactic outflow parameters with redshift at $z\lesssim 6$ based on the observed CSFH (see Fig.~\ref{fig:csfh} and Appendix~\ref{apdx:csfh_alt}), which favors decreasing outflow efficiency at lower redshifts in relatively massive halos ($M_{\rm h}\gtrsim 10^9\ \rm M_\odot$). This evolution may also be interpreted as the variation of IMF or stellar evolution tracks (see below), reducing the number and/or SN energy output of massive stars in lower-$z$, more massive, and metal-richer galaxies. 
    
Our results are also sensitive to uncertainties in stellar evolution models regarding initial chemical composition, nuclear reaction rates, convection, mixing, winds, rotation, SN explosion physics, and binary interactions, which are still under intense investigation. Such uncertainties not only affect the radiative feedback of living massive stars but also regulate their metal yields and SN energy outputs \citep[e.g.,][]{Cescutti2010,Nomoto2013,Kobayashi2020,Marchant2020,Farmer2021,Farmer2023,Tanikawa2021,Briel2022,Jeena2023,Martinet2023,Sabhahit2023,Gabrielli2024,Lecroq2024,Roberti2024,Tsiatsiou2024,West2024,Yates2024,Boco2025,Byrne2025,Dumont2025,Higgins2025,Liu2025,Liu_CHE2025,Pepe2025,Schneider2025,Shepherd2025,Simonato2025,Wasserman2025,Xin2025,Torniamenti2026}. Our finding that $m_{\max}\gtrsim 200\ \rm M_\odot$ (for a Kroupa-like IMF with a high-mass-end power-law slope of $\alpha_{\rm IMF}=2.3$) is favored by the observed MZSFR is based on the specific SN models and non-rotating single star evolution tracks discussed in Sec.~\ref{sec:stellar}. The key requirement to reproduce observations is an IMF-averaged metal yield (per unit stellar mass) {$\mathcal{M}_{Z}\sim 0.025$ with an IMF-averaged SN energy output (per unit stellar mass) $\mathcal{E}_{\rm SN}\sim 1.2-1.5\times 10^{49}\ \rm erg\ M_\odot^{-1}$} (see Fig.~\ref{fig:mzesn_mmax}). Alternative stellar evolution models and/or IMF forms can also provide similar conditions. For instance, if the upper mass limit is fixed to a lower value $m_{\max}=120\ (150)\ \rm M_\odot$ while the power-law slope $\alpha_{\rm IMF}$ at the high-mass end of the IMF ($m_\star\in [0.5\ {\rm M_\odot},m_{\max}]$) is allowed to vary, {$\mathcal{M}_{Z}\sim 0.025$ can be achieved by $\alpha_{\rm IMF}\sim 1.8$ (2.1) with a moderately larger SN energy output $\mathcal{E}_{\rm SN}\sim 2\times 10^{49}\ \rm erg\ M_\odot^{-1}$ for our stellar evolution and SN models}. More detailed observational data on multiple elements beyond the bulk metallicity traced by O can hopefully break such degeneracy \citep[e.g.,][]{Goswami2022,Arellano-Cordova2022,Charbonnel2023,D'Antona2023,Boardman2024,Nagele2023,Vink2023,Cameron2024,Nandal2024,Nandal2024vms,Rizzuti2024,Watanabe2024,Gieles2025,Ji2025,Kelly2025,Nandal2025,Schaerer2025,Nakane2025}.

\section{Summary and conclusion}
\label{sec:summary}

We explore the impact of IMF and SN-driven galactic outflows on the chemical evolution of high-redshift galaxies using the semi-analytical galaxy evolution code \textsc{a-sloth} \citep{Hartwig2022,Hartwig2024,Magg2022b} with updated prescriptions for star formation and stellar feedback coupled to stellar evolution models covering the full metallicity range ($Z \sim 10^{-11} - 0.03$), a broad stellar mass range ($m_\star\sim2 - 600\ \rm M_\odot$), and the metal yields from stellar winds, core-collapse SNe, (pulsational) PISNe, and Type Ia SNe \citep[]{Nomoto1997,Goswami2021,Goswami2022,Costa2025}. We perform 165 runs of \textsc{a-sloth} over the merger trees constructed from a high-resolution cosmological simulation \citep{Ishiyama2016} to search for the IMF and galactic outflow parameters that can best reproduce the cosmic star formation history \citep[CSFH,][]{Bouwens2016,Donnan2023full,Donnan2023,Harikane2023} and the stellar mass-metallicity-star formation rate relation \citep[MZSFR,][]{Nakajima2023,Curti2024,Sarkar2025} 
inferred from observations at $z\sim 4-10$. Here are our main findings:
\begin{itemize}
    \item An IMF-averaged metal yield per unit stellar mass {$\mathcal{M}_{Z}\sim 0.025$, combined with an IMF-averaged SN energy per unit stellar mass $\mathcal{E}_{\rm SN}\sim 1.2-1.5\times 10^{49}\ \rm erg\ M_\odot^{-1}$}, is required to reproduce the normalization of the observed MZSFR within $1\sigma$ {assuming that the anti-correlation between SFR and metallicity seen in the fundamental metallicity relation (FMR) at $z\lesssim 3$ \citep{Andrews2013} holds up to $z\sim 8$ as implied by the weak redshift evolution of FMR's median in JWST observations}. {If the IMF follows a Kroupa-like shape with a varying upper mass limit $m_{\max}$, this requirement can only be fulfilled when $m_{\max} \gtrsim 200\ \rm M_\odot$, which provides sufficient metal yields from PISNe (see Fig.~\ref{fig:mzesn_mmax}).} 
    This indicates that very massive ($\gtrsim 200\ \rm M_\odot$) stars and PISNe play important roles in galaxy chemical evolution in terms of overall metal production, consistent with the findings of recent studies based on abundance patterns \citep[e.g.,][]{ Blackwell2025,Gieles2025,Nandal2025,Rey2025,Shi2025}. 
    \item The CSFH inferred from observations involving UV-luminous ($M_{\rm UV}\lesssim -17$) galaxies at $z\gtrsim 6$ favors a galactic outflow model where the mass loss rate through outflows is proportional to the SN energy injection rate divided by the halo binding energy. This model is also preferred by the slope of MZSFR in observations. {Our results support the conclusion in previous studies \citep[e.g.,][]{Boardman2025,Koller2025,Nishigaki2025} that galactic metal enrichment is strongly regulated by the gravitational potential (captured by the binding energy in our case).} 
\end{itemize}
Given the IMF and galactic outflow parameters favored by existing observations of UV-luminous ($M_{\rm UV}<-17$) galaxies at $z\sim 4-10$ (Sec.~\ref{sec:obs}), our best-fit model makes several predictions for metal-poor, faint, low-mass galaxies (down to $M_\star\sim 1000\ \rm M_\odot$) at higher redshifts with interesting implications for future observations:
\begin{itemize}
    \item {The need for PISNe to provide enough metals by the observed MZSFR leads to high PISNe rates. The predicted all-sky event rate per redshift per unit solid angle peaks at $d\mathcal{\dot{N}}_{\rm PISN}/dz\sim 10\ \rm yr^{-1}\ deg^{-2}$ around $z\sim 5$, which is dominated by Pop~II PISNe at $z\lesssim 15$. Without very massive Pop~II stars, as in the case of $m_{\max}=100\ \rm M_\odot$, the PISN rate from massive Pop~III stars alone has a much lower peak of $0.1\ \rm yr^{-1}\ deg^{-2}$ around $z\sim 10-15$ and decreases towards lower $z$ down to $\sim 0.01\ \rm yr^{-1}\ deg^{-2}$ at $z\sim 5$.}
    \item {We predict a large population of UV-faint ($M_{\rm UV}\gtrsim -17$) galaxies, which comprise $\gtrsim 90\%$ of the total SFRD at $z\gtrsim 12$ and remain an important component ($\sim 30-50\%$) in the SFRD at $z\sim 5-7$. This prediction is consistent with the observational evidence for significant contributions to reionization by (unseen) faint galaxies \citep[e.g.,][]{Asthana2025,Atek2024,Kakiichi2025,Korber2025}.} 
    \item {A significant fraction of UV-faint galaxies host metal-poor ($Z\lesssim0.1\ \rm Z_\odot$) stars that are promising progenitors of binary black hole mergers. 
    Such metal-poor stars dominate the total SFRD at $z\gtrsim 8$ and even exceeds the SFRD of UV-luminous ($M_{\rm UV}<-17$) galaxies inferred from observations (Table~\ref{tab:sfrd}) by up to one order of magnitude at $z\sim 8- 13.5$. This highlights the important role of UV-faint, low-mass, metal-poor galaxies in making high-$z$ binary black hole mergers.}
\end{itemize}

In conclusion, we find that very massive stars (with initial masses $m_\star\gtrsim 200\ \rm M_\odot$, assuming a Kroupa-like IMF) and the relevant PISNe are needed to explain the star formation and chemical enrichment histories of high-redshift ($z\gtrsim 4$) galaxies observed by JWST, {leading to profound implications for reionization, electromagnetic transients (e.g., PISNe) and gravitational wave events (e.g., binary black hole mergers)}.  
More insights into the abundances and effects of very massive stars can be obtained from detailed chemical abundances of extragalactic galaxies/star clusters/clouds and metal-poor stars in our Galaxy. 
Benefiting from the flexibility and efficiency of \textsc{a-sloth}, this work establishes the foundation for future explorations of these topics.

We use the open-source semi-analytic code \textsc{a-sloth} (\url{https://gitlab.com/thartwig/asloth}) for our models \citep{Hartwig2022}. The new model for star formation and stellar feedback (Sec.~\ref{sec:gal_evo}) is implemented in the branch \texttt{cob\_sevn}. 
We use the open-source population synthesis code \textsc{sevn} (\url{https://gitlab.com/sevncodes/sevn}) to generate compact remnant catalogs \citep{Spera2019,Mapelli2020,Iorio2023} that are used to derive metal yields. 
The authors thank Gustavo Bruzual, Stephane Charlot, and Marie Lecroq for helpful discussion on stellar radiation. BL thanks Alex M. Garcia for inspiring discussion on the redshift evolution of metal scaling relations in simulations. 
      The authors acknowledge the Texas Advanced Computing Center (TACC) for providing HPC resources under allocation AST22003. The authors acknowledge support by the state of Baden-W\"urttemberg through bwHPC and the German Research Foundation (DFG) through grants INST 35/1597-1 FUGG and INST 35/1503-1 FUGG. BL, MM, RSK, and LB acknowledge the funding of the Deutsche Forschungsgemeinschaft (DFG, German Research Foundation) under Germany's Excellence Strategy EXC 2181/1 - 390900948 (the Heidelberg STRUCTURES Excellence Cluster). MM acknowledges financial support
from the European Research Council for the ERC Consolidator grant DEMOBLACK, under contract no. 770017. 
RSK acknowledges financial support from the ERC via Synergy Grant `ECOGAL' (project ID 855130), and from the German Ministry for Economic Affairs and Climate Action in project `MAINN' (funding ID 50OO2206). RSK also thanks the 2024/25 Class of Radcliffe Fellows for highly interesting and stimulating discussion. GC acknowledges partial financial support from the European Union—Next Generation EU, Mission 4, Component 2, CUP: C93C24004920006, project `FIRES'. 

\bibliography{ref}{} 
\bibliographystyle{aasjournal} 


\begin{appendix}

\section{Implementation details of the galaxy evolution model in \textsc{a-sloth}}\label{apdx:asloth}

The baryon mass budget in each halo is divided into four components: cold gas ($M_{\rm cold}$), hot gas ($M_{\rm hot}$), stars ($M_\star$), and outflows ($M_{\rm out}$). Here, cold gas refers to star-forming gas in a central region, while hot gas denotes all gas bound to the halo that is not cold including the warm and hot phases of the ISM/CGM. 
The outflow mass is the cumulative mass of gas that is unbound from the halo by stellar feedback and released to the IGM. 

In each global timestep $\Delta t_{j}$ when the merger tree goes from level $j+1$ to $j$, every (post-merger) parent halo at redshift $z_{j}$ inherits the mass of each baryonic component from its (progenitor) child halo(s) at $z_{j+1}>z_{j}$, which sets the initial condition ($M_{\rm cold}^{0}$, $M_{\rm hot}^{0}$, $M_\star^{0}$, and $M_{\rm out}^{0}$) at $z_{j+1}$. If the initial total mass of baryons is smaller than $M_{\rm h}\Omega_{\rm b}/\Omega_{\rm m}$ in the parent halo, additional hot gas with a total mass of $\Delta M_{\rm acc,hot} = M_{\rm h}\Omega_{\rm b}/\Omega_{\rm m}-(M_{\rm cold}^{0}+M_{\rm hot}^{0}+M_\star^{0}+M_{\rm out}^{0})$ is added to the halo, corresponding to smooth accretion from the IGM. Here, $M_{\rm h}$ is the halo (virial) mass, $\Omega_{\rm m}$ and $\Omega_{\rm b}$ are the cosmic average matter and baryon fractions. The IGM accretion is turned off for satellite galaxies and for any halos in ionized regions of the IGM, whose child halos have a maximum virial temperature $T_{\rm vir}<10^4$~K \citep[see sec. 2.5.2 in][]{Hartwig2022}.  
For simplicity, we do not include additional effects of galaxy mergers (i.e., enhanced star formation efficiency) beyond gas mass supplies and also ignore any additional loss of gas from tides or ram pressure in satellite galaxies other than shutting down the IGM accretion. 

To accurately follow star formation and stellar feedback, sub-cycling is introduced within $\Delta t_{j}$ using adaptive sub-timesteps \citep[see sec. 2.4 in][]{Hartwig2022}. In each sub-timestep $\delta t_{i}$, the baryon masses are updated by
\begin{align}
\begin{split}
    M_{\rm hot}^{i+1}&=M_{\rm hot}^{i}-\frac{\delta t_{i}M_{\rm hot}^{i}}{t_{\rm con}^{i}}-\delta M_{\rm out,hot}^{i}+\delta M_{\rm heat}^{i} +\delta M_{\rm acc,hot}^{i}+\delta M_{\rm loss,\star}^{i}\ ,\\
    M_{\rm cold}^{i+1}&=M_{\rm cold}^{i}+\frac{\delta t_{i}M_{\rm hot}^{i}}{t_{\rm con}^{i}}-\delta M_{\rm out,cold}^{i}-\delta M_{\rm heat}^{i}-\delta M_{\star}^{i}\ ,\\
    M_{\rm out}^{i+1}&=M_{\rm out}^{i}+\delta M_{\rm out,cold}^{i}+\delta M_{\rm out,hot}^{i}\ ,\\
    M_{\star}^{i+1}&=M_{\star}^{i}+\delta M_{\star}^{i}\simeq M_{\star}^{i}+\eta M_{\rm cold}^{i}\frac{\delta t_{i}}{t_{\rm cold,ff}^{i}}\ .
\end{split}\label{eq:cycle}
\end{align}
Here, $\delta t_{i}M_{\rm hot}^{i}/t_{\rm con}^{i}$ is the mass of hot gas converted to cold gas (on a timescale of $t_{\rm con}^{i}$), $\delta M_{\rm heat}^{i}$ is the mass of cold gas heated up by photo-ionization, $\delta M_{\rm acc,hot}^{i}$ is the mass of hot gas accreted from the IGM, $\delta M_{\rm loss,\star}^{i}$ is the mass lost by stars (via winds and SNe), $\delta M_\star$ is the mass of newly-formed stars, and $\delta M_{\rm out,hot/cold}^{i}$ is the mass of hot/cold gas blown away by galactic outflows. 
In the following subsections, we describe the derivations of these terms, which are classified into the inflow and outflow parts closely related to star formation (Sec.~\ref{apdx:sf}) and stellar feedback (Sec.~\ref{apdx:fdbk}), respectively. 

\subsection{Star formation}\label{apdx:sf}
Star formation is only possible when gas can cool rapidly. In \textsc{a-sloth}, this requirement is embodied by a halo mass threshold determined by primordial thermochemistry, streaming motion between baryons and dark matter, and dissociation of $\rm H_2$ by Lyman-Werner radiation \citep[see their sec. 2.3.1]{Hartwig2022}. The baryon cycle (Eq.~\ref{eq:cycle}) is only evolved for halos above this mass threshold, and each halo initially only contains hot gas $M_{\rm hot}=M_{\rm h}\Omega_{\rm b}/\Omega_{\rm m}$. 

The terms $\delta M_{\rm acc,hot}^{i}$, $\delta t_{i}M_{\rm hot}^{i}/t_{\rm con}^{i}$, and $\delta M_{\star}^{i}$ in Eq.~\ref{eq:cycle} embody the inflow cascade that leads to star formation. 
The accretion rate of hot gas from the IGM is assumed to be constant within $\Delta t_{j}$, such that $\delta M_{\rm acc, hot}^{i}=\Delta M_{\rm acc,hot}\delta t_{i}/\Delta t_{j}$ for each sub-cycle $\delta t_i$. For halos with $T_{\rm vir}>10^4$~K where atomic cooling is efficient, the condensation of hot gas down to the central region occupied by cold gas and stars is governed by the dynamical timescale $t_{\rm dyn}^{i}$, i.e., $t_{\rm con}^{i}=t_{\rm dyn}^{i}$, given
\begin{align}
    t_{\rm dyn}^{i}=\min\{R_{\star}^{3/2}/[G(M_{\star}^{i}+M_{\rm cold}^{i})]^{1/2}, R_{\rm vir}^{3/2}/(G M_{\rm h})^{1/2}\}\ ,
\end{align}
where $R_{\star}$ is the characteristic radius of the galaxy (see Sec.~\ref{sec:sf}), and $R_{\rm vir}$ is the halo virial radius. The second term is meant to capture the initial collapse phase when the mass of cold gas and stars is very small. For minihalos with $T_{\rm vir}<10^4$~K where cooling is typically driven by molecular hydrogen, the condensation is also limited by the cooling timescale $t_{\rm cool}^{i}$ as $t_{\rm con}^{i}=\max(t_{\rm dyn}^{i},t_{\rm cool}^{i})$. 

In a star-forming halo, the collapse of cold gas is governed by {the mean free-fall timescale of cold gas $t^{i}_{\rm cold,ff}=(G\rho_{\rm cold}^{i})^{-1/2}\sim 2t_{\rm dyn}^{i}$}, where $\rho_{\rm cold}^{i}=(M_{\rm cold}^{i}+M_{\star}^{i})/[(4\pi/3)R_{\star}^{3}]$. Note that the original model of \citet{Hartwig2022} uses $\rho_{\rm cold}^{i}=M_{\rm cold}^{i}/[(4\pi/3)R_{\star}^{3}]$ considering only the self-gravity of cold gas, here we further include the gravity of stars. The total mass of stars formed in this timestep is estimated as $\delta M_{\star,\rm est}=\eta M_{\rm cold}^{i}\delta t_i/t_{\rm cold,ff}^{i}$, where $\eta$ is the star formation efficiency (SFE) per mean free-fall time, 
which is treated as a constant free parameter. Individual stars are then drawn from the IMF (see Sec.~\ref{sec:stellar}) with Poisson sampling according to $\delta M_{\star,\rm est}$, as detailed in Sec.~2.3.2 in \citet{Hartwig2022}. Their initial masses $m_\star$ are summed up as $\delta M_{\star}^{i}=\sum m_{\star}\simeq \delta M_{\star,\rm est}$ and removed from the cold gas. Note that here, the mean free-fall time $t_{\rm cold,ff}^{i}$ is defined for cold gas in the central region within $R_{\star}$ as a whole. In reality, star formation and the subsequent feedback can happen in smaller/denser clumps on a timescale much shorter than the mean free-fall time. In this case, $\eta$ can be larger than unity. This is particularly true for first star formation on small timescales $\sim 0.1-1\ \rm Myr$ in minihalos ($M_{\rm h}\lesssim 10^8\ \rm M_\odot$), where $t_{\rm cold,ff}^{i}\sim 10$~Myr \citep{Hartwig2024}.

As discussed in Sec.~\ref{sec:stellar}, the IMFs of Pop~III and II stars are modeled separately. Similarly, we consider different SFE values for these two populations. Pop~III stars are unimportant in this work because the observational data considered here only involve relatively massive halos with $M_{\rm h}\gtrsim 10^{9.5}\ \rm M_\odot$ hosting mostly Pop~II stars. Therefore, we fix the Pop~III SFE to the best-fit value $\eta_{\rm III}=8.15$ of \citet[see their table~1]{Hartwig2024} for simplicity. To avoid recalibration of Pop~III parameters, we still use the old approach $R_\star=R_{\rm s}$ to model the star formation in molecular-cooling minihalos with $T_{\rm vir}<10^4$~K ($M_{\rm h}\lesssim 10^8\  M_\odot$) that typically host Pop~III stars. 
Besides, the Pop~II SFE is also fixed to the best-fit value $\eta_{\rm II}=0.237$, since its effects are mostly degenerate with those of galactic outflow parameters that are explored in detail in this work (see below and Sec.~\ref{sec:fdbk}). 

\subsection{Stellar feedback}\label{apdx:fdbk}

In addition to the inflow cascade discussed in the previous subsection, we must consider the outflow cascade driven by stellar feedback to complete the baryon cycle, which is characterized by $\delta M_{\rm heat}^{i}$, $\delta M_{\rm out,cold}^{i}$, and $\delta M_{\rm out,hot}^{i}$ in Eq.~\ref{eq:cycle}. Individual massive stars with $m_{\star}>5\ \rm M_\odot$ are tracked in \textsc{a-sloth} to compute these feedback terms. Note that the feedback from active galactic nuclei (AGN) can also regulate galaxy evolution, especially in massive galaxies. Since we focus on relatively small (dwarf) galaxies ($M_\star\sim 10^{7}-10^{10}\ \rm M_\odot$, $M_{\rm h}\sim 10^{9.5}-10^{11.5}\ \rm M_\odot$) at $z\gtrsim 4$, AGN feedback is ignored for simplicity \citep[for discussion on AGN feedback in dwarf galaxies, see, e.g.,][]{Koudmani2022,Koudmani2024,Aravindan2023,Sharma2023,Hazenfratz2025,Ivey2025}.

When the massive stars are alive, they power photoheating feedback, which is captured by $\delta M_{\rm heat}^{i}$: the mass of cold gas heated up to $\sim 10^4$~K in HII regions. It is calculated from the production rates $\dot{q}_{\rm ion}$ of (hydrogen) ionizing photons of active massive stars in the current sub-timestep $\delta t_{i}$ as detailed in sec. 2.3.3 in \citet{Hartwig2022}. {For simplicity, we adopt a constant $\dot{q}_{\rm ion}$ throughout the lifetime of each massive star as the lifetime-average value calculated from detailed stellar spectra evolution \citep{Klessen2023}.} For minihalos with $T_{\rm vir}<10^{4}$~K that are unable to bind ionized gas, we add the heated gas directly to outflows ($\delta M_{\rm out,cold}^{i}$, see below) rather than the hot phase ($\delta M_{\rm heat}^{i}$). 

A massive star dies within $\delta t_{i}$ if the time past after its birth exceeds its lifetime $t_{\star}$. The dying stars can undergo SN explosions that drive galactic outflows, as denoted by the $\delta M_{{\rm out},k}^{i}$ terms, where $k={\rm cold}$,~hot. The \textit{expected} outflow mass is estimated by
\begin{align}
\begin{split}
    \delta \hat{M}_{{\rm out}, k}^{i}=\frac{E_{\rm SNe}^{i}f_{k}}{E_{{\rm bind},k}\gamma_{\rm out}}M_{k}^{i}\ ,\quad 
    f_{k} = \frac{E_{{\rm bind},k}^{i}M_{k}^{i}}{E_{\rm bind,hot}^{i}M_{\rm hot}^{i} + E_{\rm bind,cold}^{i}M_{\rm cold}^{i}}\ .
\end{split}\label{eq:dmout}
\end{align}
Here, $E_{\rm SNe}^{i}=\sum e_{{\rm SN}}$ is the total energy of SNe exploded within $\delta t_{i}$, $E_{{\rm bind},k}$ is the binding energy of component $k$ given by eqs. 24 and 25 in \citet{Hartwig2022} based on \citet{Chen2022}, and $\gamma_{\rm out}$ is the outflow efficiency, which is a function of halo mass as described in Sec.~\ref{sec:fdbk}.

Given the expected value from Eq.~\ref{eq:dmout}, the true outflow mass is limited by the gas mass immediately available $\delta M_{{\rm out},k,\max}^{i}$ in practice, i.e., $\delta M_{{\rm out},k}^{i}=\min(\delta\hat{M}_{{\rm out},k}^{i},\delta M_{{\rm out},k,\max}^{i})$. For cold gas, we have $\delta M_{{\rm out,cold},\max}^{i}= 0.99(M_{\rm cold}^{i}-\delta M_{\rm heat}^{i}-\delta M_{\star}^{i})$. 
The excess mass $\delta M_{\rm out,cold,exc}^{i}=\max(\delta \hat{M}_{\rm out,cold}^{i}-\delta M_{\rm out,cold,max}^{i},0)$ is added to hot outflows $\delta \hat{M}_{\rm out,hot}^{i}$, whose mass limit is $\delta M_{\rm out,hot,max}^{i}=0.99[M_{\rm hot}^{i}(1-\delta t_{i}/t_{\rm con}^{i})+\delta M_{\rm acc,hot}^{i}+\delta M_{\rm heat}^{i}]$. {We have found by numerical experiments that the detailed treatments of the distribution of SN energy ($f_k$) and outflow mass budget among hot and cold gas have minor effects on our results. }

In addition to radiative and mechanical feedback, we also consider chemical feedback, i.e., metal enrichment of the ISM/CGM/IGM by stellar winds and SNe from massive stars. Similar to the overall baryon mass budget, the mass of gas-phase 
metals in a halo is divided into the bound ($M_{\rm gal,metal}^{i}$) and unbound ($M_{\rm out,metal}^{i}$) components, which are evolved by
\begin{align}
\begin{split}
    M_{\rm gal,metal}^{i+1} &= \frac{(M_{\rm gal,metal}^{i}+\delta M_{\rm metal}^{i})(M_{\rm cold}^{i+1}+M_{\rm hot}^{i+1})}{M_{\rm cold}^{i+1}+M_{\rm hot}^{i+1}+\delta M_{\rm out}^{i}+\delta M_\star^{i}}\ ,\\
    M_{\rm out,metal}^{i+1}&=M_{\rm out,metal}^{i}+\frac{(M_{\rm gal,metal}^{i}+\delta M_{\rm metal}^{i})\delta M_{\rm out}^{i}}{M_{\rm cold}^{i+1}+M_{\rm hot}^{i+1}+\delta M_{\rm out}^{i}+\delta M_\star^{i}}\ ,
\end{split}\label{eq:mmetal}
\end{align}
where $\delta M_{\rm out}^{i}\equiv \delta M_{\rm out,cold}^{i}+\delta M_{\rm out,hot}^{i}$ is the total mass of outflows generated in $\delta t_{i}$, and $\delta M_{\rm metal}^{i}$ is the total mass of metals gained by the halo:
\begin{align}
    \delta M_{\rm metal}^{i}=\delta M_{\rm metal,\star}^{i}+c_{\rm ZIGM}\delta M_{\rm acc,hot}^{i}Z_{\rm IGM}\ .\label{eq:delta_mmetal}
\end{align}
Here, by metals we mean elements with atomic numbers no smaller than 6, ignoring those between helium and Carbon that are relatively unimportant for galaxy and stellar evolution, whose abundances are fixed to the primordial values from Big Bang Nucleosynthesis.  
The first term of Eq.~\ref{eq:delta_mmetal} accounts for the in-situ metals ejected by massive stars (with $m_{\star}>5\ \rm M_\odot$), $\delta M_{\rm metal,\star}^{i}=\sum m_{Z}$ (see Sec.~\ref{sec:stellar}). Similarly, $\delta M_{\rm loss,\star}^{i}=\sum \delta m_{\rm loss,\star}$ in Eq.~\ref{eq:cycle} denotes the overall mass of baryons lost by stars, which is returned to the hot phase of ISM/CGM. For simplicity, the metal/total mass return of each massive star is modeled as a single event at its death. Therefore, the summations above go over dying stars, where $m_Z$ is the total mass of metals dispersed via winds and SN by a star throughout its lifetime. We further make the approximation $\delta m_{\rm loss,\star}=m_{\star}$. Note that in the high-$z,$ low-mass galaxies considered in our work, recycling of stellar baryons makes minor ($\lesssim 10\%$) contributions to the total gas mass budget. 

The second term of Eq.~\ref{eq:delta_mmetal} accounts for the metals accreted from the IGM, where $Z_{\rm IGM}$ is the IGM metallicity at the location of the halo, and $c_{\rm ZIGM}\ge 1$ is a clumping factor. The latter is introduced by \citet{Hartwig2024} motivated by the centrally concentrated distribution of metals in SN bubbles around early protogalaxies found in 3D hydrodynamic simulations \citep[e.g.,][]{Ritter2015,Magg2022a}, and the fact that accretion flows favor denser, colder pockets of gas with higher metallicities than the IGM average. We adopt $c_{\rm ZIGM}=3.32$ according to \citet{Hartwig2024}, which is chosen to reproduce the metallicity distribution of metal-poor stars and the mass-metallicity relation of satellite galaxies in the Milky Way (MW). {We have found by numerical experiments that the clumping factor is only important for low-mass galaxies ($M_\star\lesssim 10^{7}\ \rm M_\odot$) that are mostly too faint to show up in high-$z$ observations, and therefore, does not change our conclusions, as long as it is not too large ($c_{\rm ZIGM}\lesssim 3$).} The assumption underlying Eq.~\ref{eq:mmetal} is that the metals are mixed uniformly into the current gas reservoir of the halo and that outflows carry out metals proportionally from this reservoir. 

Given $M_{\rm gal,metal}^{i}$, we derive the average gas-phase metallicity of the galaxy as $Z_{\rm ave}=M_{\rm gal,metal}^{i}/(M_{\rm cold}^{i}+M_{\rm hot}^{i})$. 
The metallicity of star-forming gas can differ from this average value significantly due to inhomogeneous enrichment \citep{Xu2016,Ritter2016,Magg2022a}. To capture this effect, we apply a shift factor $d\log Z$ to estimate the metallicity of newly-formed stars ($\delta M_\star^{i}$) as $Z_\star = Z_{\rm ave}\times 10^{d\log Z}$. Here, $d\log Z$ is drawn randomly from a normal distribution calibrated to the results of cosmological simulations \citep{Tarumi2020}, as detailed in appendix~A.4 in \citet{Hartwig2022}. On the other hand, $M_{\rm out,metal}^{i}$ is added to the enriched region (outflow bubble) around the halo, whose expansion is modeled as a pressure-driven snowplow. The local IGM metallicity $Z_{\rm IGM}$ is estimated from the properties of all outflow bubbles that enclose the halo. The reader is referred to sec. 2.5.1 in \citet{Hartwig2022} for details on the calculation of bubble properties and $Z_{\rm IGM}$. The same approach is used to predict the abundance of any specific element with the same shift factor $d\log Z$, assuming element-independent metal mixing. {In reality, different metal elements in star-forming gas can experience different degrees of inhomogeneous metal mixing due to their distinct cooling rates \citep{Hartwig2019}. This effect is ignored for simplicity because here we are mainly concerned with the oxygen abundance that closely traces the bulk metallicity.} 

\section{Star formation main sequence}
\label{apdx:sfms}

\begin{figure}
	\centering
	\includegraphics[width=0.495\linewidth]{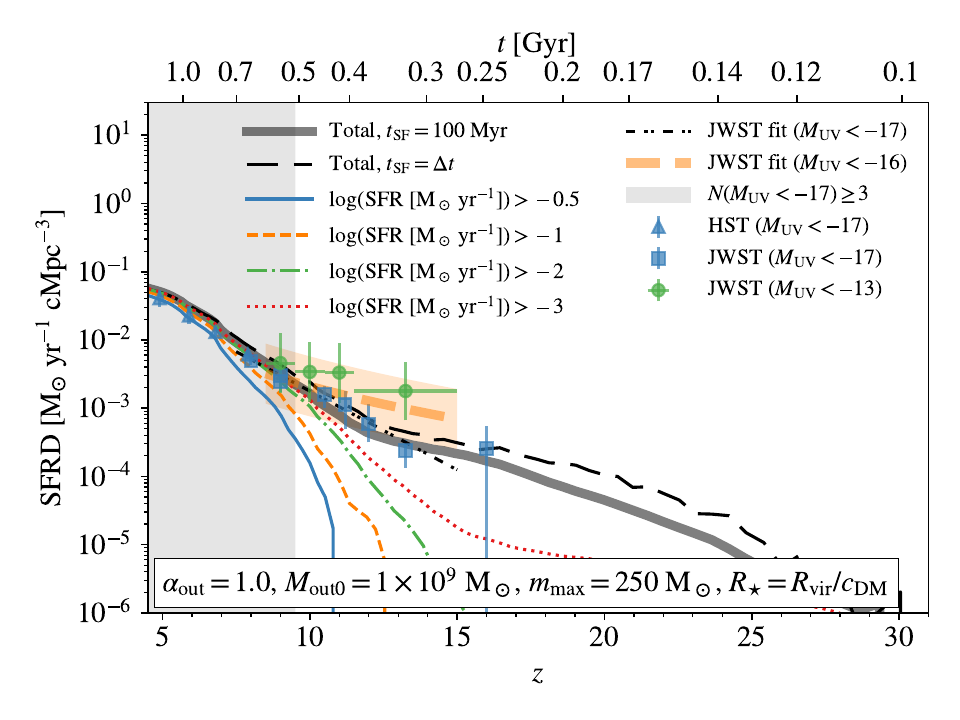}
    \includegraphics[width=0.495\linewidth]{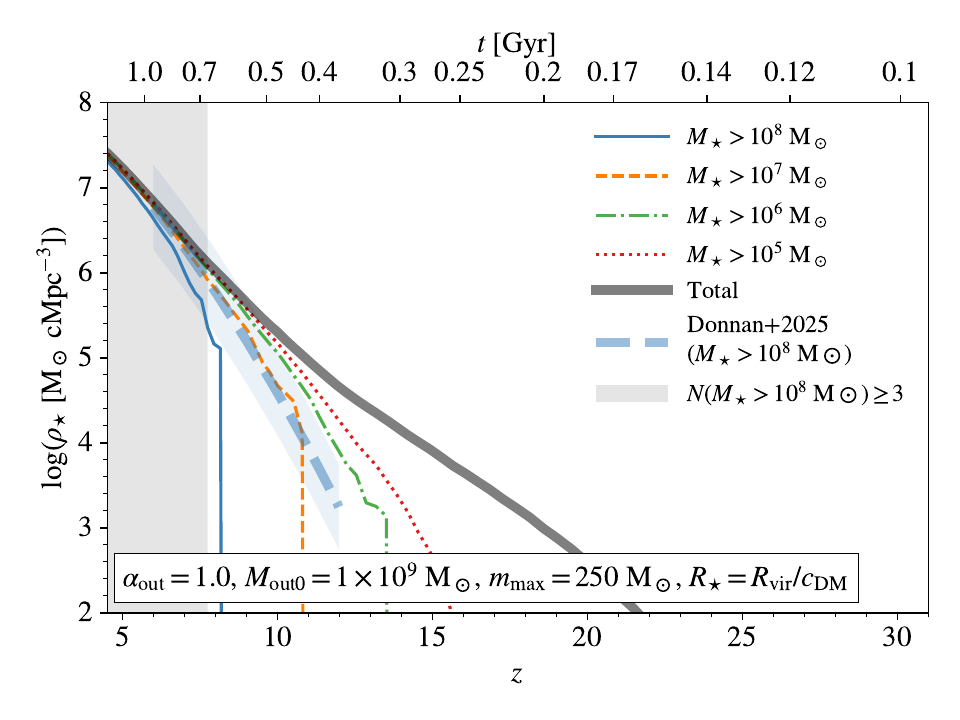}
	\caption{Same as Fig.~\ref{fig:csfh} but for an exemplar run using the old star formation prescription from \citet{Hartwig2022}with $R_{\star}=R_{\rm vir}/c_{\rm DM}$, given $\alpha_{\rm out}=1$, $M_{\rm out0}=10^{9}\ \rm M_\odot$, and $m_{\max}=250\ \rm M_\odot$.}
	\label{fig:sfrd_old}
\end{figure}

As explained in Sec.~\ref{sec:sf}, the star formation prescription in \textsc{a-sloth} is updated with reduced galaxy sizes based on the results of cosmological simulations and observations. With this update, our simulations can better reproduce the scatter around the star formation main sequence (SFMS) in observations. In this section, we compare the results of our new model and those of the original/old model of \citet{Hartwig2022} in the context of observational data at $z\sim 4-10$ \citep{Salmon2015,Rinaldi2022,Heintz2023,Nakajima2023,Clarke2024,Curti2024,Sarkar2025}. {In fact, the observational results for high-$z$ SFMS \citep[see also][]{Clarke2025,DiCesare2025,Rinaldi2025,Simmonds2025} show significant differences in both normalization and slope, which imply that the SFMS is highly sensitive to selection effects and uncertainties in SFR and stellar mass measurements \citep[e.g.,][]{Choe2025,Kramarenko2025,Wang2025}.} Therefore, we do not include SFMS in our likelihood analysis.

\begin{figure*}
    \centering
    \includegraphics[width=0.495\linewidth]{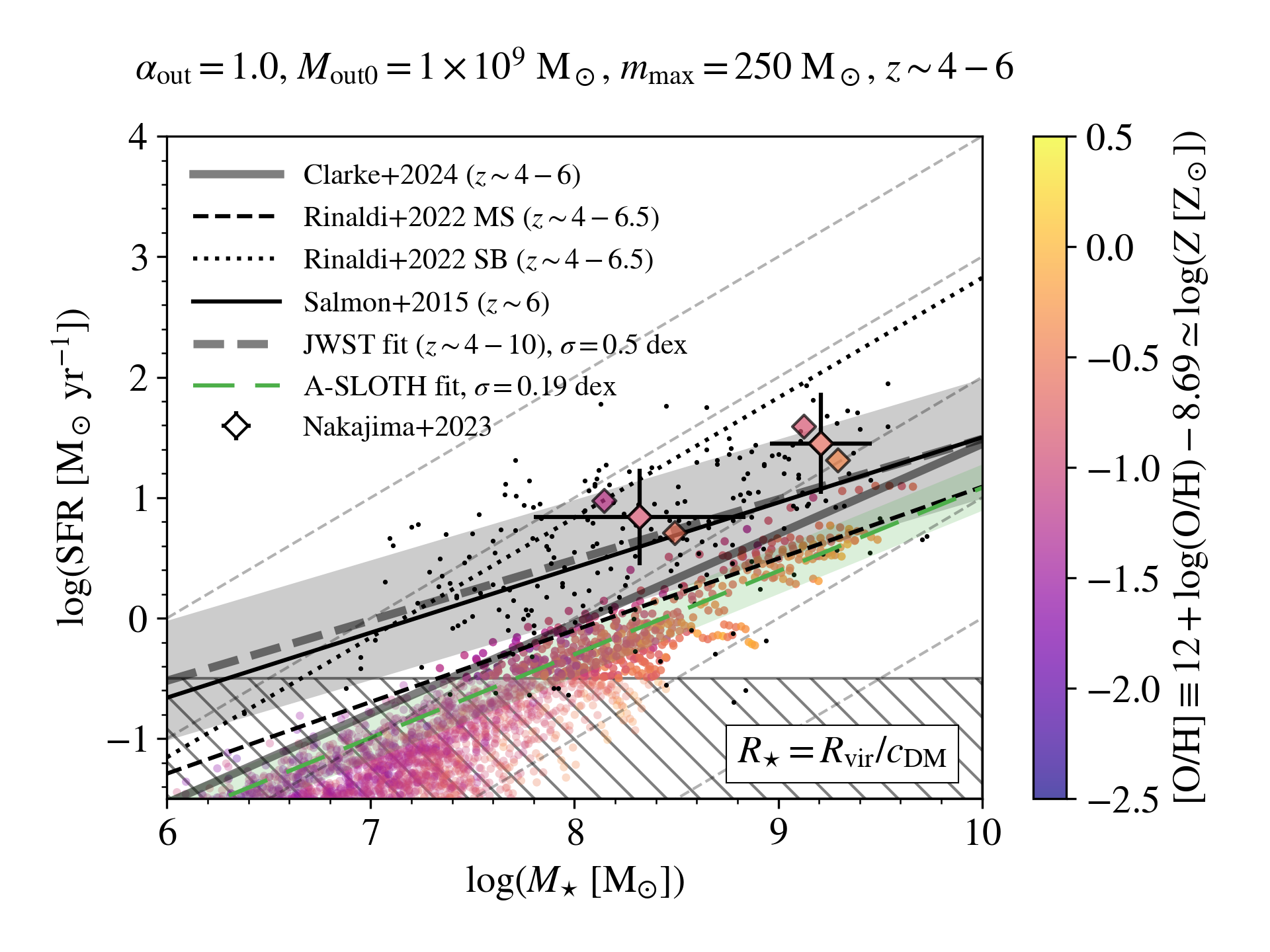}
    \includegraphics[width=0.495\linewidth]{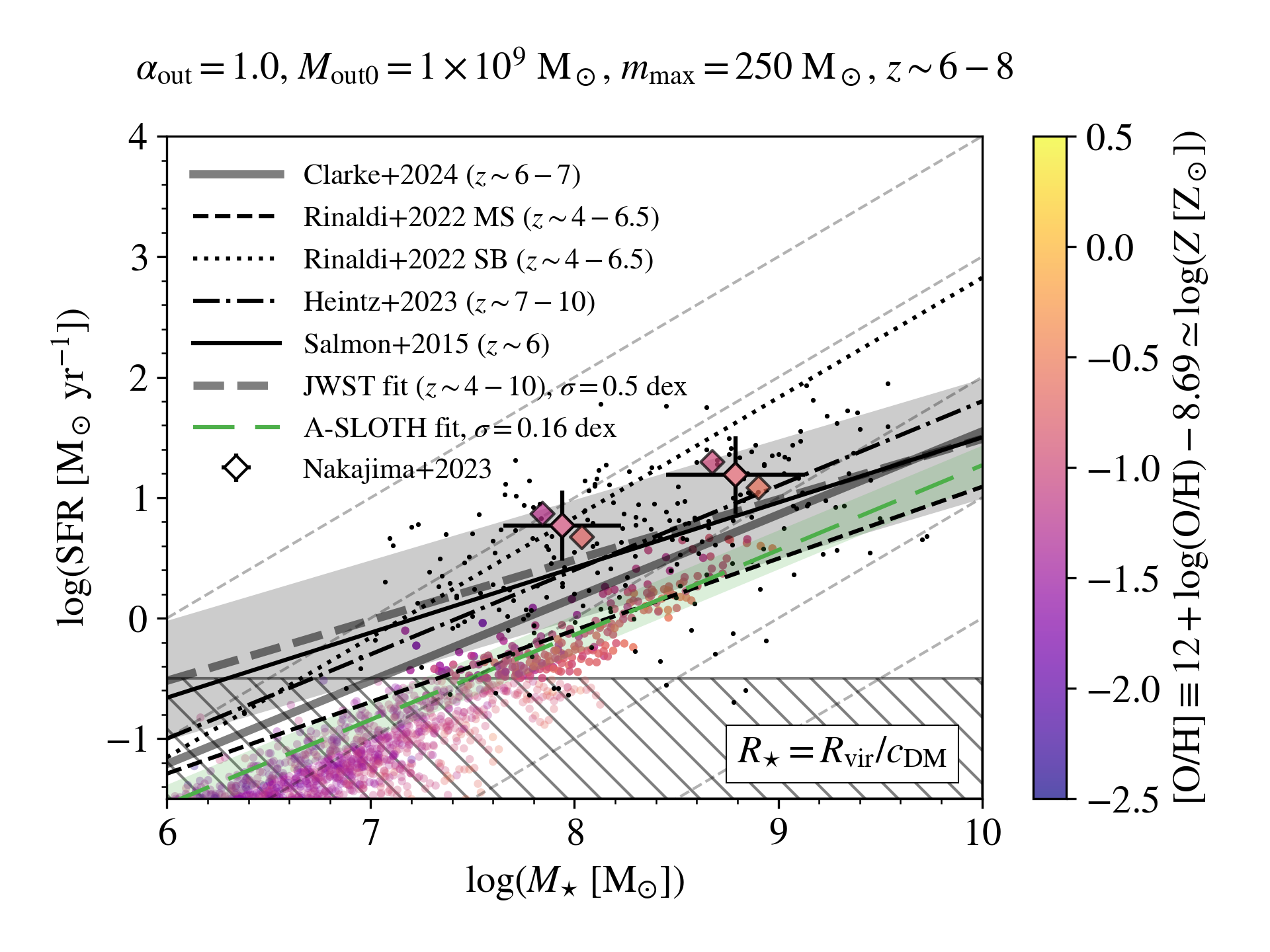}
    \caption{SFMS at $z\sim 4-6$ (left) and $z\sim 6-8$ (right) from an exemplar run using the old star formation prescription of \citet{Hartwig2022} with $R_{\star}=R_{\rm vir}/c_{\rm DM}$ for $\alpha_{\rm out}=1$, $M_{\rm out0}=10^{9}\ \rm M_\odot$, and $m_{\max}=250\ \rm M_\odot$. Individual simulated galaxies are shown as dots color-coded by metallicity, while the JWST galaxies compiled by \citet[see their fig.~3]{Sarkar2025} including those from \citet{Nakajima2023} and \citet{Curti2024} are shown as the smaller black dots. We fit a linear relation (in log-log space) to the simulated galaxies with $\log\rm (SFR\ [\rm M_\odot\ yr^{-1}])>-0.5$ outside the hatched region as the long dashed line. The relevant scatter is shown by the  green shaded region. Similarly, the thick dashed line and gray shaded region show the fit and scatter for the JWST galaxies. We also show the average values of SFR, $M_\star$, and metallicity of the JWST galaxies of \citet[see their table~2]{Nakajima2023} with the diamonds. Each data point involves three diamonds color-coded by the mean and $1\sigma$ upper and lower limits of metallicity. The relevant spreads in SFR and $M_\star$ are shown by errorbars. The thick solid, dashed, solid, and dash-dotted lines show the observed SFMS from \citet{Clarke2024}, \citet{Rinaldi2022}, \citet{Salmon2015}, and \citet{Heintz2023}. The dotted line shows the fit for starburst (SB) galaxies from \citet{Rinaldi2022}. The thin dashed lines mark 5 specific SFR values in the range ${\rm SFR}/M_{\star}\in [10^{-10},~10^{-6}]\ \rm yr^{-1}$ with 1 dex spacing.}
    \label{fig:sfms_old}
\end{figure*}

For the old model with $R_{\star}=R_{\rm vir}/c_{\rm DM}$, we run a simulation with $\alpha_{\rm out}=1$, $M_{\rm out0}=10^{9}\ \rm M_\odot$, and $m_{\max}=250\ \rm M_\odot$ chosen to reproduce the observed star formation history at $z\lesssim 7$ with optimal IMF-averaged metal yield, as shown in Fig.~\ref{fig:sfrd_old}. The resulting relation between SFR and stellar mass is shown in Fig.~\ref{fig:sfms_old} for two redshift bins: $z\sim 4-6$ (top) and $z\sim 6-8$ (bottom). The simulated SFMS is defined as a linear fit in the $\log\rm SFR$--$\log M_\star$ space for simulated galaxies with $\rm SFR > 10^{-0.5}\ \rm M_\odot\ yr^{-1}$. Interestingly, for both redshift bins, the simulated SFMS agrees with the main sequence (MS) fit of \citet{Rinaldi2022}, which is however below the fits from \citet{Salmon2015} and \citet{Heintz2023} by $\sim 0.6$~dex. The latter two are consistent with the SFMS found in the JWST galaxies  at $z\sim 4-10$ \citep[]{Nakajima2023,Curti2024,Sarkar2025}. The SFMS from \citet{Clarke2024} appears to be an in-between case. The scatter around the SFMS is $\sigma\sim 0.16-0.19$~dex in the simulation, which is significantly lower than that $\sigma\sim 0.5$~dex for the JWST galaxies. A similar intrinsic scatter of $\sigma\sim 0.42-0.49$~dex is also found by \citet{Clarke2024} from observations at $z\sim 4-7$. 
Clearly, the stochasticity of star formation (at the timescale of 10~Myr) is significantly underestimated by the old model of \citet{Hartwig2022}. Besides, there are no galaxies from the simulation that is close to the starburst track with ${\rm SFR}/M_{\star}\sim 10^{-7}\ \rm yr^{-1}$ found by \citet{Rinaldi2022}, while many JWST galaxies appear to be in the starburst phase. {In fact, such starbursts are needed to explain the low dust contents of JWST galaxies by efficiently clearing out gas and dust with bursty feedback \citep[e.g.,][]{Stiavelli2023,Tsuna2023,Ziparo2023,Topping2024}. }

\begin{figure*}
    \centering
    \includegraphics[width=0.495\linewidth]{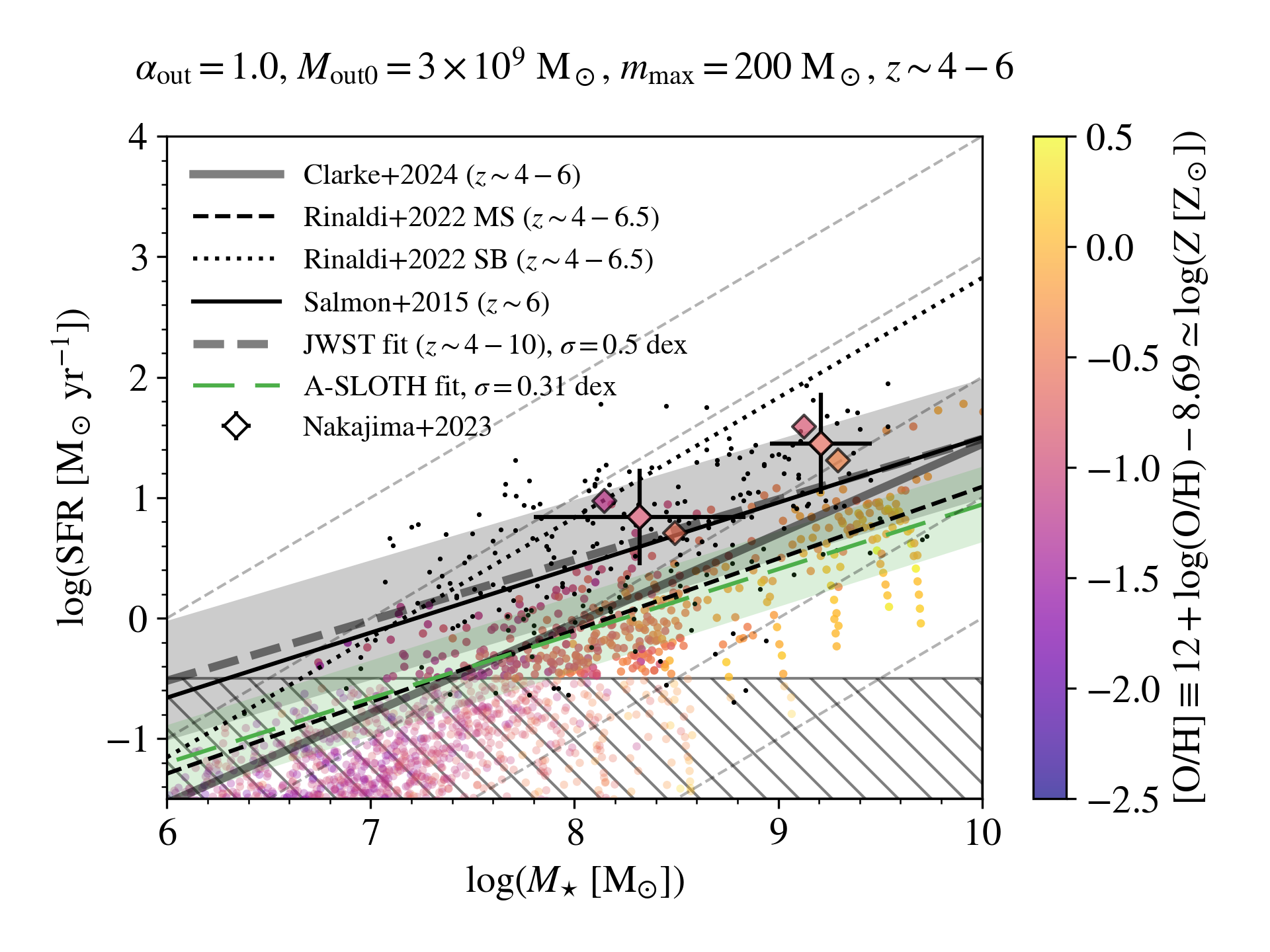}
    \includegraphics[width=0.495\linewidth]{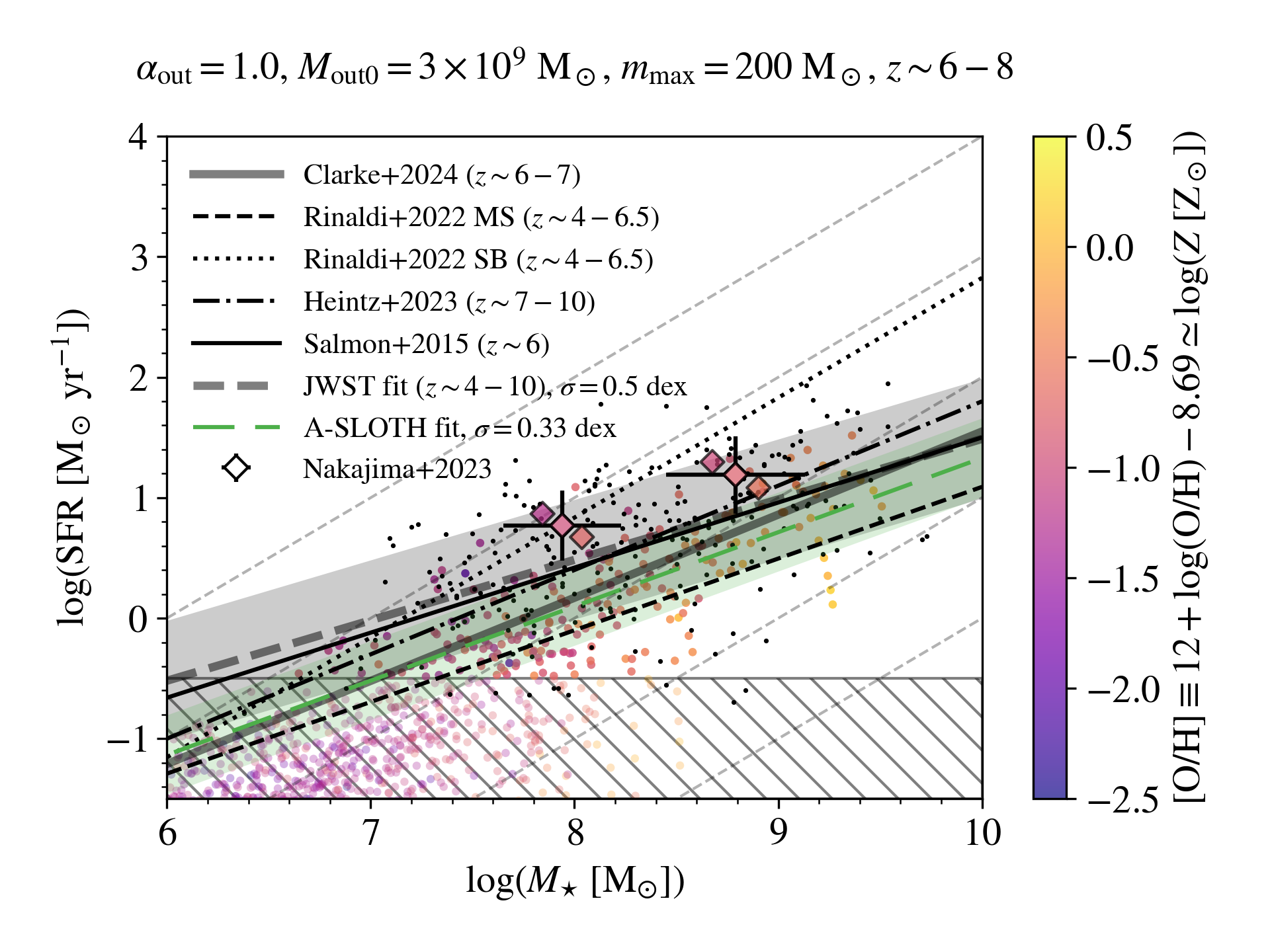}\\
    \includegraphics[width=0.495\linewidth]{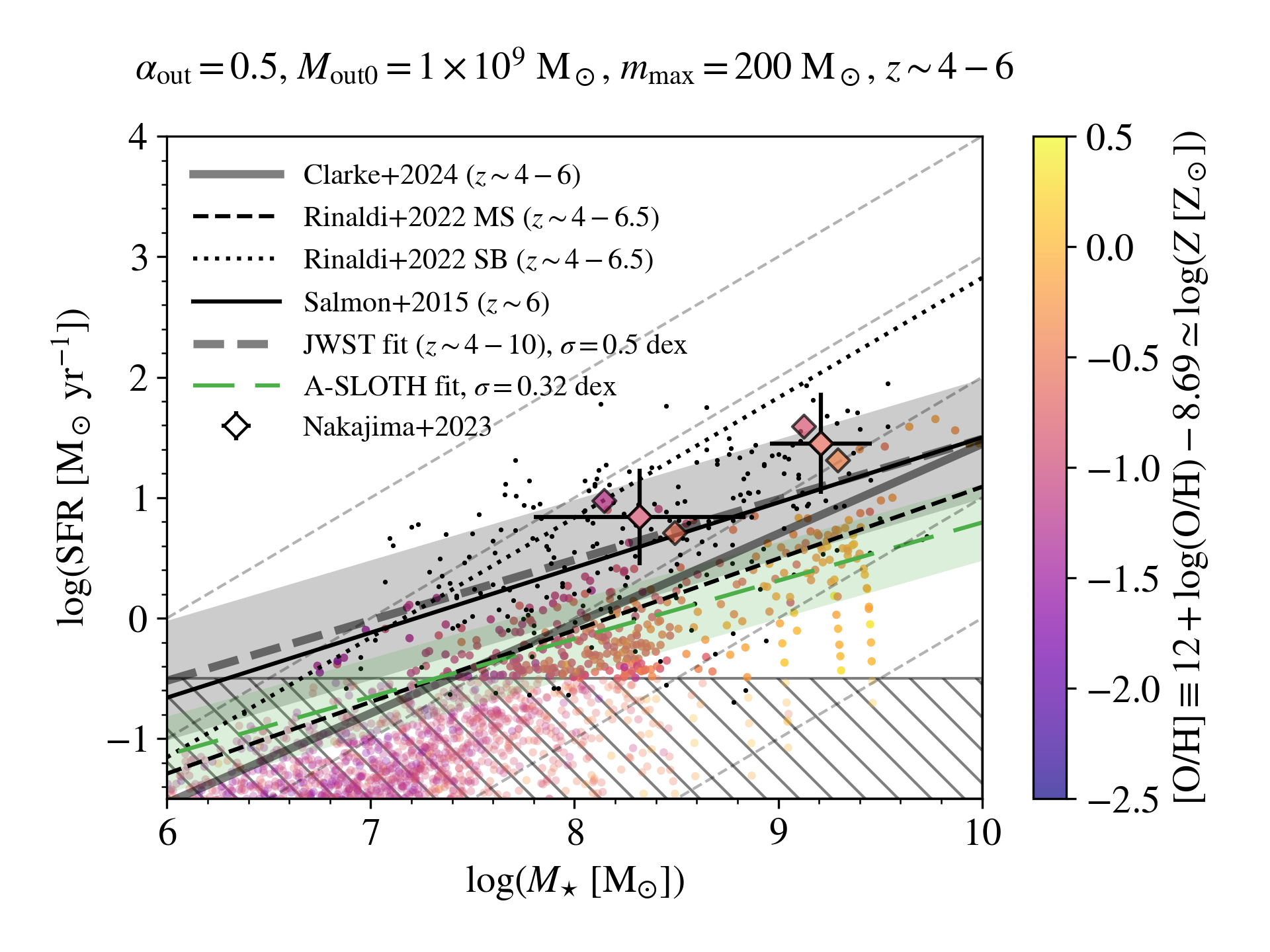}
    \includegraphics[width=0.495\linewidth]{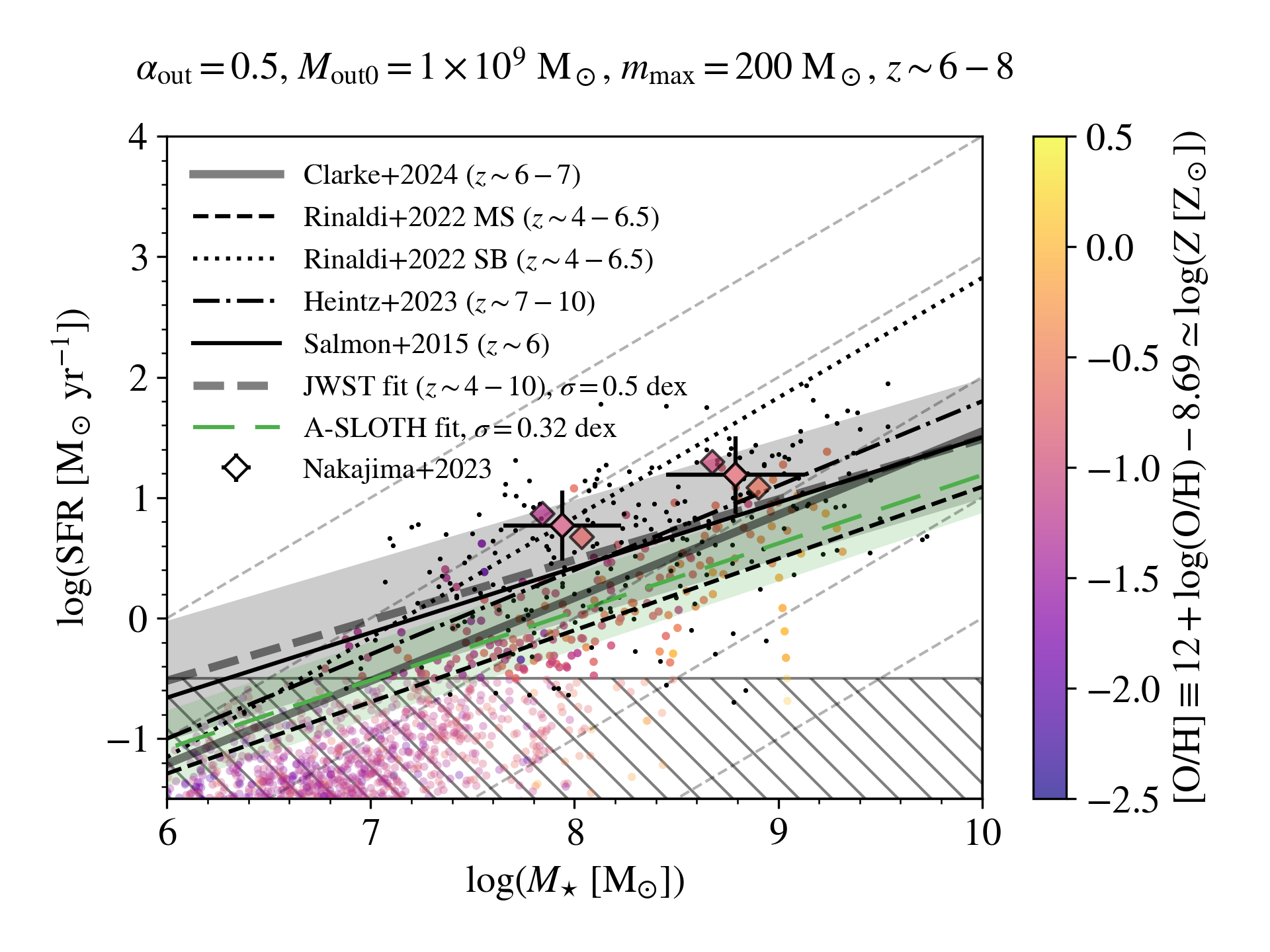}\\
    \includegraphics[width=0.495\linewidth]{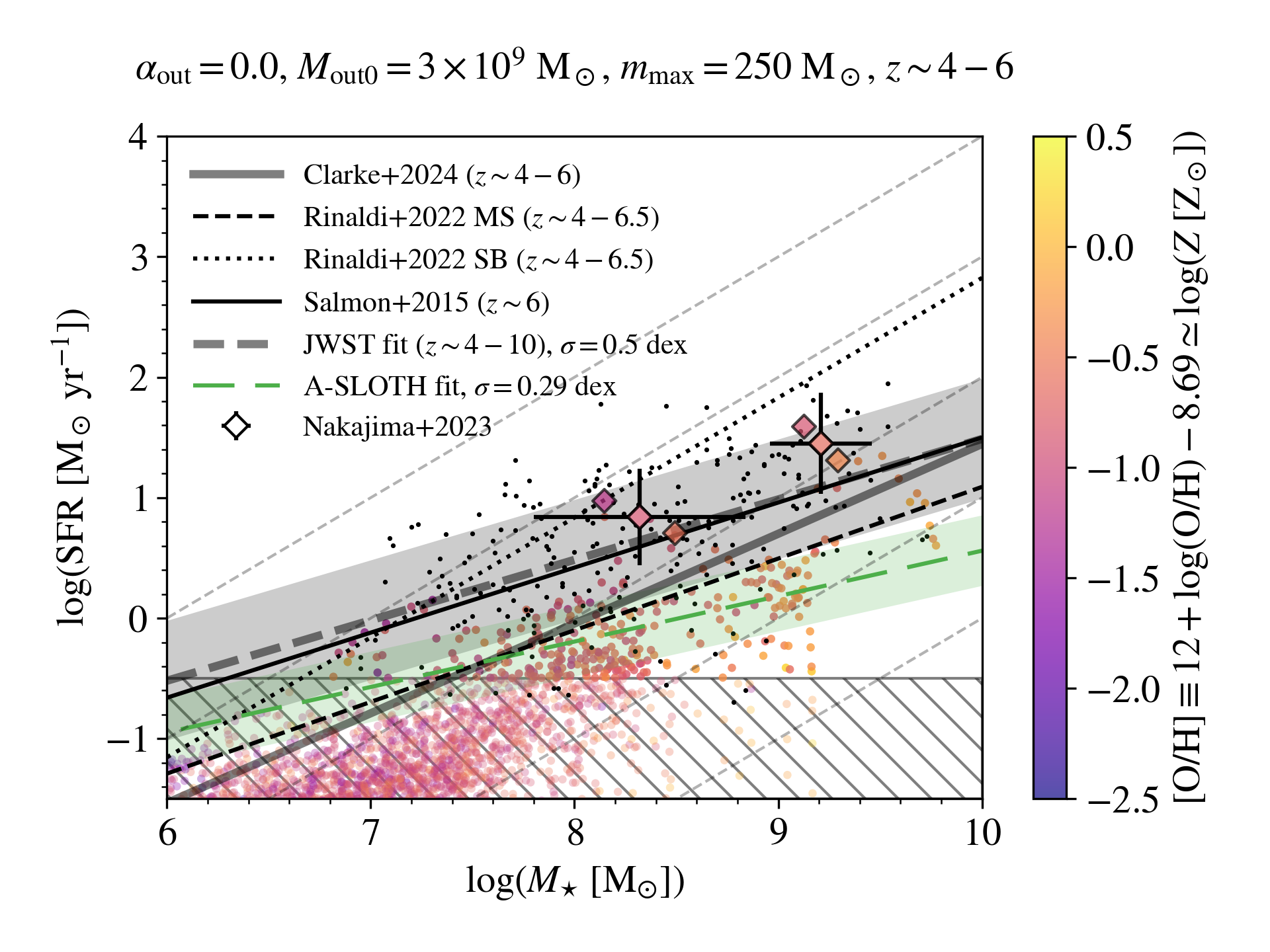}
    \includegraphics[width=0.495\linewidth]{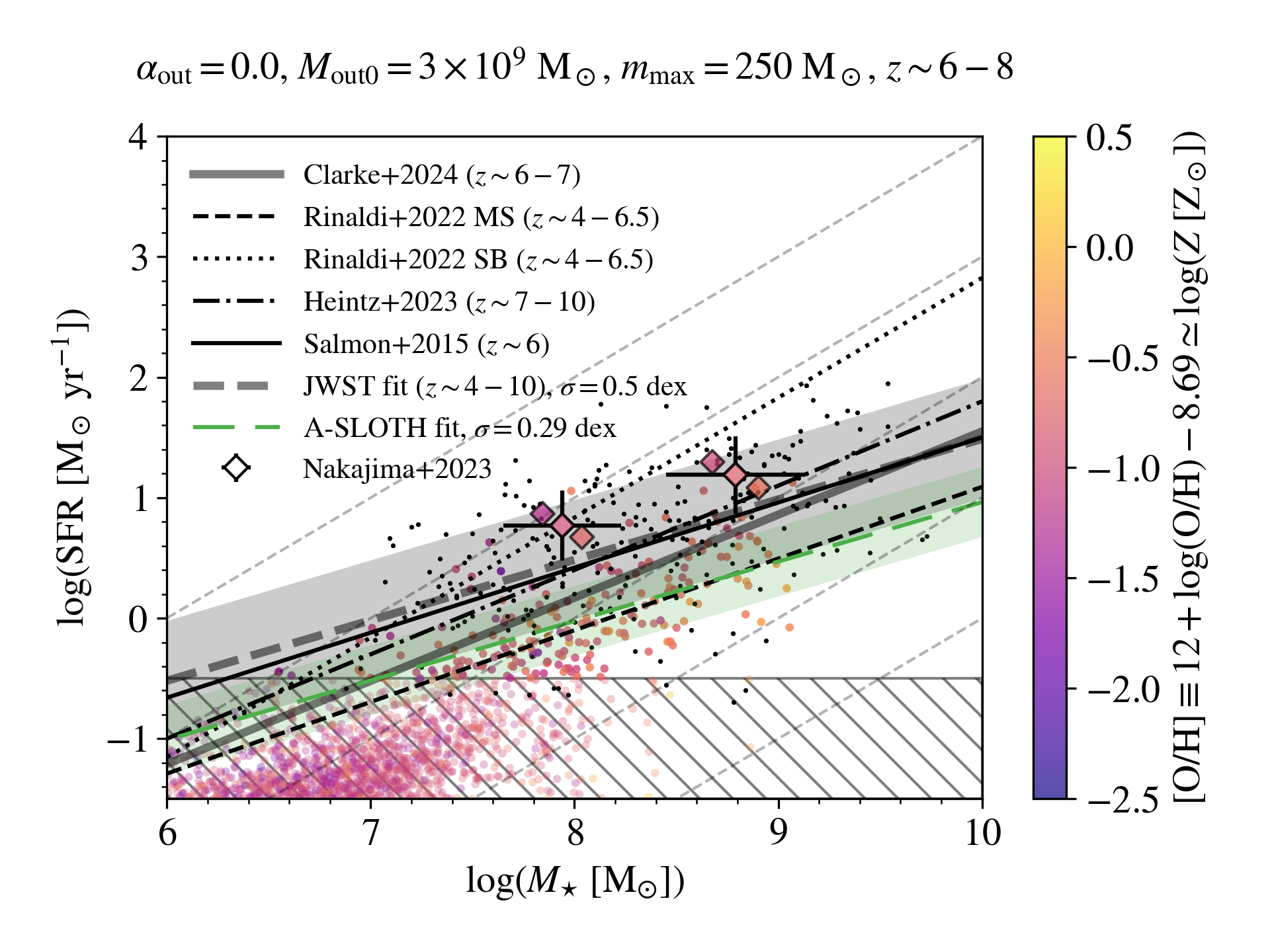}
    \caption{Same as Fig.~\ref{fig:sfms_old} but for the SFMS from simulations using the new star formation prescription. As representative examples, we consider the three best-fit models (see Sec.~\ref{sec:best}) chosen for $\alpha_{\rm out}=1.0$ (top), 0.5 (middle), and 0 (bottom) at $z\sim 4-6$ (left) and $z\sim 6-8$ (right).}
    \label{fig:sfms}
\end{figure*}

In Fig.~\ref{fig:sfms}, we show the SFMS at $z\sim 4-6$ (left) and $z\sim 6-8$ (right) from simulations using the new star formation prescription with smaller galaxy sizes $R_\star=2.5R_{50}$ (Eq.~\ref{eq:rs}) for the best-fit models (see Fig.~\ref{fig:lh_all}) chosen for $\alpha_{\rm out}=1.0$ (top), 0.5 (middle), and 0 (bottom). Similar to the case of the old prescription, the simulated SFMS is generally in agreement with that from \citet{Rinaldi2022} except for the $\alpha_{\rm out}=0$ case at $z\sim 4-6$, where the simulated SFMS has a smaller slope. So, the median SFR of the simulated galaxies is generally lower than that of the JWST galaxies by $\sim 0.6$~dex, which is a persistent feature of our small simulation volume with poor statistics of luminous objects. In the $\alpha_{\rm out}=0$ case, star formation in galaxies with $M_{\star}\gtrsim 10^{9.2}\ \rm M_\odot$ is significantly quenched by outflows, while the quenching is less rapid in the other two cases with weaker outflows. This is consistent with the finding in Sec.~\ref{sec:best} that the SFRD at $z\sim 5$ is underestimated by a factor of $\sim 2$ in the $\alpha_{\rm out}=0$ best-fit model. With the new star formation prescription, the scatter around SFMS increases by a factor of $\sim 2$ to $\sigma\sim 0.29-0.33$~dex, {which is comparable or slightly smaller than those inferred from observations $\sigma\sim 0.3-0.5$~dex \citep[]{Clarke2024,Clarke2025,Sarkar2025,Simmonds2025}}. Starburst galaxies with ${\rm SFR}/M_{\star}\gtrsim 10^{-7}\ \rm yr^{-1}$ also show up in these simulations, alongside with galaxies on quenching tracks with dropping SFR and increasing $Z${, similar to the dormant galaxies discovered by JWST \citep[e.g.,][]{Covelo-Paz2025}, indicating a feedback-regulated `breathing' mode of star formation}. Note that a similar trend of increasing SFR dispersion with decreasing $R_\star$ is found in local observations as well \citep[e.g.,][]{He2025}. 


{In conclusion, the enhanced galaxy compactness in the new star formation prescription is needed to better capture the bursty nature of star formation implied by observations of high-$z$ (dwarf) galaxies.} {
Moreover, such compactness and the resulting bursty star formation regulates metal enrichment, which can manifest in metallicity gradients \citep{Garcia2025sim,Li2025apjs} and offsets between gas-phase and stellar metallicities \citep[][]{Menon2025} in high-$z$ galaxies, as well as in their metal-scaling relations.} {In fact, the enhanced galaxy compactness helps reproduce the observed MZSFR (Sec.~\ref{sec:mzsfr}) in our simulations: Under the old scheme with larger $R_\star$, the simulated galaxies are always too metal-poor, unless the outflow efficiency is significantly reduced ($M_{\rm out0}<10^{9}\ \rm M_\odot$), which then overpredicts the SFRD and CSMD. This trend is consistent with the negative correlation between (the offsets of) galaxy size (from the galaxy mass-size relation) and metallicity (from the MZSFR) found in observations \citep[e.g.,][]{Langeroodi2023size,Jia2025,Wang2025metal}.} 

\section{CSFH in the best-fit models for $\alpha_{\rm out}=0.5$ and 1}
\label{apdx:csfh_alt}

\begin{figure*}
    \centering
    \includegraphics[width=0.495\linewidth]{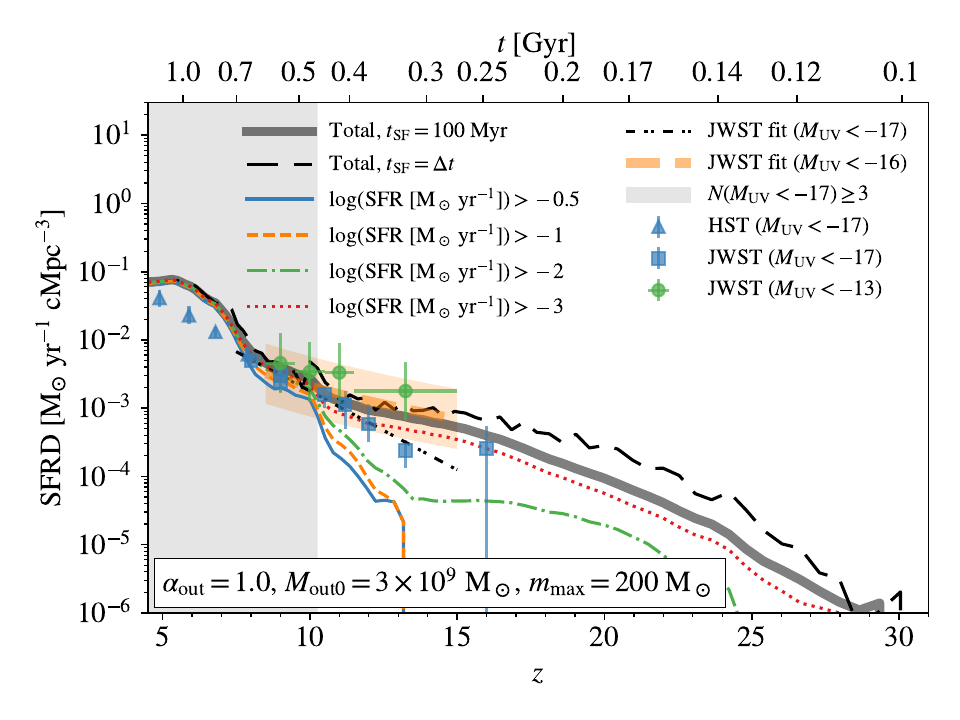}
    \includegraphics[width=0.495\linewidth]{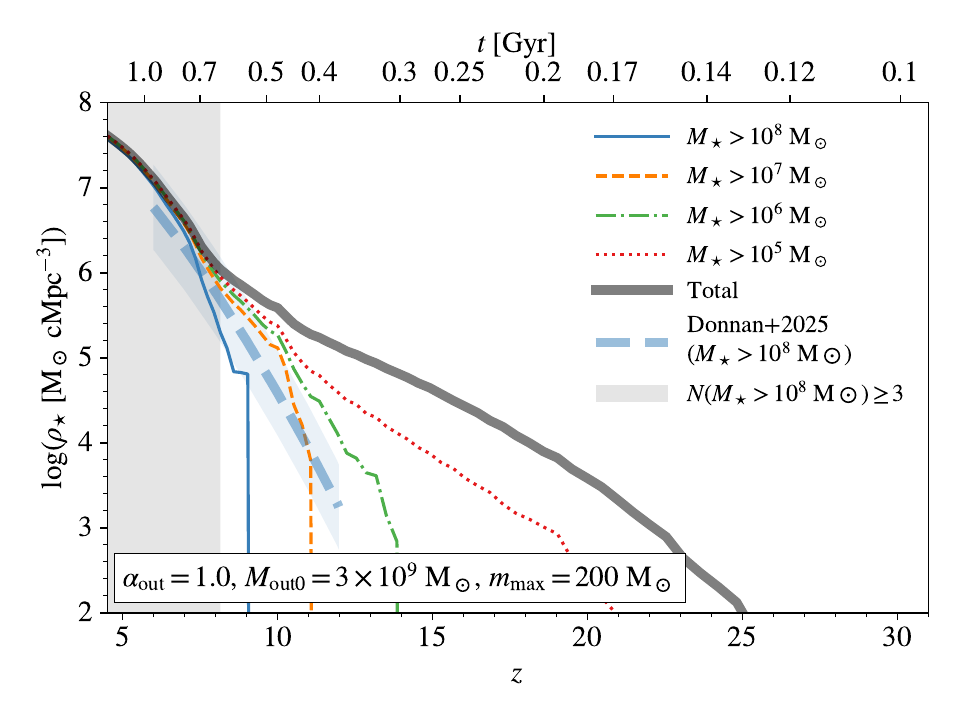}\\
    \includegraphics[width=0.495\linewidth]{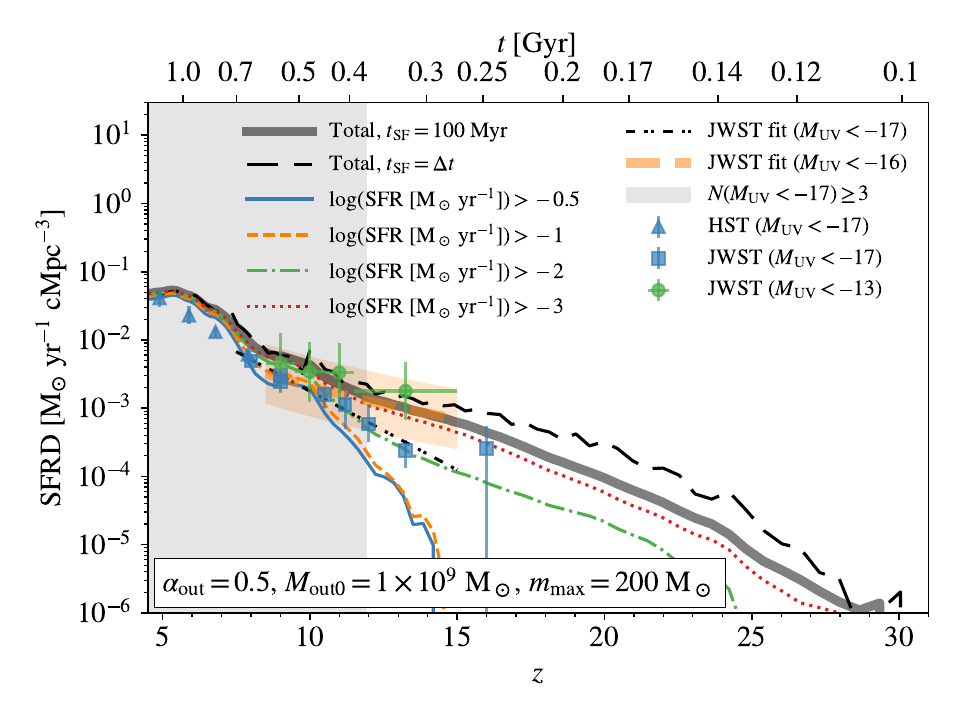}
    \includegraphics[width=0.495\linewidth]{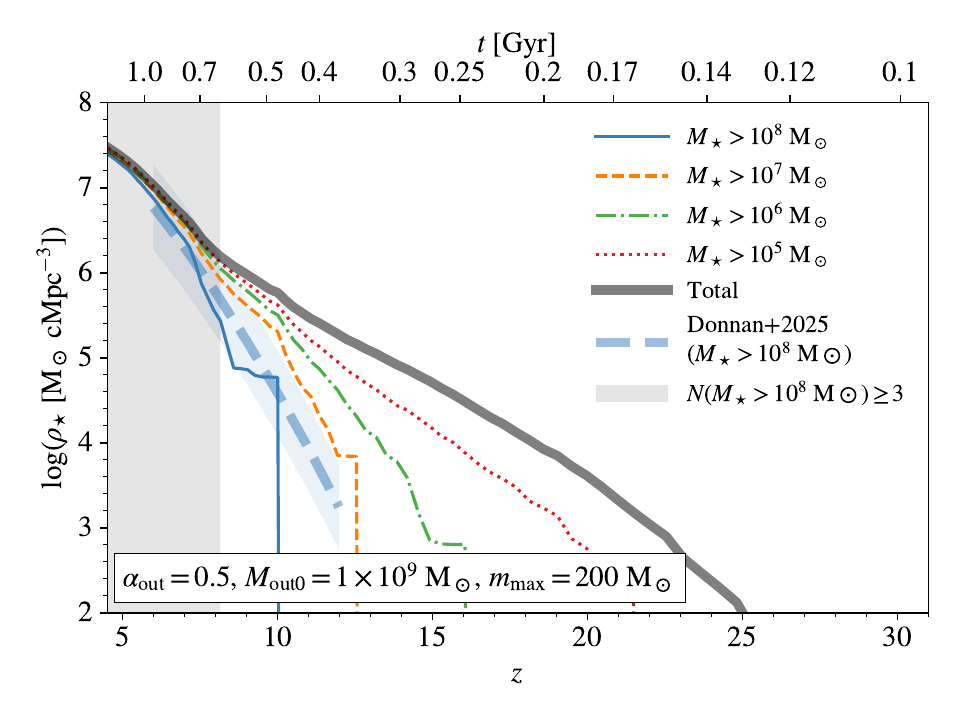}
    \caption{Same as Fig.~\ref{fig:csfh} but for the best-fit models for $\alpha_{\rm out}=1.0$ (top) and 0.5 (bottom). }
    \label{fig:csfh_alt}
\end{figure*}

In Fig.~\ref{fig:csfh_alt}, we show the CSFH in the best-fit models for $\alpha_{\rm out}=1.0$ (top) and 0.5 (bottom) in terms of SFRD (left) and CSMD (right). Here, the agreement between simulation results and observational data (detailed in Sec.~\ref{sec:sfh}) is worse than the case of the best-fit model for $\alpha_{\rm out}=0$ (Fig.~\ref{fig:csfh}). It turns out that star formation in detectable galaxies with $\rm SFR>10^{-0.5}\ \rm M_\odot\ yr^{-1}$ ($M_{\rm UV}\lesssim -17$) in these models is generally faster (slower) than that inferred from observations at $z\lesssim 8$ ($\gtrsim 10$). {The discrepancy between simulation results and SFRD measurements from observations is even larger (up to a factor of $\sim 10$ at $z\sim 13$) when fainter galaxies down to $M_{\rm UV}=-13$ are considered \citep{Chemerynska2025}.} A similar trend can be seen in the CSMD of galaxies with $M_{\star}>10^{8}\ \rm M_\odot$. The reason is that for $\alpha_{\rm out}>0$, the outflow efficiency is suppressed in massive halos with $M_{\rm h}\gtrsim M_{\rm out0}$ by Eq.~\ref{eq:gamma_out}, leading to a rapid rise of SFRD towards lower redshift where such halos become more important. On the other hand, the outflow efficiency is enhanced in smaller halos. Therefore, when $\alpha_{\rm out}$ increases, the contribution of faint galaxies with $\rm SFR\lesssim 10^{-0.5}\ \rm M_\odot\ yr^{-1}$ to the total SFRD is smaller, which becomes sub-dominant ($\lesssim 50\%$) and negligible at $z\lesssim 8$ for $\alpha_{\rm out}=0.5$ and 1. In contrast, faint galaxies still account for $\sim 30-50\%$ of the total SFRD at $z\sim 5-7$ in the best-fit model for $\alpha_{\rm out}=0$ (Fig.~\ref{fig:csfh}). 
Interestingly, in the $\alpha_{\rm out}=0.5$ best-fit model, the simulated SFRD at $z\sim 5$ matches very well with the observed value, while it is underestimated by a factor of $\sim 2$ in the $\alpha_{\rm out}=0$ best-fit model that shows good agreement with observations at higher $z$ up to $z\sim 13$. This hints for non-negligible evolution of outflow parameters with redshift at $z\lesssim 6$. 

\section{Type Ia supernovae}
\label{apdx:snia}

We implement a phenomenological model for Type Ia SNe based on the method of \citet{Deng2024}. This model is described by four parameters: the number of Type Ia SNe per unit stellar mass formed $\mathcal{N}_{\rm Ia}$, the upper bound $t_{\rm Ia,up}$, lower bound $t_{\rm Ia,low}$, and slope $\alpha_{\rm DTD}$ of the delay time distribution (DTD): $p_{\rm DTD}(t)\propto t^{-\alpha_{\rm DTD}}$, $\int_{t_{\rm Ia,low}}^{t_{\rm Ia, up}}p_{\rm DTD}(t)dt=1$. Here, we adopt $\mathcal{N}_{\rm Ia}=1.3\times10^{-3}\ \rm M_\odot^{-1}$, $t_{\rm Ia,up}=14$~Gyr, $t_{\rm Ia,low}=40$~Myr (corresponding to the lifetime of a $8\rm\ M_\odot$ star), and $\alpha_{\rm DTD}=1.12$ following \citet{Maoz2012,Vogelsberger2013}. With these choices of parameters, the solar abundance of Fe can be reproduced in MW-like galaxies at $z\sim 0$. 

To reduce computational cost and memory usage, a hybrid approach is adopted to keep track of (1) individual progenitors of Type Ia SNe and (2) progenitor populations. At each star formation timestep $i$, we first estimate the number of SN Ia progenitors expected to form as $N_{\rm Ia,P,est}^{i}=\mathcal{N}_{\rm Ia}\delta M_{\star}$ given the mass $\delta M_{\star}^{i}$ of stars formed in this step. If (1) $N_{\rm Ia,P,est}^{i}\le 10$, we sample the number $N_{\rm Ia,P}$ of SN Ia progenitors formed in this step from a Poisson distribution ($p_{N_{\rm Ia,P}}=\lambda^{N_{\rm Ia,P}}e^{-\lambda}/N_{\rm Ia,P}!$) with parameter $\lambda=N_{\rm Ia,P,est}^{i}$. Then, each progenitor is assigned a delay time $t_{\rm Ia}$ randomly drawn from the DTD. The clock of each progenitor is checked by \textsc{a-sloth} in each adaptive sub-timestep, and the SN Ia event is triggered when $t_{\rm Ia}$ has past since its formation. 

If (2) $N_{\rm Ia,P,est}^{i}>10$, we instead create a population of SN Ia progenitors. Thereafter, at each star formation timestep $l$ with $t_{l+1}>t_{\rm Ia,low}$, we estimate the number of Type Ia SNe expected to go off from this population as $N_{\rm Ia,est}=N_{\rm Ia,P,est}^{i}\int_{\max(t_{\rm Ia,low},t_{l}-t_{i})}^{t_{l+1}-t_{i}}p_{\rm DTD}(t)dt$. The actually number of SN Ia events is again drawn from a Poisson distribution with $\lambda=N_{\rm Ia,est}$. Such sampling of SN Ia events continues until the total number of Type Ia SNe sampled exceeds $N_{\rm Ia,P,est}^{i}$. 

The first method for $N_{\rm Ia,P,est}^{i}\le 10$ ensures that we do not create too many SN Ia progenitor populations for galaxies with low SFR, while the second method for $N_{\rm Ia,P,est}^{i}>10$ ensures that we do not track too many individual SN Ia progenitors in \textsc{a-sloth}. Each SN Ia event adds $e_{\rm SN}=10^{51}\ \rm erg$ to the energy budget of SNe $E_{\rm SN}^{i}$. Following \citet{Deng2024}, the `W7' model from \citet{Nomoto1997} is adopted for the SN Ia metal yields.

\end{appendix}

\end{document}